\documentclass[12pt]{article}
\usepackage{latexsym}
\usepackage{float}
\usepackage{amsmath,bm,amssymb}

\usepackage[font=footnotesize]{caption}
\usepackage{colordvi}
\usepackage{multicol,multirow}
\usepackage{color}
\usepackage{tabu}
\usepackage{booktabs}
\usepackage{rotating}
 \usepackage{url,hyperref}
\usepackage[ruled]{algorithm2e}
\usepackage{pdflscape}
\usepackage{subfigure}
\usepackage{graphicx}
\usepackage[utf8]{inputenc}
\usepackage{longtable}
\usepackage{tabularx}
\usepackage{array}
\usepackage[super,sort&compress]{natbib}
\def\doublespace{\baselineskip=22pt}
\usepackage{lscape}

\begin{document}
\doublespace
\baselineskip 2.8ex
\begin{center}
{\bf \LARGE Robust Bayesian high-dimensional variable selection and inference with the horseshoe family of priors}\\

\doublespace

{\bf \large Kun Fan$^1$, Srijana Subedi$^2$, Vishmi Ridmika Dissanayake Pathiranage$^2$ and  Cen Wu$^{2\ast}$}\\

	{ $^1$ Department of Health Data Science and Biostatistics, Peter O’Donnell Jr School of Public Health, UT Southwestern Medical Center, Dallas, TX}\\
	
{ $^2$ Department of Statistics, Kansas State University, Manhattan, KS \\

	}

\end{center}

{\bf $\ast$ Corresponding author}:
Cen Wu, wucen@ksu.edu\\

\noindent  {\bf\Large Abstract}\\

\noindent Frequentist robust variable selection has been extensively investigated in high-dimensional regression. Despite success, developing the corresponding statistical inference procedures remains a challenging task. Recently, tackling this challenge from a Bayesian perspective has received much attention. In literature, the two-group spike-and-slab priors that can induce exact sparsity have been demonstrated to yield valid inference in robust sparse linear models. Nevertheless, another important category of sparse priors, the horseshoe family of priors, including horseshoe, horseshoe+, and regularized horseshoe priors, has not yet been examined in robust high-dimensional regression by far. Their performance in variable selection and especially statistical inference in the presence of heavy-tailed model errors is not well understood. In this paper, we address the question by developing robust Bayesian hierarchical models utilizing the horseshoe family of priors along with an efficient Gibbs sampling scheme. We show that compared with competing methods with alternative sampling strategies such as slice sampling, our proposals lead to superior performance in variable selection, Bayesian estimation and statistical inference. In particular, our numeric studies indicate that even without imposing exact sparsity, the one-group horseshoe priors can still yield valid Bayesian credible intervals under robust high-dimensional linear regression models. Applications of the proposed and alternative methods on real data further illustrates the advantage of the proposed methods. \\


\newpage
\section{Introduction}

In bioinformatics studies, robust variable selection has been extensively developed and conducted to identify important genomics features associated with complex disease traits such as cancer outcomes \cite{wu2015selective}. The robustness of variable selection procedures lies in the robust loss functions, such as Huber loss, quantile check loss and rank based loss, among many others, that can downweigh the influence of outliers in obtaining shrinkage estimators compared to the non-robust penalized least square loss. In high-dimensional scenarios, as penalized estimation and variable selection are conducted simultaneously, statistical inference has played a critical role in ensuring the superiority of shrinkage estimators and identification results \cite{dezeure2015high}. Nevertheless, establishing robust variable selection methods with valid inferential procedures that can be validated on finite samples is challenging in frequentist studies \cite{fan2024seeing}. Surveys show that in the frequentist framework, robust variable selection with inference procedures have been mainly investigated under sparse linear models that cannot account for complicated data structure in cancer genomics studies \cite{fan2024seeing}.

The limitation of frequentist robust variable selection methods has motivated us to tackle the problem from an Bayesian perspective. Fully Bayesian analysis is well acknowledged to provide uncertainty quantification measures to enable exact statistical inference. Even on finite samples, as long as the full posterior distribution is accessible via MCMC, marginal credible intervals can be readily computed for inference. In literature, a vast amount of shrinkage priors have been proposed \cite{o2009review}. Representative ones include the spike-and-slab priors \cite{mitchell1988bayesian,george1993variable,bai2021spike} and horseshoe priors \cite{carvalho2010horseshoe,polson2010shrink,bhadra2019lasso} among many others. The spike-and-slab prior is a two-group model that treats zero and non-zero regression coefficients differently a priori, typically with a point mass at zero and a continuous density away from zero. In contrast, horseshoe priors form a one-group model that globally shrinks all regression coefficients toward zero, while allowing some to escape away from zero through the “shrink globally, act locally” mechanism \cite{polson2010shrink}. 

Although spike-and-slab priors are computationally more demanding than horseshoe priors, their two-group structure is consistent with the sparse nature of high-dimensional models. In particular, the recently proposed robust Bayesian LASSO with spike-and-slab priors (RBLSS) has been shown to provide valid inference procedures \cite{ren2023robust, fan2024seeing}. Built on the robust Laplace likelihood, it induces exact sparsity (i.e., exact zeros) in the posterior estimates. Empirical results demonstrate that, in addition to superior estimation and variable selection performance, it yields marginal Bayesian credible intervals with nominal coverage probabilities even under heavy-tailed model errors \cite{fan2024seeing}. 

A natural next question is whether the one-group horseshoe priors can yield superior shrinkage estimation, variable selection, and, in particular, valid statistical inference under robust hierarchical models. The main empirical argument from Fan et al. (2024) \cite{fan2024seeing} in support of RBLSS for achieving valid inference on finite samples is that the use of spike-and-slab priors induces exact sparsity, which aligns with the sparsity assumption in high-dimensional regression. Although one-group  horseshoe priors do not yield exact zero posterior estimates directly, published literature has shown that they can still provide valid inference under sparse normal means models \cite{van2017uncertainty}. This suggests that their favorable inferential properties may also extend to robust settings.

In this article, we address the above question by developing robust hierarchical models utilizing the horseshoe prior family, including horseshoe, horseshoe+ and regularized horseshoe priors \cite{carvalho2010horseshoe, bhadra2017horseshoe+, piironen2017sparsity} for Bayesian shrinkage estimation and inference. Robust likelihood functions are usually built based on heavy-tailed distributions or Bayesian nonparametrics methodology such as Dirichlet process mixture models \cite{kottas2009bayesian}. In addition,  it has been shown that the robust generalized framework can be formulated by replacing the likelihood with a robust loss function \cite{bissiri2016general}. In our study, we adopt the Laplace likelihood to impose robustness for important computational considerations. As the Laplace distribution can be expressed as a conditional scaled normal distribution \cite{kozumi2011gibbs}, fast Gibbs sampling with horseshoe priors can be readily extended to robust hierarchical models\cite{wand2011mean,makalic2015simple} . 

In fact, the benefits arising from the connection between a normal likelihood and an (asymmetric) Laplace likelihood for robust analysis have rarely been acknowledged, or at least explicitly discussed, in the published literature. We show two pieces of evidence here. First, in terms of estimation, unlike the striking contrast between the non-differentiable L1 loss and the smooth least squares loss in frequentist studies, Liu et al. (2024) \cite{liu2024spike} demonstrates that updates of sparse regression coefficients follows the soft-thresholding rule under spike-and-slab quantile LASSO utilizing the Laplacian working likelihood, which is rarely observed in frequentist robust penalization studies but similar to its non-robust counterpart in the Bayesian realm \cite{rovckova2018spike}. Second, in terms of inference, Bayesian LASSO with spike-and-slab priors, or BLSS, the non-robust cousin of RBLSS, yields valid credible intervals with nominal frequentist coverage probabilities under normal model errors \cite{fan2024seeing}. It is not surprising that RBLSS also performs well, given the excellent inferential properties already demonstrated by the non-robust BLSS. With our interest in robust Bayesian shrinkage estimation and particularly inference with the horseshoe family of priors, we adopt the Laplace working likelihood to leverage the strengths of horseshoe priors developed under the sparse normal means model for their robust counterpart.

We explore the relevant published literature, as summarized in Table \ref{hslist}, to further demonstrate how our study differs from existing work. Table \ref{hslist} is consistent with the survey by Bhadra et al. (2019) \cite{bhadra2019lasso} in that horseshoe prior sparse regressions are mostly focused on non-robust, normal settings and ignore heavy-tailed model errors. Although van der Pas et al. (2017) \cite{van2017uncertainty} have theoretically established uncertainty quantification properties of horseshoe prior regression under the sparse normal means model, it's theoretical results and numeric study are tailored for low dimension settings (e.g., $n$=400, $p$=20  $\&$ 200). In addition, Kohns and Szendrei (2024) \cite{kohns2024horseshoe} appears to be the only published study on robust horseshoe regression. Primarily focused on forecasting, it does not address high-dimensional variable selection and inference questions. Furthermore, our numerical studies show that the slice sampling proposed in  Kohns and Szendrei (2024) \cite{kohns2024horseshoe} leads to much slower computational speed and inferior performance in variable selection, shrinkage estimation and especially statistical inference compared to our methods. 

\begin{table}[h!]
	\def\arraystretch{1.5}
	\begin{center}
		\caption{An incomplete list of baseline Bayesian sparse models with horseshoe (HS) priors }
		\label{hslist}
		\centering
		\fontsize{9}{10}\selectfont{
			\begin{tabu}{X[3.0] X[2.0] X[2.2] X[2.4] X[2.8]}
			\hline
			
\textbf{Reference} & \textbf{Likelihood (robustness)} & \textbf{Prior} & \textbf{Inference} & \textbf{Implementation} \\
\hline
Carvalho et al. (2010) \cite{carvalho2010horseshoe} & Normal (No)     & HS              & Not examined                & MCMC (R) \\
Bhadra et al. (2017) \cite{bhadra2017horseshoe+} &     Normal (No)             &    HS+                &     Not examined              &   MCMC (R) \\
Piironen and Vehtari (2017) \cite{piironen2017sparsity} & Normal (No)     & Regularized HS              & Not examined                & NUTS (stan) \\
van der Pas et al. (2017) \cite{van2017uncertainty} & Normal (No)     & HS                & Yes (low-dim)     & MCMC (R) \cite{van2022package}  \\
Kohns and Szendrei (2024) \cite{kohns2024horseshoe} & ALD (Yes)        & HS                 & Not examined                & Slice sampling (R) \\

Fan et al. (2025) & ALD (Yes)        & HS, HS+, Regularized HS   & Yes (high-dim)    & Gibbs sampling (C++) \\
\hline
\end{tabu}
		}
	\end{center}
\end{table}

In summary, the major contribution of our study are as follows:

${\bullet}$ First, unlike the slice sampling adopted for horseshoe prior quantile regression \cite{kohns2024horseshoe}, we have developed Gibbs samplers for robust regression with the family of horseshoe priors, including horseshoe, horseshoe+ and regularized horseshoe priors, by utilizing the auxiliary mixture sampling strategy \cite{wand2011mean} that is the key to set up Gibbs samplers for non-robust horseshoe regression \cite{makalic2015simple}. We present detailed Gibbs samplers that have not been published in any existing work before.

${\bullet}$ Second, we demonstrate, for the first time, the variable selection performance of robust Bayesian regression with representative horseshoe priors. Different from the forecasting-oriented horseshoe prior quantile regression \cite{kohns2024horseshoe} where variable selection has not been investigated, we have conducted comprehensive simulations to assess the performance of robust Bayesian  regression with the family of horseshoe priors in variable selection and shrinkage estimation. Our results demonstrate that the horseshoe family of priors yields a distinctive variable selection pattern under aggressive shrinkage, even though it does not induce exact sparsity in the parameter estimates.

${\bullet}$ Third, among the first time, we report high-dimensional inference results for robust horseshoe regression to address the aforementioned question that whether the one-group robust horseshoe models can yield valid inference results when ``exact sparsity" does not hold. Our numerical analysis indicates that horseshoe type of priors, especially horseshoe and horseshoe+ priors, can produce promising inference results under robust sparse models commonly adopted in published inference studies \cite{dezeure2015high,fan2024seeing}, facilitating readily cross-comparison with existing high-dimensional inference results.

${\bullet}$ Fourth, we have investigated the connection between posterior sampling schemes and robust high-dimensional inference, which has been largely overlooked in published studies. Robust horseshoe models with alternative sampling strategies have been explored in horseshoe prior quantile regression \cite{kohns2024horseshoe}, or briefly mentioned in software documentation such as the \textit{Bayesreg} package \cite{makalic2016high},  but without further investigation. We show that, compared to these competitors, the proposed Gibbs sampling leads to superior performance especially in terms of accurate statistical inference in sparse robust models.

To ensure fast computation and reproducible research, we implement all the proposed and non-robust alternatives in C++ which will be available from Package \textit{pqrBayes} soon. We conclude the paper with a case study on high dimensional eQTL data.

\section{Statistical Methods}
\subsection{Robust Bayesian Regression}

Consider a linear model of the form:
\begin{equation}\label{eq1}
    y_i = \beta_0+\boldsymbol{x}_i^\top \boldsymbol{\beta} + \epsilon_i, \quad \text{for}\quad i= 1,\dots, n,
\end{equation}
where \( \{y_i\}_{i=1}^n \) denotes a scalar response variable, \( \{\boldsymbol{x}_i\}_{i=1}^n \) represents a known \( p \times 1 \) vector of covariates and random errors \( \epsilon_i \) are assumed to be $i.i.d.$. The intercept is denoted as $\beta_0$. Estimated regression coefficient \( \boldsymbol{\hat{\beta}} \) can be obtained by minimizing the following robust least absolute deviation (LAD) loss:
\begin{equation*}
\boldsymbol{\hat{\beta}}= \arg\min_{\boldsymbol{\beta}} \sum_{i=1}^{n} |y_i- \beta_0-\boldsymbol{x_i}^\top \boldsymbol{\beta}|,
\end{equation*}
which is resistant to outliers due to the L1 form of the LAD loss. More broadly, the LAD loss is a special case of the check loss function in quantile regression evaluated at the 50\%th quantile level. Model fitting procedures to obtain the corresponding minimizer have been widely discussed in frequentist literature. In our study, to motivate robust Bayesian estimation and inference, we consider the Bayesian median regression corresponding to model (\ref{eq1}) by assuming that the error \( \epsilon_i \) follow $i.i.d.$ Laplace distribution, or $ f(\epsilon_{i}|\tau) = \frac{\tau}{4} \text{exp}\{-\frac{\tau}{2}|\epsilon_{i}|\},  i=1,\dots, n,$ where $(\frac{\tau}{2})^{-1}$ is the scale parameter of the Laplace distribution. This can be viewed as the asymmetric Laplace distribution with 50$\%$th quantile level examined in Bayesian quantile regression\cite{yu2001bayesian}. Under the aforementioned probabilistic assumption on the model errors, the conditional distribution of \( y_i \), given the regression coefficients specified in model (\ref{eq1}), can be formally expressed as
\begin{equation}\label{eq4}
f(y_i\mid\beta_0,\boldsymbol{x_i}^\top \boldsymbol{\beta},\tau) = \frac{\tau}{4} \exp \left( -\frac{\tau}{2}|y_i-\beta_0-\boldsymbol{x_i}^\top \boldsymbol{\beta}| \right).
\end{equation}

To facilitate the setup of a robust Bayesian hierarchical model, we utilize an alternative representation of the Laplace distribution, by expressing it as a mixture of an exponential distribution and a scaled normal distribution \cite{kozumi2011gibbs}. In particular, by setting $\xi = \sqrt{8}$, the error term $\epsilon_{i}$, which follows a Laplace distribution with scale parameter $2\tau^{-1}$, admits the following hierarchical representation: $\epsilon_{i} = \tau^{-1} \xi \sqrt{v_{i}} z_{i}$, where $v_{i} \sim \text{Exp}(1)$ and $z_{i} \sim \text{N}(0,1)$. Accordingly, model (\ref{eq1}) can be reformulated as:
\begin{equation*}
    y_i = \beta_0+\boldsymbol{x}_i^\top \boldsymbol{\beta} + \tau^{-1} \xi \sqrt{v_{i}} z_{i}, \quad \text{for}\quad i= 1,\dots, n,
\end{equation*}
Let $\tilde{v}_{i} = \tau^{-1} v_{i}$, then $\tilde{v}_{i}\stackrel{ind}{\thicksim}\text{Exp}(\tau^{-1})$. The Bayesian hierarchal model has the following representation:

\begin{equation*}
\begin{aligned}
y_{i} &= \beta_0+\boldsymbol{x}_i^\top \boldsymbol{\beta}+\tau^{-1/2}\xi\sqrt{\tilde{v}_{i}}z_{i},\\
\tilde{v}_{i}|\tau&\stackrel{ind}{\thicksim}\text{Exp}(\tau^{-1}), \\
z_{i}&\stackrel{ind}{\thicksim}\text{N}(0,1).
\end{aligned}
\end{equation*}

\textbf{Remark}: As pointed out by Yang et al. (2016) \cite{yang2016posterior}, the above reprametrization of Laplace distribution leads to a working likelihood that matches its maximum likelihood estimation to the minimization of LAD loss under the model (\ref{eq1}). In addition to its widespread use in Bayesian regularized quantile regression, this formulation has recently been noted to bridge frequentist and Bayesian paradigms of robust variable selection \cite{liu2024spike}.

\subsection{Robust Bayesian regression with the horseshoe family of priors}

With high-dimensional regression, sparse priors play a critical role in ensuring superior performance in variable selection, shrinkage estimation and especially statistical inference. In fact, the limitation of adopting the aforementioned likelihood has been pointed out in Yang et al. (2016) \cite{yang2016posterior}, showing that with the (asymmetric) Laplce based working likelihood, a posterior variance adjustment step is needed to retain valid inference in low-dimensional settings. Although the variance correction is not available when $p > n$, it has been demonstrated that if the sparse prior is chosen carefully, such as the point-mass spike-and-slab priors, robust Bayesian variable selection can yield valid inference procedures that can be verified on finite samples \cite{fan2024seeing,ren2023robust}. An important question we attempt to address is whether the horseshoe family of priors can achieve comparable or even superior performance, particularly in inference under robust high-dimensional linear models.

\subsubsection{Robust Horseshoe prior regression}

Carvalho et al. (2010) \cite{carvalho2010horseshoe} and Polson and Scott (2010) \cite{polson2010shrink} have developed the horseshoe prior for continuous shrinkage and sparse signal recovery. The hierarchical form of the horseshoe prior under the robust likelihood specified in Section 2.1 can be expressed as follows.

\begin{equation}\label{eqq}
\begin{aligned}
\beta_{j} \mid s_j^2, \lambda^2 &\sim \text{N}(0, s_j^2 \lambda^2), \quad j = 1, \ldots, p, \\
s_j &\sim \mathcal{C}^{+}(0, 1), \quad j = 1, \ldots, p, \\
\lambda &\sim \mathcal{C}^{+}(0, 1),
\end{aligned}
\end{equation}
which slightly differs from the formulation under normal mean models since the variance parameter that exists in those models is no longer applicable here. The hierarchical model specifically utilizes half-Cauchy distributions for both the local scale parameter $s_j$ and the global scale parameter $\lambda$, characterized by the density function $p(x) = \frac{2}{\pi(1 + x^2)}, \quad x > 0$.

The horseshoe prior has a global-local shrinkage mechanism in that the global shrinkage parameter $\lambda$ applies uniform shrinkage over all $\beta_{j}$ to push regression coefficients toward zero globally. Meanwhile, the heavy-tailed half-Cauchy priors on the local scales $s_j$ allow individual coefficients to resist the shrinkage and remain large when necessary. By adjusting $\lambda$, one can control the level of sparsity in the model: a large value of $\lambda$ leads to weak shrinkage and diffuse priors for all variables, whereas letting $\lambda \to 0$ forces all coefficients $\beta_j$ toward zero.

To better understand how the shrinkage is imposed under the half-Cauchy priors which concentrate substantial mass near zero with heavy tails, we establish the following two proportions on the shrinkage factor $\kappa_j$ from robust horseshoe regression models. The details of derivations can be found in the Appendix.

\textbf{Proposition 1.1} Consider the full conditional posterior distribution of $\beta_j$ given the data and other parameters: $\beta_j \mid \cdot \sim \text{N}(\mu_{\beta_j}, \sigma_{\beta_j}^2)$, then the posterior mean can be expressed as
\begin{equation*}
\mu_{\beta_j} = (1 - \kappa_j) \hat{\beta}_j, 
\end{equation*}
where the shrinkage coefficient $\kappa_j$ is defined as $\kappa_j = \left( 1 + \lambda^2 s_j^2 a_j \right)^{-1} \in (0, 1)$, with $a_j = \sum_{i=1}^n \frac{x_{ij}^2}{\tau^{-1} \xi^2 \tilde{v}_i}$ and $\hat{\beta}_j = a_j^{-1}
\sum_{i=1}^n \frac{x_{ij}(y_i - \beta_0 - \sum_{k \ne j} x_{ik} \beta_k)}{\tau^{-1} \xi^2 \tilde{v}_i}.$

This decomposition illustrates how the shrinkage factor $\kappa_j$ governs the degree to which $\hat{\beta}_j$ is pulled towards zero. Specifically, when the scale parameters  $\lambda$ or $s_j$ approach 0, the shrinkage coefficient $\kappa_j $ goes to 1. Therefore, the posterior mean $\mu_{\beta_j} \to 0$ to induce almost zero regression coefficient. On the other hand, signals are retained, i.e., $\mu_{\beta_j} \to \hat{\beta}_j$ as $\lambda \to \infty$ or $s_j \to \infty$.


\textbf{Proposition 1.2} For the horseshoe prior hierarchical form specified in equations (\ref{eqq}), the density $\kappa_j \mid \lambda,\tau^{-1},\tilde{v}_i$ is given as
\begin{equation*}
p(\kappa_j \mid \lambda,\tau^{-1},\tilde{v}_i)
=
\frac{1}{\pi} \frac{k_j}{ \left( k_j^2 - 1 \right) \kappa_j + 1 }\frac{1}{ \sqrt{ \kappa_j (1 - \kappa_j) } },
\quad \kappa_j \in (0, 1).
\end{equation*}
where $k_j = \lambda \sqrt{ \sum_{i=1}^n \dfrac{x_{ij}^2}{\tau^{-1} \xi^2 \tilde{v}_i} }.$

Carvalho et al. (2010) \cite{carvalho2010horseshoe} and follow-up papers have demonstrated a close connection between the horseshoe shrinkage weights $(1 - \kappa_{j})$ as ``pseudo-posterior" probabilities and posterior inclusion probabilities (PIP) from the two-group mixture models such as spike-and-slab priors. They also suggest selecting all features with $\kappa_{j} > $ 0.5, although Piironen and Vehtari (2017) \cite{piironen2017sparsity} have found that such a criterion would fail in the case of a large number of correlated features. Nevertheless, a cutoff other than 0.5 may be adopted and it is still reasonable to consider the cutoff for determining selected features as a judgment call under the one-group Horseshoe priors. On the other hand, with two-group spike-and-slab priors, a median probability threshold of 0.5 is commonly used to compare with the PIPs for selecting features, which is typically not the case of a judgement call \cite{barbieri2004optimal,barbieri2021median,fan2024seeing}. The difference in interpreting and applying the cutoffs to determine selected features lies in dintinct meachnisms of imposing shrinkage between one-group and two-group sparse priors. 

The auxiliary mixture sampling techniques developed in the context of mean field variational Bayes \cite{wand2011mean} turn out to be crucial for setting up efficient Gibbs samplers under normal mean models \cite{makalic2015simple}. Specifically, via a auxiliary variable $\nu_j$, the half-Cauchy distribution that the local shrinkage parameter $s_j$ follows can be expressed as a scale mixture of inverse-Gamma distributions:
\begin{equation}\label{horse}
\begin{aligned}
s_j^2 \mid \nu_j &\sim \text{Inverse-Gamma}(1/2, 1/\nu_j), 
\quad j = 1, \ldots, p, \\
\nu_j &\sim \text{Inverse-Gamma}(1/2, 1).
\end{aligned}
\end{equation}
A similar hierarchical form for the global shrinkage parameter $\lambda$ can also be obtained. Combined, the robust Bayesian horseshoe prior regression can be hierarchically formulated as:

\begin{equation}
\begin{aligned}
y_i \mid \beta_0,\boldsymbol{x_i}, \boldsymbol{\beta}, \tilde{v}_i &\sim \text{N}(\beta_0+\boldsymbol{x_i}^\top\boldsymbol{\beta}, \tau^{-1}\xi^2\tilde{v}_i), \\
\beta_0&\sim\text{N}(0,\sigma^2_{\beta_0}),\\
\tilde{v}_{i}|\tau&\sim\text{Exp}(\tau^{-1}), \\
\tau &\sim \text{Gamma}(e, f),\\
\beta_j \mid s_j, \lambda &\sim \text{N}(0, \lambda^2 s_j^2),\\
s_j^2 \mid \nu_j &\sim \text{Inverse-Gamma}(1/2, 1/\nu_j), \\
\lambda^2 \mid \xi_1 &\sim \text{Inverse-Gamma}(1/2, 1/\xi_1), \\
\nu_j &\sim \text{Inverse-Gamma}(1/2, 1),\\
\xi_1 &\sim \text{Inverse-Gamma}(1/2, 1).\\
\end{aligned}
\end{equation}

\subsubsection{Robust Horseshoe+ prior regression}

The horseshoe+ prior proposed by Bhadra et al. (2017) \cite{bhadra2017horseshoe+} has the following hierarchical form:
\begin{equation}\label{horse+}
\begin{aligned}
s_j \sim \mathcal{C}^+(0, \phi_j), \\
\phi_j \sim \mathcal{C}^+(0, 1),\\
\lambda \sim \mathcal{C}^{+}(0, 1),
\end{aligned}
\end{equation}
where independent half-Cauchy priors centered at zero with an half-Cauchy mixing variable $\phi_j$ have been assigned on the local shrinkage hyperparameters $(s_1, \ldots, s_p)$. Compared with the original horseshoe hierarchical formulation in (\ref{eqq}), incorporating $\phi_j$ leads to an extra level of local shrinkage that makes $s_j$'s conditionally independent given $\phi_j$.

The horseshoe+ estimator is a refinement of the original horseshoe estimator, tailored for ultra-sparse signals. Bhadra et al. (2017) \cite{bhadra2017horseshoe+} have shown that horseshoe+ achieves superior performance by reducing the posterior mean squared error and accelerating posterior concentration under the Kullback--Leibler divergence. The extra level of hierarchy in (\ref{horse+}) induces a shrinkage prior with a sharper spike at zero, thus resulting in more aggressive shrinkage on noises near zero. Meanwhile, the nested Cauchy priors produce a heavier tail compared to the horseshoe prior, reducing shrinkage on larger signals. Together, it facilitates better recovery of strong signals and greater separation of signals and noise in ultra-sparse models. This is in spirit very close to the ``theoretically ideal'' and ``gold standard" spike-and-slab priors \cite{bai2021spike,rovckova2018bayesian,carvalho2010horseshoe}but has a computational advantage due to its one-group nature.

Marginalizing over $\phi_j$ results in the following density for $s_j$:
\begin{equation*}
p(s_j) = \frac{4}{\pi^2} \frac{\log(s_j)}{s_j^2 - 1}.
\end{equation*}

Propositions 1.1 and 1.2 on the shrinkage factor $\kappa_j$ under robust horseshoe regression models naturally extend to the horseshoe+ setting. Since the full conditional  distribution of $\beta_j$ remain unchanged, Proposition 1.1 holds for horseshoe+ prior as well. Derivations of the following proposition are available in the Appendix. 

\textbf{Proposition 1.3} For the horseshoe+ prior formulated in (\ref{horse+}), the density $\kappa_j \mid \lambda,\tau^{-1},\tilde{v}_i$ is given by
\begin{equation*}
p(\kappa_j \mid \lambda,\tau^{-1},\tilde{v}_i)
=
\frac{2}{\pi^2}
\frac{
\log \left( \sqrt{\frac{1 - \kappa_j}{\kappa_j  k_j^2} }\right)
}{
1 - \kappa_j \left(1 + k_j^2 \right)
}
\frac{k_j}{\sqrt{ \kappa_j(1 - \kappa_j)}
},
\quad \kappa_j \in (0, 1).
\end{equation*}
where $k_j = \lambda \sqrt{ \sum_{i=1}^n \dfrac{x_{ij}^2}{\tau^{-1} \xi^2 \tilde{v}_i} }.$

Again, by following the auxiliary mixture sampling \cite{wand2011mean} and resulting Gibbs sampling under sparse normal means models \cite{makalic2015simple}, we can express the robust Bayesian hierarchical model with horseshoe+ prior as:
\begin{equation*}
\begin{aligned}
y_i \mid \beta_0,\boldsymbol{x_i}, \boldsymbol{\beta}, \tilde{v}_i &\sim \text{N}(\beta_0+\boldsymbol{x_i}^\top\boldsymbol{\beta}, \tau^{-1}\xi^2\tilde{v}_i), \\
\beta_0 &\sim \text{N}(0, \sigma^2_{\beta_0}), \\
\tilde{v}_{i} \mid \tau &\sim \text{Exp}(\tau^{-1}), \\
\tau &\sim \text{Gamma}(e, f),\\
\beta_j \mid s_j, \lambda &\sim \text{N}(0, \lambda^2 s_j^2),\\
s_j^2 \mid \nu_j &\sim \text{Inverse-Gamma}(1/2, 1/\nu_j), \\
\nu_j \mid \phi^2_j &\sim \text{Inverse-Gamma}(1/2, 1/\phi^2_j),\\
\phi^2_j \mid \zeta_j &\sim \text{Inverse-Gamma}(1/2, 1/\zeta_j),\\
\zeta_j &\sim \text{Inverse-Gamma}(1/2, 1),\\
\lambda^2 \mid \xi_1 &\sim \text{Inverse-Gamma}(1/2, 1/\xi_1), \\
\xi_1 &\sim \text{Inverse-Gamma}(1/2, 1).\\
\end{aligned}
\end{equation*}

\subsubsection{Robust Bayesian regularized horseshoe regression}

The hierarchical models for Horseshoe and Horseshoe+ prior regression depend on extremely heavy tails of the half-Cauchy prior, which has infinite variance. Therefore, regression parameters with very large magnitude are not regularized at all and can still be sampled, leading to potential numeric instability in MCMC. In addition, there is no principled way to incorporate sparsity belief in specifying the global shrinkage parameters $\lambda$. To overcome the limitations, Piironen and Vehtari (2017) \cite{piironen2017sparsity} have proposed to regularize Horseshoe prior as follows:

\begin{equation}\label{eq5}
\begin{aligned}
\beta_j \mid s_j, \lambda, b &\sim \text{N}\left(0, \lambda^2 \tilde{s}_j^2\right), \quad
\tilde{s}^2_j = \frac{b^2 s_j^2}{b^2 + \lambda^2 s_j^2},\\
s_j &\sim \mathcal{C}^+(0, 1), \quad j = 1, \ldots, p,
\end{aligned}
\end{equation}
where $\tilde{s}^2_j$ is a modified local shrinkage parameter with a positive tuning parameter $b$. Under robust Horseshoe prior regression, Proposition 1.1 indicates that the relative amount of shrinkage can be expressed as $\kappa_j = \left( 1 + \lambda^2 s_j^2 a_j \right)^{-1} $. Correspondingly, the shrinkage factor for robust regression with regularized Horseshoe prior is $\tilde{\kappa}_j = \left( 1 + \lambda^2 \tilde{s}_j^2 a_j \right)^{-1}$. Therefore, if $b \to \infty$, regularized Horseshoe induces the same amount of shrinkage imposed by Horseshoe. On the other hand, $\tilde{\kappa}_j$=1 as $b \to 0$, so the regression coefficient is not penalized. When $0 < b < \infty$, regularized Horseshoe performs a more aggressive shrinkage estimation as $\tilde{\kappa}_j > \kappa_j $.

In fact, regularized Horseshoe has a conceptual connection to Bayesian elastic net \cite{li2010bayesian}, although it has not been explicitly mentioned in Piironen and Vehtari (2017) \cite{piironen2017sparsity}. Recall that a Bayesian elastic net prior is given as:
\[
\pi(\beta_j \mid \lambda_1, \lambda_2) \propto \exp\left( -\lambda_1 |\beta_j| - \lambda_2 \beta_j^2 \right),
\]
where $\lambda_1$ and $\lambda_2$ are positive tuning parameters associated with Laplacian and Gaussian priors, respectively. The hierarchical structure of regularized Horseshoe in (\ref{eq5}) can be viewed as replacing the Laplace prior using a Horseshoe prior while keeping the Gaussian kernel. Consequently, a regularized horseshoe prior can also be expressed as:
\[
\pi(\beta_j \mid s_j, \lambda, b) \propto \text{N}(0, \lambda^2 s_j^2) \, \text{N}(0, b^2) \propto \text{N}(0, \lambda^2 \tilde{s}_j^2),
\]
where a Gaussian prior $\text{N}(0, b^2)$ on $\beta_j$ is multiplied to the horseshoe prior in order to mimic the role that a ridge penalty plays in elastic net \cite{zou2005regularization}. Therefore, regularized Horseshoe is an extension of the original horseshoe prior to account for structured sparsity in terms of multicollinearity. Zou and Hastie (2005) \cite{zou2005regularization} have provided a detailed rationale on why incorporating ridge term in penalized variable selection can accommodate correlations among predictors, which has partially justified the advantage of analyzing correlated genomics features using regularized Horseshoe demonstrated in Piironen and Vehtari (2017) \cite{piironen2017sparsity}. The connection between regularized horseshoe prior and Bayesian elastic net is also important for addressing the central question in our study, that is, how essential is exact sparsity for facilitating exact robust Bayesian inference in high-dimensional regression.We will demonstrate more details in numerical studies of this paper. 

It has been recommended to specify either a fixed value or a prior distribution for $b$. In the absence of strong prior knowledge about the scale of the relevant coefficients, we generally advise assigning a prior to $b$ rather than fixing it. A common and practical choice is to place an inverse-gamma prior on $b^2$, such that $b^2 \sim \text{Inv-Gamma}(\frac{c}{2}, \frac{d}{2})$\cite{piironen2017sparsity}. 

Our limited literature mining indicates that the hierarchical model for robust Bayesian regression with regularized horseshoe priors has not been established yet. We specify the hierarchical model below: 
\begin{equation*}
\begin{aligned}
y_i \mid \beta_0,\boldsymbol{x_i}, \boldsymbol{\beta}, \tilde{v}_i &\sim \text{N}(\beta_0+\boldsymbol{x_i}^\top\boldsymbol{\beta}, \tau^{-1}\xi^2\tilde{v}_i), \\
\beta_0&\sim\text{N}(0,\sigma^2_{\beta_0}),\\
\tilde{v}_{i}|\tau&\sim\text{Exp}(\tau^{-1}), \\
\tau &\sim \text{Gamma}(e, f),\\
\beta_j \mid s_j, \lambda, b &\sim \text{N}(0, \lambda^2 s_j^2) \, \text{N}(0, b^2),\\
s_j^2 \mid \nu_j &\sim \text{Inverse-Gamma}(1/2, 1/\nu_j), \\
\lambda^2 \mid \xi_1 &\sim \text{Inverse-Gamma}(1/2, 1/\xi_1), \\
\nu_j &\sim \text{Inverse-Gamma}(1/2, 1),\\
\xi_1 &\sim \text{Inverse-Gamma}(1/2, 1),\\
b^2 &\sim \text{Inverse-Gamma}(c/2, d/2).\\
\end{aligned}
\end{equation*}

\subsection{Summary of all methods under comparison}

We summarize the proposed robust methods with the horseshoe family of priors as follows: robust Bayesian regression with the horseshoe prior (RBHS), robust Bayesian regression with the horseshoe+ prior (RBHS+), and robust Bayesian regression with the regularized horseshoe prior (RBRHS). Their non-robust counterparts are referred to as BHS, BHS+, and BRHS \cite{piironen2017sparsity,makalic2015simple}, respectively. A concise summary is presented in Table \ref{methods1} below. Detailed Gibbs samplers for posterior inferences for all six methods under comparison are shown in Section C of the Appendix. 


\begin{table}[h!]
	\def\arraystretch{1.2}
	\begin{center}
		\caption{Summary of all methods.}
		\label{methods1}
		\centering
		\fontsize{10}{12}\selectfont{
			\begin{tabu}{X[1.0] X[2.6] X[0.8] X[1.2] X[2.4] X[1.8]}
				\hline
				\textbf{Methods} & \textbf{Names} &\textbf{Robust} & \textbf{Likelihood} & \textbf{Variable Selection} & \textbf{Reference} \\ 
				\hline
				RBHS & Robust Bayesian regression with horseshoe prior & Yes & Laplacian & Horseshoe prior & newly proposed \\
				RBHS+ & Robust Bayesian regression with horseshoe+ prior & Yes & Laplacian & Horseshoe+ prior & newly proposed \\ 
				RBRHS & Robust Bayesian regression with regularized horseshoe prior & Yes & Laplacian & Regularized horseshoe prior & newly proposed \\
				BHS & Bayesian regression with horseshoe prior & No & Gaussian & Horseshoe prior & Carvalho et al. (2010)\cite{carvalho2010horseshoe} \\
				BHS+ & Bayesian regression with horseshoe+ prior & No & Gaussian & Horseshoe+ prior & Bhadra et al. (2017)\cite{bhadra2017horseshoe+} \\
				BRHS & Bayesian regression with regularized horseshoe prior & No & Gaussian & Regularized horseshoe prior & Piironen and Vehtari (2017)\cite{piironen2017sparsity} \\
				\hline
			\end{tabu}
		}
	\end{center}
\end{table}

\section{Simulation}

We perform a thorough evaluation of the robust Bayesian regression with horseshoe, horseshoe+, and regularized horseshoe priors, which are termed as RBHS, RBHS+ and RBRHS, respectively. They are non-robust alternatives are BHS, BHS+ and BRHS. Please refer to Section 2.3 for a detailed summary. The datasets are simulated based on model (\ref{eq1}) under two configurations: one with sample size \(n = 200\) and dimensionality \(p = 600\), and the other with a larger scale of \(n = 400\) and \(p = 600\). The predictors are drawn from multivariate normal distributions with (1) an autoregressive (AR(1)) correlation structure where the $i$th and $j$th genes have correlation $0.5^{|i - j|}$, and (2) a banded correlation structure where the $i$th and $j$th genes have correlation $0.5$ if $|i - j| = 1$, and zero otherwise. The true regression coefficient vector is defined as $\boldsymbol{\beta} = (\beta_1, \beta_2, \dots, \beta_{p})^\top$, where 15 components are nonzero and independently drawn from a $\text{Uniform}(0.4, 0.9)$ distribution, and the remaining entries are set to zero. The intercept is fixed at one, i.e., $\beta_0$=1. We consider five error distributions in simulation: $\text{N}(0,1)$ (Error 1), $t(2)$ (Error 2), $\text{Laplace}(0,1)$ (Error 3); 80\% $\text{N}(0,1)$ + 20\% $\text{N}(0,3)$ (Error 4), and $\text{Lognormal}(0,1)$ (Error 5). We additionally examine the scenario with $non-i.i.d.$ random errors by adopting the following data-generating model: 
\begin{equation*}
    y_i = 1+\boldsymbol{x}_i^\top \boldsymbol{\beta} + (1+x_{i2})\epsilon_i, \quad \text{for}\quad i= 1,\dots, n,
\end{equation*}
The $i.i.d.$ error terms $\epsilon_i$ in model (\ref{eq1}) are replaced with heteroscedastic errors of the form $(1 + x_{i2})\epsilon_i$, while the regression coefficients remain unchanged from the $i.i.d.$ error setting.

The simulation study is performed over 100 independent replicates. For each of the six Bayesian methods, posterior samples were obtained from 10,000 MCMC iterations, following the removal of the initial 5,000 iterations used as burn-in. In addition, a predictor is considered identified if its 95\% credible interval excludes zero. The identified features are subsequently used to calculate the number of true positives (TPs), false positives (FPs), F1 score (F1) and Matthews correlation coefficient (MCC). We use the posterior medians $\hat{\boldsymbol{\beta}}$ to compute the $L_1$ estimation error as $\sum_{j=1}^{p} \lvert \beta_j - \hat{\beta}_j \rvert$. 

\textbf{Variable selection and estimation} Table \ref{id0.1} shows the identification and estimation results under the AR(1) correlation structure with \((n, p) = (200, 600)\) across all the five error distributions. The three robust Bayesian procedures, RBRHS, RBHS+, and RBHS, consistently demonstrate superior performance over their non-robust counterparts (BRHS, BHS+, and BHS), especially under heavy-tailed and skewed errors. For instance, under the $t$(2) error (Error 2), all three robust methods demonstrate much better performance than their non-robust counterparts. RBHS+ achieves the highest identification accuracy with TPs of 8.182 (sd 3.332) and FPs of 0.009 (sd 0.095), followed by RBHS with TPs of 7.659 (sd 3.368) and FPs of 0.012 (sd 0.109), and RBRHS with TPs of 5.878 (sd 3.763) and FPs of 0.010 (sd 0.100). In contrast, the non-robust methods show inferior performance: BHS yields TPs of 3.390 (sd 3.128) with FPs of 0.159 (sd 2.234); BHS+ gives TPs of 3.350 (sd 3.144) with FPs of 0.182 (sd 2.860); and BRHS reports TPs of 3.074 (sd 3.037) with FPs of 0.100 (sd 1.906). In terms of F1 score and MCC, which provide a more comprehensive assessment of variable selection accuracy, the robust methods also outperform their non-robust counterparts. For example, under the $t$(2) error (Error 2), RBHS+ leads to the largest F1 score of 0.676 (sd 0.208) and an MCC of 0.722 (sd 0.160), while RBHS has a slightly smaller F1 score of 0.644 (sd 0.215) and an MCC of 0.696 (sd 0.167). RBRHS yields an F1 score of 0.515 (sd 0.270) and an MCC of 0.614 (sd 0.196). In comparison, the non-robust approaches result in lower metric values. Specifically, BHS attains an F1 score of 0.324 (sd 0.256) and an MCC of 0.489 (sd 0.187). BHS+ shows an F1 score of 0.320 (sd 0.258) and an MCC of 0.487 (sd 0.190), while BRHS records an F1 score of 0.297 (sd 0.253) and an MCC of 0.477 (sd 0.184). In terms of L1 error, Table~\ref{id0.1} clearly indicates that robust methods generally leads to smaller estimation errors. 

\begin{figure}[h!]
	\centering
	\includegraphics[angle=0,origin=c,width=0.85\textwidth]{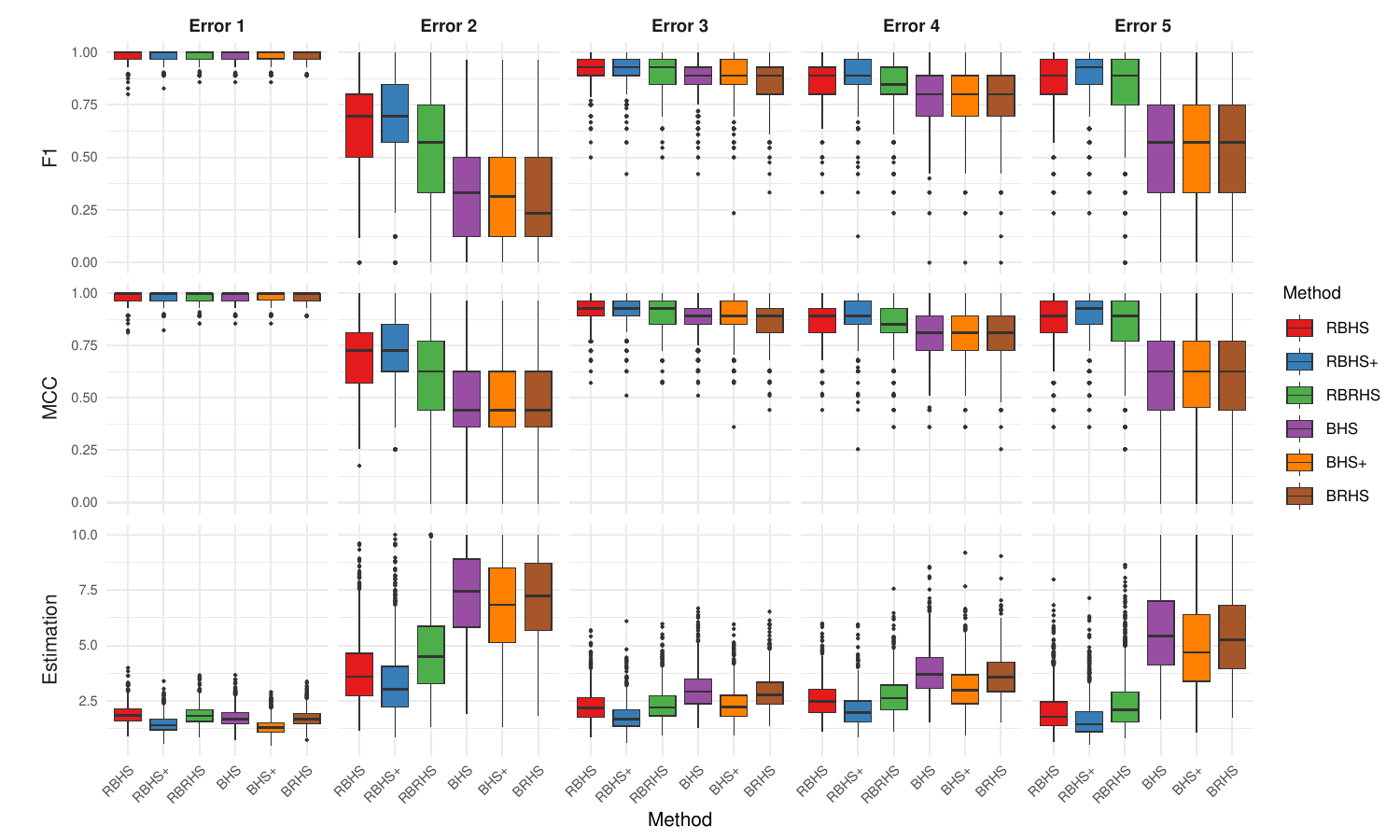}
	\caption{F1 score,  MCC score and Estimation error under AR(1) correlation, $i.i.d.$ error and $(n,p) = (200,600)$ corresponding to Table~\ref{id0.1}.}
	\label{fig:1}
\end{figure}

In addition to reporting numerical results in Table~\ref{id0.1}, we further present boxplots of the F1 score, MCC, and $L_1$ estimation error from Table~\ref{id0.1} in Figures \ref{fig:1}. The visualization provides a complementary summary of method performance under five $i.i.d.$ model error. Figure \ref{fig:1} displays the results for $i.i.d.$ errors. Under N(0,1) error (Error 1), all six methods perform well, with F1 and MCC scores tightly clustered near 1, and small estimation errors. As the error distributions become heavy-tailed or skewed (Errors 2–5), we observe a clear difference in performance. The robust RBHS, RBHS+, and RBRHS all maintain high identification accuracy (F1 and MCC) and low estimation error, with RBHS+ generally achieving the narrowest interquartile ranges (IQR) and lowest medians across all metrics. In contrast, non-robust BHS, BHS+, and BRHS demonstrate increased variability and deteriorated performance, particularly under the contaminated normal (Error 4) and log-normal error (Error 5) settings. These trends indicate the competitive performance of robust regression with horseshoe family of priors to outliers and heavy-tailed errors. 

In total, there are eight settings under different sample size ($n$=200 or 400, $p$=600), correlation structure (AR(1) vs. banded), and error types ($i.i.d.$ vs. non-$i.i.d.$). Except Table ~\ref{id0.1}, identification and estimation performance for the other seven settings is listed in Tables \ref{id0.2} through \ref{id0.8} in the appendix, respectively. Specifically, Tables \ref{id0.2}, \ref{id0.3}, and \ref{id0.4} correspond to $n = 200$, with Tables \ref{id0.2} and \ref{id0.3} focusing on banded correlation under $i.i.d.$ and non-$i.i.d.$ errors, respectively, and Table \ref{id0.4} reporting AR(1) with non-$i.i.d.$ errors. Tables \ref{id0.5}–\ref{id0.8} extend these scenarios to $n = 400$, with Tables \ref{id0.5} and \ref{id0.6} addressing $i.i.d.$ errors under AR(1) and banded structures, and Tables \ref{id0.7} and \ref{id0.8} covering non-$i.i.d.$ errors. Combined, these tables demonstrate the superior variable selection and estimation performance of three robust Bayesian approaches (RBHS, RBHS+, and RBRHS) under  heteroscedastic model errors. Besides, in the Appendix, we choose to visualize Table \ref{id0.3} using Figures \ref{fig:2}, which clearly illustrates that the robust methods are competitive.

\textbf{Remark on distinct identification and prediction patterns} Table \ref{id0.1} in the main text and Tables \ref{id0.2} -- \ref{id0.8} in the appendix suggests that, in general, the number of false positives is very low, often close to zero, which is sometimes achieved at the cost of identifying significantly fewer true positives. For example, under the $t$(2) error in Table \ref{id0.1}, the number of false positives for all three robust horseshoe models is nearly zero, but their true positives are also very low, which is only about half of the total number of true signals. This phenomenon suggests that horseshoe prior regressions aggressively shrinks regression coefficients toward zero, even if exact sparsity is not naturally accommodated. 

Under sparse normal mean models, Bhadra et al. (2017) \cite{bhadra2017horseshoe+} have theoretically established that horseshoe+ priors lead to better prediction performance in terms of posterior mean squared error compared to horseshoe priors. Table \ref{id0.1} shows that RBHS+ consistently results in the smallest L1 estimation errors. Prediction and estimation performance are typically related, although they are two different aspects assessing model fitting. Our results partially justify the superior prediction performance of horseshoe+ priors with the corresponding theoretical results established in Bhadra et al. (2017) \cite{bhadra2017horseshoe+}.

\begin{table} [ht!]
\def\arraystretch{1.2}
\begin{center}
\caption{Identification and estimation results of RBHS, RBHS+, RBRHS, BHS, BHS+ and BRHS for the datasets with AR(1) correlation under $i.i.d.$ errors, $n = 200$ and $p = 600$. mean(sd) of true positives (TP), false positives (FP), F1 score (F1), MCC and  L1 error based on 100 replicates.}\label{id0.1}
\centering
\fontsize{9}{10}\selectfont{
\begin{tabular}{ c c c c c c c c  }
\hline

  &Methods& TP &FP  & F1 & MCC &L1 error                 \\
\hline
Error 1  & RBHS   &14.663(0.618) &0.059(0.260) &0.986(0.024)&0.986(0.024)&1.887(0.424)\\
   N(0,1)          & RBHS+   &14.665(0.611) & 0.079(0.298) &0.986(0.024)&0.986(0.024)&1.456(0.376)\\ 
     & RBRHS    & 14.639(0.625) &0.050(0.240)&0.986(0.023)&0.986(0.023)&1.863(0.394)\\
         & BHS   &14.698(0.575) &0.039(0.209)&0.988(0.022)&0.988(0.022)&1.745(0.409) \\
         & BHS+&14.713(0.552)&0.051(0.246)&0.988(0.021)&0.988(0.021)&1.324(0.349)\\
         & BRHS&14.660(0.596)&0.023(0.150)&0.987(0.022)&0.987(0.022)&1.724(0.372)\\
   
\hline
Error 2         &     RBHS   &7.659(3.368)&0.012(0.109) &0.644(0.215)&0.696(0.167)&3.832(1.507)\\
   $t$(2)          & RBHS+   &8.182(3.330) & 0.009(0.095) &0.676(0.208)&0.722(0.160)&3.345(1.545)\\ 
     & RBRHS    & 5.878(3.763) &0.010(0.100)&0.515(0.270)&0.614(0.196)&4.766(1.910)\\
         & BHS   &3.390(3.128) &0.159(2.234)&0.324(0.256)&0.489(0.187)&8.633(33.848)\\
         & BHS+&3.350(3.144)&0.182(2.860)&0.320(0.258)&0.487(0.190)&8.123(35.855)\\
         & BRHS&3.074(3.037)&0.100(1.906)&0.297(0.253)&0.477(0.184)&8.083(28.971)\\
\hline
Error 3    & RBHS   &12.818(1.742)&0.030(0.187)&0.916(0.073)&0.919(0.067)&2.286(0.721)\\
 Laplace(0,1)   & RBHS+ &13.021(1.671) &0.034(0.221)&0.924(0.070)&0.927(0.065)&1.797(0.644)\\  
          & RBRHS  &12.578(1.918) &0.011(0.130)  &0.906(0.082)&0.911(0.075)&2.339(0.734)\\
           & BHS    &11.982(1.987) &0.071(0.286) &0.880(0.089)&0.886(0.080)&3.017(0.853)\\
           & BHS+&12.009(2.041)&0.061(0.256)&0.881(0.092)&0.887(0.082)&2.352(0.778)\\
         & BRHS&11.709(2.095)&0.040(0.211)&0.868(0.096)&0.876(0.085)&2.907(0.790)\\
         
\hline
Error 4  &RBHS &11.679(2.144)&0.027(0.162)&0.867(0.099)&0.875(0.087)&2.577(0.809)\\
 80$\%$N(0,1)
 &RBHS+ &11.978(2.065)&0.039(0.209)&0.880(0.095)&0.886(0.085)&2.093(0.729)\\
+20$\%$N(0,3) &RBRHS&11.124(2.462)&0.018(0.133)&0.840(0.121)&0.852(0.104)&2.746(0.860)\\
 &BHS&10.012(2.488)&0.085(0.319)&0.785(0.130)&0.804(0.108)&3.835(1.051)\\
 &BHS+&10.021(2.514)&0.071(0.293)&0.786(0.132)&0.805(0.110)&3.102(1.012)\\
 &BRHS&9.629(2.585)&0.046(0.232)&0.766(0.139)&0.788(0.115)&3.671(10.992)\\
 
\hline
Error 5  &RBHS &11.843(2.739)&0.005(0.071)&0.869(0.133)&0.879(0.115)&2.058(0.990)\\
 Lognormal(0,1) &RBHS+&12.251(2.538)&0.006(0.077)&0.888(0.120)&0.896(0.105)&1.700(0.891)\\
 &RBRHS&11.102(3.415)&0.007(0.095)&0.827(0.186)&0.848(0.149)&2.464(1.311)\\
 &BHS&6.290(3.722)&0.076(0.290)&0.545(0.257)&0.624(0.195)&5.632(1.963)\\
 &BHS+&6.257(3.749)&0.060(0.269)&0.542(0.260)&0.621(0.200)&4.980(2.039)\\
 &BRHS&5.859(3.731)&0.056(0.259)&0.514(0.265)&0.603(0.200)&5.479(1.928)\\
 
\hline 
\end{tabular} }
\end{center}
\centering
\end{table}
\clearpage

\textbf{High-dimensional inference for robust Bayesian horseshoe regression} We have also evaluated the six methods based on their ability to quantify uncertainty. The data have been generated from model(\ref{eq1}) under two settings: $(n, p) = (100, 500)$ and $(n, p) = (200, 1000)$. The predictors are generated from multivariate normal distributions with an autoregressive (AR(1)) correlation structure where the $i$th and $j$th genes have correlation $0.5^{|i - j|}$. The true regression coefficient vector is $\boldsymbol{\beta} = (1, 1.5, 2, 0, \ldots, 0)^\top$, where the first three entries are nonzero ($\boldsymbol{\beta}_1 = (1, 1.5, 2)^\top$) and the remaining coefficients are zero ($\boldsymbol{\beta}_2 = (0, \ldots, 0)^\top$). The intercept is set to zero. Two error distributions are considered for $\epsilon$: a standard normal distribution N(0,1) and a heavy-tailed $t(2)$ distribution. The simulation study is based on 1,000 replicates. We note that this data generating model has been adopted to assess validness of inference for two-group spike-and-slab priors \cite{song2023nearly,fan2024seeing}. Therefore, high-dimensional inference results from different studies can be directly compared. 

Table \ref{id1.3} shows 95\% empirical coverage probabilities for all the six methods under N(0,1) and $t$(2) model errors. With N(0,1) error, RBHS, RBHS+, and RBRHS yield coverage probabilities for $\beta_1$(=1), $\beta_2$(=1.5) and $\beta_3$(=2) ranging from 0.933 to 0.940, all close to the 0.95 nominal level, except that RBRHS leads to a low coverage probability 0.853 for $\beta_3$. We also observe competitive performance in inference for non-robust BHS, BHS+ and BRHS as the empirical coverage probabilities are slightly below 0.95, except that RBHS provides an under-coverage of 0.825 for $\beta_3$ with its marginal credible intervals. We also observe that the lengths of the credible intervals across all six methods are similar, fluctuating around 0.5. For the zero coefficients $\boldsymbol{\beta}_2$, Table \ref{id1.3} shows a super efficient phenomenon with 100\% coverage probability and much shorter interval length, which is consistent with the observation made for spike-and-slab priors under the same data generating model\cite{song2023nearly,fan2024seeing}.  

The pattern of inference results under heavy-tailed $t$(2) error is even more interesting. We now observe that robust methods distinguish from non-robust counterparts with more accurate coverage probabilities and narrower marginal credible intervals. Although heavy-tailed model errors lead to wider credible intervals compared to those under N(0,1) error for non-zero coefficients, RBHS, RBHS+ and RBRHS generally yield empirical coverage probabilities around 0.95 , with interval lengths below 1.  In contrast, the interval lengths for the non-robust BHS, BHS+, and BRHS are all much greater than 1, and their coverage probabilities deviate from the nominal level, mostly falling in the range of 0.7 to 0.8. Similar to the pattern under N(0,1) error, RBHS and RBHS+ maintain stable performance under heavy-tailed errors, RBRHS shows a notable lower coverage probability of 0.805 for $\beta_3$. 

The behavior of RBRHS under heavy-tailed model errors indicates a key challenge in high-dimensional inference: balancing satisfactory empirical coverage probability with the optimal interval length. Table \ref{id1.3} suggests that RBRHS struggles to achieve nominal coverage probabilities for some of non-zero coefficients while maintaining the narrowest interval length, compared to RBHS and RBHS+. We conjecture that this is due to the different mechanisms of constructing these priors. As shown in Section 2.2.3, regularized Horseshoe is tightly connected to Bayesian elastic net that accounts for structured sparsity in terms of multicolinearity \cite{li2010bayesian}. In high-dimensional regression, performing exact statistical inference becomes increasingly challenging when the model involves complex structures beyond predictor independence. Bayesian elastic net itself does not yield valid marginal credible intervals. In contrast, both the Horseshoe and Horseshoe+ priors have simpler structures that do not explicitly account for correlations or other forms of structured sparsity. Therefore, both lead to more accurate inference results under robust models. 

A further comparison among interval lengths shows consistency between the numerical results and prior theoretical results established in literature. Table \ref{id1.3} shows that RBHS+ typically yields shorter Bayesian credible intervals compared to RBHS, which can be explained by more aggressive shrinkage induced by construction of the Horseshoe+ prior. Note that RBRHS introduces additional amount of shrinkage through the Gaussian prior that mimics the ridge penality in elastic net, so the interval lengths are also smaller compared to RBHS. The pattern also suggests that the shrinkage imposed through the Gaussian prior under RBRHS is even more extreme than the amount  due to the additional layer of half-Cauchy prior on local shrinkage parameters under RBHS+. Besides, across all Bayesian approaches, the super-efficiency phenomenon persists for noise variables: the coverage probabilities for the zero coefficients remain at 1.000.

\begin{figure}[h!]
	\centering
	\includegraphics[angle=0,origin=c,width=0.55\textwidth]{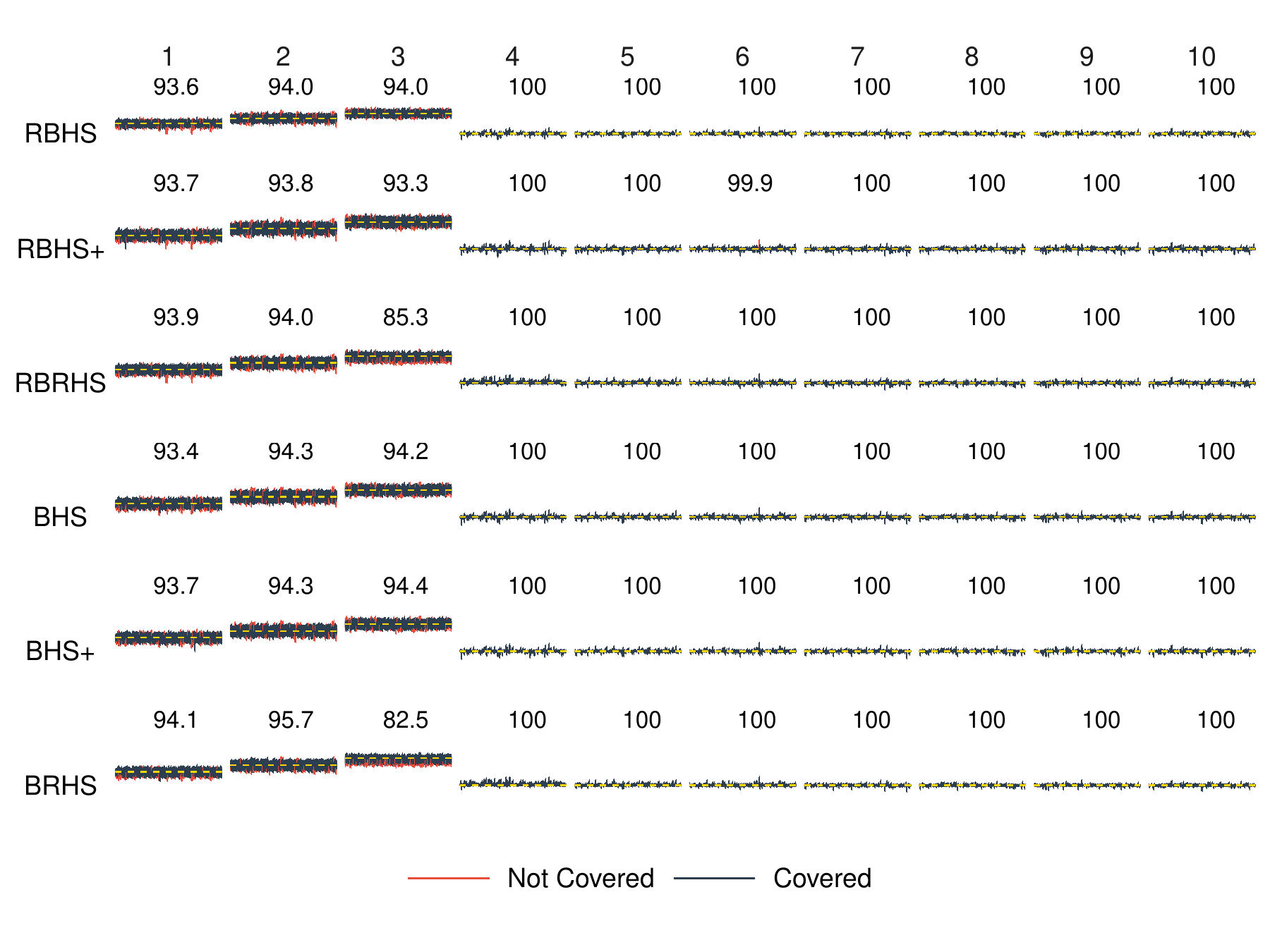}
	\caption{Credible intervals with empirical coverage probabilities under a 95\% nominal level are shown for the first 10 predictors, with the first 3 having true nonzero coefficients. Results are based on 1{,}000 simulated datasets under a homogeneous model with AR(1) covariate correlation, $(n, p) = (100, 500)$, and N(0,1) error.}
	\label{fig:4}
\end{figure}
\begin{figure}[h!]
	\centering
	\includegraphics[angle=0,origin=c,width=0.55\textwidth]{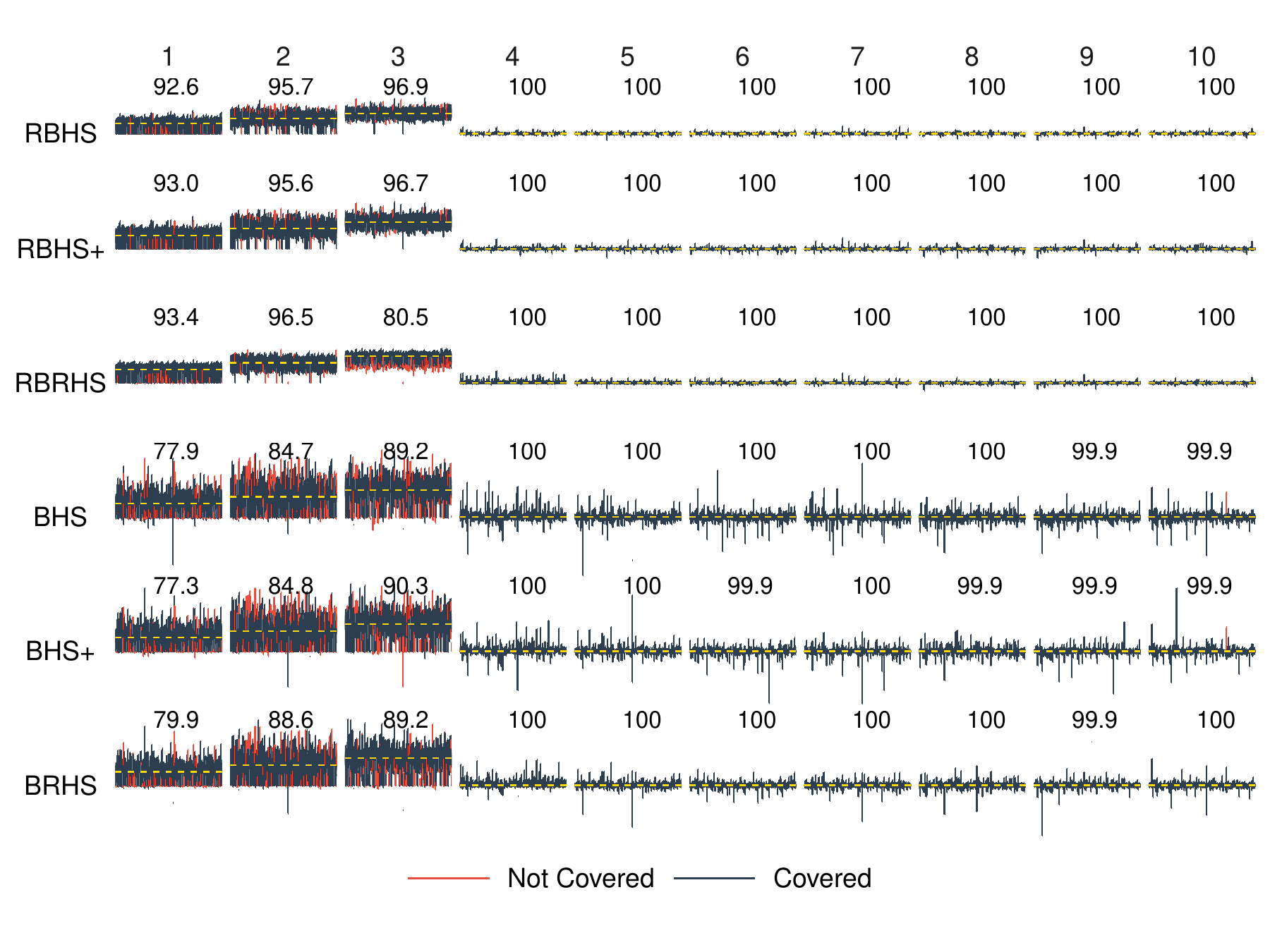}
	\caption{Credible intervals with empirical coverage probabilities under a 95\% nominal level are shown for the first 10 predictors, with the first 3 having true nonzero coefficients. Results are based on 1{,}000 simulated datasets under a homogeneous model with AR(1) covariate correlation, $(n, p) = (100, 500)$, and $t(2)$ error.}
	\label{fig:5}
\end{figure}

Figure \ref{fig:4} and Figure \ref{fig:5} provide visual summaries of the inference results reported in Table \ref{id1.3} , corresponding to the N(0,1) and $t(2)$ error settings, respectively. For illustration, we display results for the first 10 predictors, where the first 3 have nonzero coefficients and the remaining 7 are truly zero. The true coefficient values are indicated by black horizontal bars. Each plot overlays the intervals obtained from 1,000 replicates, with black segments denoting intervals that include the true value and red segments indicating non-coverage. The empirical coverage probability for each coefficient is shown above its corresponding bar. Figures \ref{fig:4} and \ref{fig:5} clearly highlight the inferential advantage of using robust Bayesian sparse regression with the horseshoe family of priors.

We also evaluate the inference performance under a much larger scale setting with  $(n, p)$= (200, 1000), and present detailed results in Figures \ref{fig:6} and \ref{fig:7} and Table \ref{id3.3} in the Appendix. We observe a similar pattern, where the robust Bayesian methods with the family of horseshoe priors continue to outperform their non-robust counterparts in terms of coverage and size of the interval. In particular, a larger sample size leads to slightly better empirical coverage probabilities and narrower marginal credible intervals under non-zero coefficients, reflecting improved stability and more accurate posterior inference. These results in supplementary demonstrate the utility of the proposed approaches for inference in both moderate- and high-dimensional settings.

\textbf{Additional Evaluations} We assess MCMC convergence using the potential scale reduction factor (PSRF) \cite{gelman1992inference, brooks1998general}, which indicates convergence when values are close to 1. In line with the recommendation by Gelman et al.\cite{gelman1995bayesian}, we adopt 1.1 as the convergence threshold. In our analysis, the PSRF values for all nonzero parameters fall below this benchmark after discarding burn-in samples, confirming that the chains have successfully converged to the stationary distribution. Figures \ref{fig:curve1} to  \ref{fig:curve3}  in the Appendix B illustrate the convergence behavior of the robust methods under $i.i.d.$ Error 2 with $n=200$ and $p=600$. To further demonstrate the advantage of our methods over published robust Bayesian variable selection methods including horseshoe prior quantile regression \cite{kohns2024horseshoe}, we have performed comparison studies as shown  in Table \ref{id0.9} and Table \ref{id1.00} in the Appendix. Table \ref{tab:comp_cost_ar1_t2} in the Appendix B shows the computational time and further establishes the advantage of proposed methods. 

\begin{table}[ht!]
\def\arraystretch{1.2}
\begin{center}
\caption{95\% empirical coverage probabilities assessed over 1000 replications for RBHS, RBHS+, RBRHS, BHS, BHS+ and BRHS using the datasets with AR(1) covariates, $n = 100$ and $p = 500$.}
\label{id1.3}
\centering
\fontsize{9}{11}\selectfont{
\begin{tabular}{ c c c c c c c c }
\hline
&&&&Methods\\
\cline{3-8}
  && RBHS & RBHS+ & RBRHS & BHS & BHS+ & BRHS \\
\hline
N(0,1) & Coverage of $\boldsymbol{\beta_1}$ \\
& $\beta_1$ & 0.936 & 0.937 & 0.939 & 0.934 & 0.937 & 0.941 \\
& $\beta_2$ & 0.940 & 0.938 & 0.940 & 0.943 & 0.943 & 0.957 \\
& $\beta_3$ & 0.940 & 0.933 & 0.853 & 0.942 & 0.944 & 0.825 \\
& Average length \\
& $\beta_1$ & 0.500 & 0.496 & 0.472 & 0.491 & 0.486 & 0.460 \\
& $\beta_2$ & 0.552 & 0.550 & 0.515 & 0.546 & 0.542 & 0.503 \\
& $\beta_3$ & 0.492 & 0.489 & 0.470 & 0.487 & 0.484 & 0.464 \\
\cline{2-8}
& Coverage of $\boldsymbol{\beta_2}$ & 1.000 & 1.000 & 1.000 & 1.000 & 1.000 & 1.000 \\
& Average length & 0.101 & 0.100 & 0.090 & 0.273 & 0.094 & 0.076 \\
\hline
$t$(2) & Coverage of $\boldsymbol{\beta_1}$ \\
& $\beta_1$ & 0.926 & 0.930 & 0.934 & 0.779 & 0.773 & 0.799 \\
& $\beta_2$ & 0.957 & 0.956 & 0.965 & 0.847 & 0.848 & 0.886 \\
& $\beta_3$ & 0.969 & 0.967 & 0.805 & 0.892 & 0.903 & 0.892 \\
& Average length \\
& $\beta_1$ & 0.917 & 0.891 & 0.793 & 1.230 & 1.211 & 1.166 \\
& $\beta_2$ & 0.972 & 0.943 & 0.784 & 1.506 & 1.491 & 1.400 \\
& $\beta_3$ & 0.808 & 0.795 & 0.720 & 1.328 & 1.321 & 1.251 \\
\cline{2-8}
& Coverage of $\boldsymbol{\beta_2}$ & 1.000 & 1.000 & 1.000 & 1.000 & 1.000 & 1.000 \\
& Average length & 0.101 & 0.080 & 0.083 & 0.273 & 0.210 & 0.206 \\
\hline
\end{tabular}}
\end{center}
\centering
\end{table}

\newpage

\section{Real Data Analysis}

We apply all the methods under comparison to analyze an expression quantitative trait locus (eQTL) dataset generated using 120 twelve-week-old male offspring of laboratory rats for a study to analyze RNA from the eyes of these animals based on microarrays containing over 31,042 probe sets \cite{scheetz2006regulation}. Following Huang et al. (2008)\cite{huang2008adaptive} and Wang et al. (2012) \cite{wang2012quantile}, we have performed two pre-processing steps to filter out genes that are not expressed or lack variations. First, we have excluded any probe whose maximum expression across the 120 rats fall below the 25th percentile of all expression values; and second, we have excluded any probe whose expression range across the 120 rats is less than 2. The preprocessing step leads to a total of 3,057 probes. We are interested in detecting gene expressions that are associated with gene TRIM32 (represented by probe 1389163\_at) which has been reported to cause retinal disorders in humans. As noted by Scheetz et al. (2006)\cite{scheetz2006regulation}, any genetic element demonstrated to influence the expression of a particular gene or gene family implicated in a specific disease serves as a strong candidate for involvement in that disease—either as a primary contributor or as a genetic modifier. We rank all remaining probes based on their coefficient of variation (CV) and select the top 300 probes with the highest CV values. These selected probes are subsequently kept for downstream analysis.

Specifically, RBHS, RBHS+, and RBRHS, along with their non-robust counterparts, BHS, BHS+, and BRHS, have been adopted for analysis. The 95\% Bayesian credible intervals have been used to determine whether probes are selected or not. We present the number of selected probes and their pairwise overlaps of all the six methods in Table \ref{tab:gene_overlap}. The diagonal entries of Table \ref{tab:gene_overlap} are the total number of nonzero coefficients identified by each method, while the off-diagonal entries represent the number of overlapping probes selected by each pair of methods. Since all the six methods belong to the horseshoe family, it is not surprise to observe that their findings show certain degree of overlap. In addition, some findings are method-specific, reflecting differences in variable selection sensitivity between robust and non-robust models.

We have also assessed the prediction performance of all the six methods through multi-splitting of the microarray data. Specifically, we perform 50 random data splits, each consisting of 80 randomly selected rats for training and the remaining 40 rats for testing. For each of the 50 random partitions, we fit models on the training data and evaluate predictive accuracy on the held-out test set using mean absolute deviation (MAD), defined as $\text{MAD} = \frac{1}{40} \sum_{i=1}^{40} \left| y_i - \hat{y}_i \right|$. Table \ref{tab:real_data} lists the average number of nonzero regression coefficients (Ave \# nonzero) obtained across 50 random data partitions, along with the corresponding average MADs. The standard errors for both summary statistics are also provided. Overall, the results show that the robust methods (RBHS, RBHS+, RBRHS) achieve much lower prediction errors compared to the non-robust methods (BHS, BHS+, BRHS), while also maintaining a more parsimonious model with smaller model size. For example, RBHS achieves a prediction error of 0.098 (sd 0.013) with 16.620 (sd 6.227) predictors. On the other hand, the non-robust BHS yields a prediction error of 0.790 (sd 2.239) with 52.580 (sd 36.331) predictors. BHS results in a substantially larger number of selected models and greater variance in model size, further demonstrating heterogeneity of the data and benefit of robust Bayesian high-dimensional variable selection. 

As with any variable selection method, different repetitions may identify different subsets of key predictors. In Table \ref{tab:frequency_table}, the columns list the probes selected using the complete dataset, along with how frequently these probes appeared in the final models across 50 random partitions for the six methods: RBHS, RBHS+, RBRHS, BHS, BHS+, and BRHS. The probes are arranged in order of decreasing frequency within each method. From Table \ref{tab:frequency_table}, we observe that certain probes, such as 1368778\_at, are consistently selected with high frequency across both robust (RBHS, RBHS+, RBRHS) and non-robust (BHS, BRHS) methods, indicating their overall selection stability. In contrast, probes such as 1370539\_at, 1370810\_at, 1370056\_at, 1370359\_at, and 1370946\_at show high selection frequencies exclusively under robust method but are rarely or never selected by non-robust approaches, reflecting the unique sensitivity of robust frameworks in uncovering signals that may be masked by outliers or model misspecification. This suggests that while some key predictors are universally robust and consistently detected across methods, others are uniquely captured by robust modeling, offering complementary insights and highlighting the added value of robust approaches in high-dimensional variable selection.

\begin{table}[htbp]
\centering
\caption{Variable selection results for the case study. The numbers of probes identified by different Bayesian methods and their overlaps.}
\begin{tabular}{lcccccc}
\hline
        & RBHS & RBHS+ & RBRHS & BHS & BHS+ & BRHS \\
\hline
RBHS    & 18   & 9     & 6   & 11   & 3    & 5   \\
RBHS+   &      & 23    & 5     & 10   & 3   & 2   \\
RBRHS   &      &       & 25    & 9  & 7  & 9   \\
BHS     &      &       &       & 32 & 4   & 6 \\
BHS+    &      &       &       &     & 13  & 3   \\
BRHS    &      &       &       &     &      & 15  \\
\hline
\end{tabular}
\label{tab:gene_overlap}
\end{table}

\begin{table}[htbp]
	\centering
	\caption{Multi-split analysis of the microarray dataset }
	\label{tab:real_data}
	\begin{tabular}{lccc}
		\hline
		&  & \multicolumn{2}{c}{Random partition} \\
		\cline{3-4}
		Method & Model size & Average model size & Prediction error \\
		\hline
		RBHS               & 18 & 16.620(6.227) &0.098(0.013)  \\
		RBHS+              & 23 & 15.860(6.719) &0.099(0.013)  \\
		RBRHS              & 25 & 26.320(7.113) & 0.097(0.014) \\
		BHS                & 32 & 52.580(36.331) &0.790(2.239)  \\
		BHS+               & 13 & 23.020(14.098) & 0.115(0.023) \\
		BRHS  & 15 & 49.280(34.450) & 0.485(0.817) \\
		
		\hline
		\end{tabular}
	\end{table}

\begin{table}[htbp]
\def\arraystretch{1.6}
\centering
\caption{Frequency table for the real data}
\resizebox{\textwidth}{!}{%
\begin{tabular}{cccccccccccc}
\hline
\multicolumn{2}{c}{RBHS} & \multicolumn{2}{c}{RBHS+} & \multicolumn{2}{c}{RBRHS} & \multicolumn{2}{c}{BHS} & \multicolumn{2}{c}{BHS+} & \multicolumn{2}{c}{BRHS} \\
\hline
Probe & Frequency & Probe & Frequency & Probe & Frequency & Probe & Frequency & Probe & Frequency & Probe & Frequency \\
\hline
1368778\_at & 42 & 1368778\_at & 42 & 1370539\_at & 50 & 1368778\_at & 49 & 1369268\_at & 25 & 1368778\_at & 50 \\
1367555\_at & 34 & 1367555\_at & 34 & 1389573\_at & 48 & 1384681\_at & 39 & 1369654\_at & 24 & 1384681\_at & 41 \\
1368353\_at & 27 & 1397361\_x\_at & 23 & 1370810\_at & 45 & 1367555\_at & 36 & 1369144\_a\_at & 23 & 1369202\_at & 31 \\
1397361\_x\_at & 24 & 1368353\_at & 21 & 1370056\_at & 42 & 1397361\_x\_at & 32 & 1389573\_at & 15 & 1369268\_at & 27 \\
1384681\_at & 22 & 1369248\_a\_at & 20 & 1370359\_at & 37 & 1369268\_at & 31 & 1369248\_a\_at & 14 & 1386855\_at & 18 \\
1368487\_at & 18 & 1385477\_at & 20 & 1370946\_at & 34 & 1369202\_at & 29 & 1370056\_at & 12 & 1389573\_at & 16 \\
1369268\_at & 18 & 1370349\_a\_at & 12 & 1368778\_at & 33 & 1369248\_a\_at & 24 & 1397959\_at & 12 & 1378347\_at & 15 \\
1369248\_a\_at & 13 & 1368438\_at & 11 & 1369628\_at & 18 & 1370349\_a\_at & 24 & 1383052\_a\_at & 11 & 1370359\_at & 8 \\
1370349\_a\_at & 13 & 1386552\_at & 11 & 1369268\_at & 14 & 1368353\_at & 23 & 1379645\_at & 10 & 1370539\_at & 8 \\
1369202\_at & 11 & 1376226\_at & 10 & 1383052\_a\_at & 14 & 1376226\_at & 22 & 1377321\_at & 6 & 1387283\_at & 8 \\
1378366\_at & 10 & 1371182\_at & 7 & 1369502\_a\_at & 10& 1382749\_at & 19 & 1379132\_at & 5 & 1369492\_at & 7 \\
1371960\_at & 8 & 1369268\_at & 6 & 1387283\_at & 9 & 1389573\_at & 16 & 1378904\_at & 3 & 1397211\_at & 7 \\
1388746\_at & 7 & 1377774\_at & 4 & 1369248\_a\_at & 9 & 1368826\_at & 15 & 1381964\_at & 3 & 1377321\_at & 6 \\
1389573\_at & 7 & 1378518\_at & 4 & 1379023\_at & 8 & 1382340\_at & 14 & & & 1392704\_at & 6 \\
1370669\_a\_at & 6 & 1383052\_a\_at & 4 & 1384681\_at & 8 & 1387060\_at & 14 & & & 1385286\_at & 3 \\
1377762\_at & 5 & 1393773\_at & 4 & 1369202\_at & 8 & 1368718\_at & 13 & & & & \\
1377774\_at & 5 & 1398695\_at & 4 & 1369144\_a\_at & 7 & 1375118\_at & 12 & & & & \\
1384293\_at & 1 & 1375212\_at & 3 & 1369492\_at & 6& 1395714\_at & 12 & & & & \\
& & 1387060\_at & 3 & 1394972\_at & 6& 1397959\_at & 12 & & & & \\
& & 1389444\_at & 3 & 1395542\_at & 5 & 1386552\_at & 11 & & & & \\
& & 1379023\_at & 2 & 1395956\_at & 4 & 1388746\_at & 11 & & & & \\
& & 1395546\_at & 2 & 1368718\_at & 3 & 1392860\_at & 11 & & & & \\
& & 1370669\_a\_at & 1 & 1369654\_at & 2 & 1380371\_at & 10 & & & & \\
& & & & 1378765\_at & 2 & 1380669\_at & 8 & & & & \\
& & & & 1396839\_at & 1 & 1387283\_at & 8 & & & & \\
& & & & & & 1394118\_at & 8 & & & & \\
& & & & & & 1378866\_at & 7 & & & & \\
& & & & & & 1390328\_at & 7 & & & & \\
& & & & & & 1395956\_at & 7 & & & & \\
& & & & & & 1389986\_at & 6 & & & & \\
& & & & & & 1375921\_at & 4 & & & & \\
& & & & & & 1387144\_at & 4 & & & & \\
\hline
\end{tabular}
}
\label{tab:frequency_table}
\end{table}

\newpage
\section{Discussion}

In this study, we have comprehensively assessed the performance of robust Bayesian sparse regression with the horseshoe family of priors. Our numeric results indicate that the proposed RBHS, RBHS+ and RBRHS are superior in terms of variable selection, shrinkage estimation and statistical inference compared with their non-robust alternatives in the presence of heavy-tailed model errors. In particular, RBHS and RBHS+ demonstrate competitive performance in yielding marginal credible intervals with nominal coverage probabilities in high-dimensions. Our study has shown that even though horseshoe family of priors cannot lead to exact zero parameter estimates, or exact sparsity, it can still maintain valid inference under sparse linear models. While this interesting phenomenon is seemingly not consistent with the previous findings that exact sparsity facilitates valid inference in high-dimensional scenarios \cite{fan2024seeing}, a closer examination of the numerical study suggests otherwise. Table \ref{id0.1} and tables alike demonstrate that the number of false features identified by all the six methods are very close to 0, sometimes at the cost of reducing true positive findings for a uniform shrinkage on all regress coefficients, suggesting that horseshoe family of priors tends to encourage sparse findings even exact 0 posterior estimates are not obtained. 

A similar observation in inference can be made in penalized generalized estimation equation (PGEE) for longitudinal data in frequentist studies \cite{wang2012penalized}. Although the Newton-Raphson model fitting procedure cannot yield exact zero estimates for the PGEE estimator, Wang et al. (2012) \cite{wang2012penalized} have established asymptotic distribution of PGEE estimator and validated the inference procedure on finite samples. Nevertheless, beyond the linear models, longitudinal studies show that lacking exact sparsity cause issues in estimation by treating group selection as sparse-group selection \cite{zhou2019penalized,zhou2022sparse}. Therefore, robust Bayesian estimation and inference with the horseshoe family of priors worth further exploration under more complicated sparse structures. 

\section{Conflict of Interest}
The authors declare that there is no conflict of interest.

\section{Acknowledgments}
This work was partially supported by an Innovative Research Award from the Johnson Cancer Research Center at Kansas State University.

\bibliography{references}

\newpage
\appendix
\section{Additional simulation results}
\subsection{Variable selection and estimation}
\begin{figure}[h!]
	\centering
	\includegraphics[angle=0,origin=c,width=0.85\textwidth]{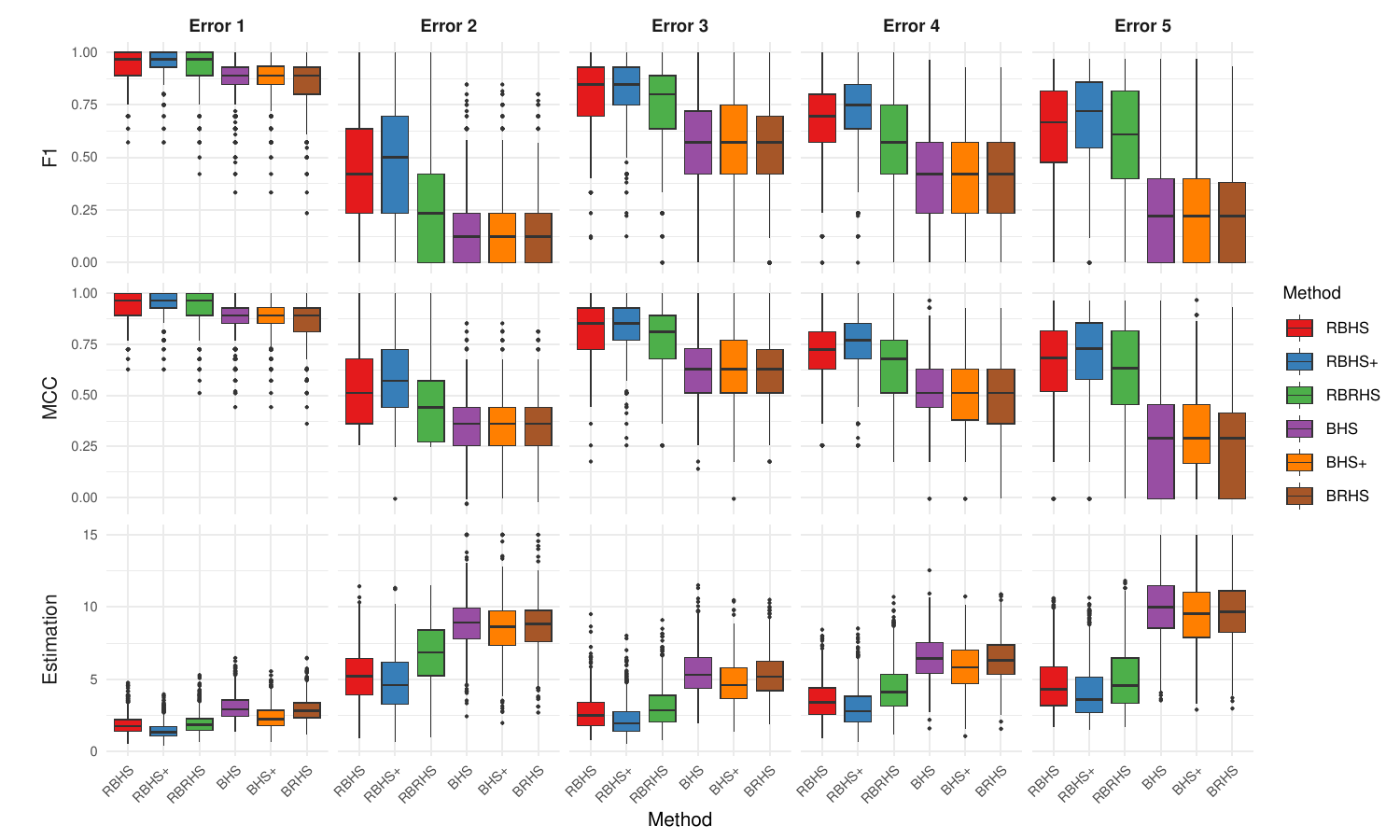}
	\caption{F1 score,  MCC score and Estimation error under AR(1) correlation, $non-i.i.d.$ error and $(n,p) = (200,600)$ corresponding to Table \ref{id0.3}.}
	\label{fig:2}
\end{figure}

\begin{table} [ht!]
\def\arraystretch{1.2}
\begin{center}
\caption{Identification and estimation results of RBHS, RBHS+, RBRHS, BHS, BHS+ and BRHS for the datasets with banded correlation under $i.i.d.$ errors, $n = 200$ and $p = 600$. mean(sd) of true positives (TP), false positives (FP), F1 score (F1), MCC and L1 error based on 100 replicates.}\label{id0.2}
\centering
\fontsize{9}{10}\selectfont{
\begin{tabular}{ c c c c c c c c  }
\hline

  &Methods& TP &FP  & F1 & MCC &L1 error                     \\
\hline
Error 1  & RBHS   &14.578(0.689) &0.071(0.313) &0.983(0.027)&0.983(0.027)&1.880(0.430)\\
   N(0,1)          & RBHS+   &14.603(0.684) & 0.089(0.321) &0.983(0.027)&0.983(0.027)&1.463(0.381)\\ 
     & RBRHS    & 14.515(0.757) &0.051(0.246)&0.981(0.029)&0.981(0.029)&1.888(0.417)\\
         & BHS   &14.617(0.662) &0.046(0.219)&0.985(0.025)&0.985(0.025)&1.740(0.395)  \\
         & BHS+&14.653(0.614)&0.049(0.238)&0.986(0.023)&0.986(0.023)&1.321(0.336)\\
         & BRHS&14.545(0.730)&0.034(0.192)&0.983(0.027)&0.983(0.027)&1.766(0.388)\\
   
\hline
Error 2         &     RBHS   &7.077(3.511)&0.021(0.144) &0.604(0.234)&0.668(0.177)&4.158(1.671)\\
   $t$(2)          & RBHS+   &7.654(3.607) & 0.025(0.174) &0.639(0.234)&0.696(0.176)&3.633(1.711)\\ 
     & RBRHS    & 4.913(3.789) &0.016(0.126)&0.440(0.286)&0.577(0.200)&5.457(2.083)\\
         & BHS   &3.146(2.995) &0.073(0.282)&0.304(0.252)&0.480(0.183)&7.432(2.033)  \\
         & BHS+&3.139(3.014)&0.104(1.106)&0.303(0.252)&0.478(0.185)&7.532(13.642)\\
         & BRHS&2.727(2.807)&0.053(0.233)&0.269(0.242)&0.458(0.177)&7.356(1.984)\\
\hline
Error 3    & RBHS   &12.619(1.910)&0.026(0.165)&0.907(0.082)&0.912(0.075)&2.364(0.780) \\
 Laplace(0,1)   & RBHS+ &12.896(1.798) &0.032(0.176)&0.919(0.077)&0.922(0.070)&1.855(0.678)   \\  
          & RBRHS  &12.237(2.218) &0.022(0.147)  &0.890(0.101)&0.896(0.090)&2.495(0.823)\\
           & BHS    &11.698(2.221) &0.056(0.247) &0.866(0.102)&0.874(0.090)&3.093(0.914)\\
           & BHS+&11.800(2.204)&0.053(0.237)&0.871(0.101)&0.878(0.090)&2.442(0.834)\\
         & BRHS&11.356(2.369)&0.029(0.174)&0.851(0.113)&0.861(0.099)&3.060(0.886)\\
         
\hline
Error 4  &RBHS &11.441(2.294)&0.027(0.168)&0.855(0.109)&0.865(0.095)&2.659(0.862)\\
 80$\%$N(0,1)
 &RBHS+ &11.792(2.134)&0.031(0.185)&0.872(0.099)&0.879(0.087)&2.140(0.767)\\
+20$\%$N(0,3) &RBRHS&10.472(2.958)&0.018(0.140)&0.803(0.162)&0.822(0.134)&3.051(1.121)\\
 &BHS&9.603(2.741)&0.068(0.264)&0.762(0.150)&0.784(0.125)&3.940(1.114)\\
 &BHS+&9.706(2.734)&0.070(0.271)&0.767(0.149)&0.789(0.125)&3.221(1.099)\\
 &BRHS&9.046(2.883)&0.047(0.221)&0.732(0.164)&0.760(0.135)&3.890(1.107)\\
 
\hline
Error 5  &RBHS &11.623(2.922)&0.009(0.105)&0.857(0.150)&0.869(0.129)&2.124(1.070)\\
 Lognormal(0,1) &RBHS+&12.147(2.640)&0.012(0.134)&0.882(0.130)&0.891(0.111)&1.721(0.929)\\
 &RBRHS&10.373(3.790)&0.009(0.105)&0.785(0.220)&0.813(0.176)&2.782(1.565)\\
 &BHS&5.632(3.716)&0.063(0.259)&0.497(0.267)&0.598(0.195)&5.880(1.976)\\
 &BHS+&5.661(3.767)&0.077(0.288)&0.498(0.269)&0.600(0.197)&5.306(2.112)\\
 &BRHS&5.113(3.678)&0.039(0.209)&0.459(0.273)&0.578(0.196)&5.831(1.961)\\
 
\hline 
\end{tabular} }
\end{center}
\centering
\end{table}
\clearpage
\begin{table} [ht!]
\def\arraystretch{1.2}
\begin{center}
\caption{Identification and estimation results of RBHS, RBHS+, RBRHS, BHS, BHS+ and BRHS for the datasets with AR(1) correlation under $non - i.i.d.$ errors, $n = 200$ and $p = 600$. mean(sd) of true positives (TP), false positives (FP), F1 score (F1), MCC and L1 error based on 100 replicates.}\label{id0.3}
\centering
\fontsize{9}{10}\selectfont{
\begin{tabular}{ c c c c c c c c  }
\hline

  &Methods& TP &FP  & F1 & MCC &L1 error                     \\
\hline
Error 1  & RBHS   &13.343(1.669) &0.013(0.130) &0.937(0.068)&0.940(0.063)&1.875(0.649)\\
   N(0,1)          & RBHS+   &13.550(1.545) & 0.010(0.100) &0.946(0.062)&0.947(0.058)&1.474(0.543)\\ 
     & RBRHS    & 13.155(1.900) &0.005(0.071)&0.929(0.082)&0.932(0.074)&1.959(0.680)\\
         & BHS   &11.911(2.136) &0.054(0.239)&0.876(0.098)&0.883(0.087)&3.051(0.865)  \\
         & BHS+&11.937(2.111)&0.048(0.228)&0.877(0.095)&0.884(0.085)&2.386(0.794)\\
         & BRHS&11.655(2.244)&0.034(0.192)&0.865(0.105)&0.873(0.092)&2.920(0.804)\\
   
\hline
Error 2         &     RBHS   &4.673(3.256)&0.011(0.104) &0.435(0.244)&0.542(0.183)&5.259(1.895)\\
   $t$(2)          & RBHS+   &5.204(3.482) & 0.016(0.141) &0.472(0.255)&0.575(0.188)&4.781(2.015)\\ 
     & RBRHS    & 2.646(3.011) &0.005(0.084)&0.258(0.250)&0.462(0.182)&6.736(2.149)\\
         & BHS   &1.408(1.865) &0.188(2.651)&0.151(0.178)&0.369(0.143)&11.077(44.292)  \\
         & BHS+&1.387(1.846)&0.188(2.294)&0.149(0.177)&0.364(0.148)&9.674(17.712)\\
         & BRHS&1.205(1.702)&0.102(2.178)&0.131(0.167)&0.359(0.136)&10.055(35.402)\\
\hline
Error 3    & RBHS   &10.317(2.841)&0.011(0.104)&0.799(0.147)&0.817(0.125)&2.739(1.225) \\
 Laplace(0,1)   & RBHS+ &10.917(2.810) &0.010(0.109)&0.827(0.141)&0.842(0.121)&2.240(1.144)   \\  
          & RBRHS  &9.475(3.362) &0.007(0.083)  &0.748(0.192)&0.778(0.155)&3.127(1.385)\\
           & BHS    &6.516(2.989) &0.085(0.332) &0.576(0.200)&0.635(0.157)&5.493(1.554)\\
           & BHS+&6.524(3.029)&0.091(0.330)&0.576(0.204)&0.635(0.161)&4.790(1.589)\\
         & BRHS&6.059(3.054)&0.054(0.259)&0.544(0.212)&0.613(0.165)&5.292(1.496)\\
         
\hline
Error 4  &RBHS &8.150(3.129)&0.009(0.095)&0.679(0.191)&0.718(0.154)&3.590(1.339)\\
 80$\%$N(0,1)
 &RBHS+ &8.728(3.131)&0.015(0.122)&0.712(0.183)&0.744(0.150)&3.051(1.343)\\
+20$\%$N(0,3) &RBRHS&6.560(3.516)&0.009(0.095)&0.570(0.241)&0.646(0.178)&4.352(1.668)\\
 &BHS&4.349(2.737)&0.071(0.268)&0.418(0.215)&0.523(0.160)&6.504(1.521)\\
 &BHS+&4.329(2.742)&0.067(0.266)&0.416(0.215)&0.517(0.164)&5.903(1.614)\\
 &BRHS&3.983(2.675)&0.042(0.201)&0.389(0.215)&0.505(0.159)&6.356(1.502)\\
 
\hline
Error 5  &RBHS &8.219(3.813)&0.997(0.095)&0.643(0.228)&0.666(0.206)&4.700(1.923)\\
 Lognormal(0,1) &RBHS+&9.058(3.785)&1.001(0.084)&0.691(0.218)&0.708(0.196)&4.121(1.869)\\
 &RBRHS&7.287(4.289)&0.991(0.130)&0.576(0.277)&0.602(0.259)&5.047(2.147)\\
 &BHS&2.495(2.603)&1.045(0.308)&0.239(0.216)&0.279(0.224)&9.917(2.168)\\
 &BHS+&2.537(2.637)&1.029(0.317)&0.243(0.217)&0.284(0.225)&9.389(2.294)\\
 &BRHS&2.345(2.535)&1.014(0.261)&0.227(0.211)&0.267(0.221)&9.577(2.121)\\
 
\hline 
\end{tabular} }
\end{center}
\centering
\end{table}
\clearpage
Figure \ref{fig:2} shows results under non-$i.i.d.$ errors, which corresponding to Table \ref{id0.3}. While all methods experience some degradation, robust procedures again stand out. RBHS+ maintains the most stable and accurate performance across all metrics, followed by RBHS and RBRHS. The non-robust methods demonstrate sharp declines in both F1 and MCC scores and inflated estimation errors, especially under heavy-tailed and skewed distributions. The boxplots show increased spread and downward shifts in identification metrics, and upward shifts in estimation error. Overall, Figures \ref{fig:1} and \ref{fig:2} consistently reinforce the superiority of the robust Bayesian methods in challenging scenarios, highlighting their effectiveness in both accurate variable selection and stable coefficient estimation.

\clearpage
\begin{table} [ht!]
\def\arraystretch{1.2}
\begin{center}
\caption{Identification and estimation results of RBHS, RBHS+, RBRHS, BHS, BHS+ and BRHS for the datasets with banded correlation under $non - i.i.d.$ errors, $n = 200$ and $p = 600$. mean(sd) of true positives (TP), false positives (FP), F1 score (F1), MCC and L1 error based on 100 replicates.}\label{id0.4}
\centering
\fontsize{9}{10}\selectfont{
\begin{tabular}{ c c c c c c c c  }
\hline

  &Methods& TP &FP  & F1 & MCC &L1 error                     \\
\hline
Error 1  & RBHS   &13.245(1.669) &0.018(0.140) &0.933(0.070)&0.936(0.065)&1.924(0.711)\\
   N(0,1)          & RBHS+   &13.530(1.548) & 0.020(0.100) &0.944(0.064)&0.946(0.060)&1.497(0.596)\\ 
     & RBRHS    & 12.960(1.900) &0.012(0.109)&0.921(0.082)&0.925(0.074)&2.089(0.747)\\
         & BHS   &11.783(2.161) &0.069(0.269)&0.870(0.098)&0.877(0.088)&3.122(0.927)  \\
         & BHS+&11.844(2.139)&0.071(0.265)&0.873(0.097)&0.880(0.087)&2.463(0.847)\\
         & BRHS&11.442(2.287)&0.037(0.189)&0.855(0.107)&0.865(0.094)&3.067(0.882)\\
   
\hline
Error 2         &     RBHS   &3.847(2.997)&0.026(0.159) &0.370(0.239)&0.500(0.177)&5.777(1.869)\\
   $t$(2)          & RBHS+   &4.339(3.305) & 0.028(0.165) &0.406(0.254)&0.529(0.186)&5.326(2.036)\\ 
     & RBRHS    & 1.863(2.520) &0.011(0.104)&0.188(0.222)&0.422(0.171)&7.462(1.952)\\
         & BHS   &1.121(1.572) &0.128(2.632)&0.123(0.158)&0.348(0.133)&9.956(31.923)  \\
         & BHS+&1.144(1.613)&0.125(1.854)&0.125(0.159)&0.346(0.137)&9.485(20.664)\\
         & BRHS&0.917(1.404)&0.031(0.179)&0.102(0.145)&0.339(0.125)&8.794(1.438)\\
\hline
Error 3    & RBHS   &9.728(3.252)&0.028(0.203)&0.763(0.182)&0.789(0.150)&2.940(1.403) \\
 Laplace(0,1)   & RBHS+ &10.516(3.214) &0.026(0.203)&0.802(0.173)&0.822(0.146)&2.389(1.296)   \\  
          & RBRHS  &8.079(3.995) &0.016(0.141)  &0.657(0.253)&0.711(0.196)&3.780(1.818)\\
           & BHS    &5.754(3.082) &0.080(0.300) &0.520(0.219)&0.594(0.170)&5.753(1.645)\\
           & BHS+&5.825(3.153)&0.086(0.339)&0.524(0.224)&0.597(0.174)&5.123(1.750)\\
         & BRHS&5.213(3.083)&0.056(0.255)&0.480(0.228)&0.569(0.172)&5.716(1.626)\\
         
\hline
Error 4  &RBHS &7.839(3.340)&0.021(0.144)&0.656(0.207)&0.700(0.167)&3.759(1.494)\\
 80$\%$N(0,1)
 &RBHS+ &8.367(3.386)&0.025(0.163)&0.687(0.202)&0.725(0.165)&3.192(1.452)\\
+20$\%$N(0,3) &RBRHS&5.470(3.853)&0.013(0.113)&0.483(0.279)&0.591(0.206)&4.988(1.990)\\
 &BHS&3.828(2.791)&0.086(0.317)&0.372(0.226)&0.493(0.169)&6.736(1.653)\\
 &BHS+&3.859(2.813)&0.090(0.323)&0.374(0.227)&0.490(0.174)&6.191(1.777)\\
 &BRHS&3.391(2.676)&0.061(0.244)&0.336(0.223)&0.470(0.167)&6.698(1.615)\\
 
\hline
Error 5  &RBHS &7.804(3.864)&0.995(0.084)&0.618(0.236)&0.642(0.216)&5.134(2.048)\\
 Lognormal(0,1) &RBHS+&8.541(3.933)&0.998(0.078)&0.659(0.218)&0.679(0.214)&4.562(2.056)\\
 &RBRHS&6.388(4.588)&0.992(0.100)&0.509(0.309)&0.536(0.295)&5.730(2.461)\\
 &BHS&2.263(2.546)&1.016(0.275)&0.218(0.215)&0.256(0.227)&10.154(2.129)\\
 &BHS+&2.231(2.518)&1.012(0.279)&0.215(0.214)&0.254(0.225)&9.695(2.270)\\
 &BRHS&2.022(2.456)&1.003(0.226)&0.196(0.210)&0.231(0.224)&9.889(2.098)\\
 
\hline 
\end{tabular} }
\end{center}
\centering
\end{table}

\begin{table} [ht!]
\def\arraystretch{1.2}
\begin{center}
\caption{Identification and estimation results of RBHS, RBHS+, RBRHS, BHS, BHS+ and BRHS for the datasets with AR(1) correlation under $i.i.d.$ errors, $n = 400$ and $p = 600$. mean(sd) of true positives (TP), false positives (FP), F1 score (F1), MCC and L1 error based on 100 replicates.}\label{id0.5}
\centering
\fontsize{9}{10}\selectfont{
\begin{tabular}{ c c c c c c c c  }
\hline
  &Methods& TP &FP  & F1 & MCC &L1 error                     \\
\hline
Error 1  & RBHS & 15.000(0.000) & 0.105(0.338) & 0.997(0.011) & 0.997(0.011) & 1.428(0.300)\\
   N(0,1)  & RBHS+   & 15.000(0.000) & 0.100(0.361) & 0.997(0.011) & 0.997(0.011) & 1.076(0.271)\\
         & RBRHS & 15.000(0.000) & 0.110(0.329) & 0.996(0.011) & 0.996(0.011) & 1.405(0.290)\\
         & BHS & 15.000(0.000) & 0.025(0.157) & 0.999(0.005) & 0.999(0.005) & 1.082(0.211)\\
         & BHS+ & 15.000(0.000) & 0.020(0.140) & 0.999(0.005) & 0.999(0.005) & 0.820(0.183)\\
         & BRHS & 15.000(0.000) & 0.030(0.171) & 0.999(0.006) & 0.999(0.006) & 1.071(0.197)\\
\hline
Error 2  & RBHS & 14.640(0.634) & 0.005(0.071) & 0.987(0.023) & 0.987(0.023) & 1.227(0.250)\\
   $t$(2)  & RBHS+   & 14.655(0.631) & 0.000(0.000) & 0.988(0.022) & 0.988(0.022) & 1.021(0.227)\\
         & RBRHS & 14.585(0.725) & 0.005(0.071) & 0.985(0.026) & 0.985(0.026) & 1.305(0.267)\\
         & BHS & 9.150(4.087) & 0.040(0.221) & 0.714(0.254) & 0.767(0.186) & 4.360(2.179)\\
         & BHS+ & 9.010(4.102) & 0.025(0.234) & 0.707(0.257) & 0.757(0.193) & 3.779(2.208)\\
         & BRHS & 9.025(4.135) & 0.040(0.221) & 0.706(0.260) & 0.756(0.197) & 4.300(2.128)\\
\hline
Error 3  & RBHS & 14.960(0.221) & 0.020(0.140) & 0.998(0.009) & 0.998(0.009) & 1.178(0.263)\\
   Laplace(0,1)  & RBHS+   & 14.970(0.171) & 0.030(0.171) & 0.998(0.009) & 0.998(0.009) & 0.926(0.225)\\
         & RBRHS & 14.965(0.184) & 0.020(0.140) & 0.998(0.008) & 0.998(0.008) & 1.174(0.256)\\
         & BHS & 14.830(0.390) & 0.020(0.140) & 0.993(0.014) & 0.993(0.014) & 1.614(0.342)\\
         & BHS+ & 14.820(0.422) & 0.015(0.122) & 0.993(0.016) & 0.993(0.016) & 1.229(0.295)\\
         & BRHS & 14.815(0.402) & 0.025(0.157) & 0.993(0.015) & 0.993(0.015) & 1.581(0.324)\\
\hline
Error 4  & RBHS & 14.890(0.329) & 0.025(0.157) & 0.995(0.013) & 0.995(0.013) & 1.308(0.271)\\
   80$\%$N(0,1)+20$\%$N(0,3)  & RBHS+   & 14.910(0.287) & 0.015(0.122) & 0.996(0.011) & 0.996(0.011) & 1.042(0.249)\\
         & RBRHS & 14.865(0.397) & 0.030(0.171) & 0.994(0.015) & 0.994(0.015) & 1.321(0.278)\\
         & BHS & 14.435(0.848) & 0.030(0.171) & 0.979(0.031) & 0.979(0.031) & 1.904(0.461)\\
         & BHS+ & 14.460(0.832) & 0.040(0.196) & 0.980(0.031) & 0.980(0.030) & 1.455(0.410)\\
         & BRHS & 14.425(0.835) & 0.010(0.100) & 0.979(0.030) & 0.979(0.030) & 1.863(0.451)\\
\hline
Error 5  & RBHS & 14.980(0.140) & 0.000(0.000) & 0.999(0.005) & 0.999(0.005) & 0.872(0.175)\\
   Lognormal(0,1)  & RBHS+   & 14.970(0.171) & 0.000(0.000) & 0.999(0.006) & 0.999(0.006) & 0.741(0.162)\\
         & RBRHS & 14.970(0.171) & 0.000(0.000) & 0.999(0.006) & 0.999(0.006) & 0.909(0.178)\\
         & BHS & 12.350(2.455) & 0.055(0.269) & 0.891(0.123) & 0.901(0.093) & 2.804(1.072)\\
         & BHS+ & 12.250(2.494) & 0.045(0.208) & 0.886(0.126) & 0.898(0.096) & 2.269(1.038)\\
         & BRHS & 12.310(2.523) & 0.045(0.208) & 0.889(0.128) & 0.900(0.098) & 2.745(1.056)\\
\hline
\end{tabular} }
\end{center}
\centering
\end{table}
\clearpage
\begin{table} [ht!]
\def\arraystretch{1.2}
\begin{center}
\caption{Identification and estimation results of RBHS, RBHS+, RBRHS, BHS, BHS+ and BRHS for the datasets with banded correlation under $i.i.d.$ errors, $n = 400$ and $p = 600$. mean(sd) of true positives (TP), false positives (FP), F1 score (F1), MCC and L1 error based on 100 replicates.}\label{id0.6}
\centering
\fontsize{9}{10}\selectfont{
\begin{tabular}{ c c c c c c c c  }
\hline
  &Methods& TP &FP  & F1 & MCC &L1 error                     \\
\hline
Error 1  & RBHS & 14.995(0.071) & 0.110(0.344) & 0.996(0.011) & 0.996(0.011) & 1.422(0.296)\\
   N(0,1)  & RBHS+   & 14.995(0.071) & 0.150(0.358) & 0.995(0.012) & 0.995(0.012) & 1.088(0.247)\\
         & RBRHS & 14.995(0.071) & 0.080(0.272) & 0.997(0.009) & 0.997(0.009) & 1.405(0.291)\\
         & BHS & 14.990(0.100) & 0.035(0.210) & 0.999(0.007) & 0.999(0.007) & 1.088(0.216)\\
         & BHS+ & 14.990(0.100) & 0.045(0.271) & 0.998(0.009) & 0.998(0.009) & 0.840(0.188)\\
         & BRHS & 14.990(0.100) & 0.045(0.231) & 0.998(0.008) & 0.998(0.008) & 1.082(0.208)\\
\hline
Error 2  & RBHS & 14.505(0.972) & 0.005(0.071) & 0.982(0.042) & 0.982(0.038) & 1.258(0.367)\\
   $t$(2)  & RBHS+   & 14.535(0.945) & 0.005(0.071) & 0.983(0.042) & 0.983(0.037) & 1.047(0.283)\\
         & RBRHS & 14.355(1.299) & 0.015(0.122) & 0.974(0.075) & 0.979(0.029) & 1.394(0.531)\\
         & BHS & 8.965(3.958) & 0.050(0.240) & 0.706(0.251) & 0.768(0.169) & 4.233(1.893)\\
         & BHS+ & 8.930(4.016) & 0.055(0.250) & 0.703(0.256) & 0.765(0.175) & 3.670(1.947)\\
         & BRHS & 8.830(3.929) & 0.055(0.250) & 0.699(0.250) & 0.762(0.169) & 4.214(1.897)\\
\hline
Error 3  & RBHS & 14.960(0.221) & 0.010(0.100) & 0.998(0.008) & 0.998(0.008) & 1.156(0.261)\\
   Laplace(0,1)  & RBHS+   & 14.980(0.140) & 0.040(0.196) & 0.998(0.008) & 0.998(0.008) & 0.908(0.217)\\
         & RBRHS & 14.950(0.240) & 0.015(0.122) & 0.998(0.009) & 0.998(0.009) & 1.166(0.258)\\
         & BHS & 14.780(0.513) & 0.050(0.218) & 0.991(0.019) & 0.991(0.019) & 1.608(0.364)\\
         & BHS+ & 14.795(0.494) & 0.055(0.229) & 0.991(0.018) & 0.991(0.018) & 1.225(0.295)\\
         & BRHS & 14.780(0.513) & 0.025(0.157) & 0.991(0.019) & 0.991(0.018) & 1.578(0.363)\\
\hline
Error 4  & RBHS & 14.900(0.317) & 0.010(0.100) & 0.996(0.012) & 0.996(0.012) & 1.265(0.268)\\
   80$\%$N(0,1)+20$\%$N(0,3)  & RBHS+   & 14.910(0.287) & 0.015(0.158) & 0.996(0.012) & 0.996(0.012) & 1.009(0.246)\\
         & RBRHS & 14.875(0.346) & 0.000(0.000) & 0.996(0.012) & 0.996(0.012) & 1.315(0.265)\\
         & BHS & 14.485(0.750) & 0.055(0.229) & 0.980(0.027) & 0.980(0.027) & 1.835(0.408)\\
         & BHS+ & 14.405(0.815) & 0.055(0.229) & 0.977(0.031) & 0.977(0.031) & 1.396(0.344)\\
         & BRHS & 14.410(0.797) & 0.040(0.196) & 0.978(0.029) & 0.978(0.029) & 1.813(0.402)\\
\hline
Error 5  & RBHS & 14.980(0.140) & 0.000(0.000) & 0.999(0.005) & 0.999(0.005) & 0.852(0.166)\\
   Lognormal(0,1)  & RBHS+   & 14.970(0.171) & 0.005(0.071) & 0.999(0.006) & 0.999(0.006) & 0.726(0.152)\\
         & RBRHS & 14.970(0.171) & 0.000(0.000) & 0.999(0.006) & 0.999(0.006) & 0.906(0.189)\\
         & BHS & 12.140(2.599) & 0.040(0.196) & 0.881(0.127) & 0.890(0.110) & 2.949(1.120)\\
         & BHS+ & 12.115(2.580) & 0.035(0.184) & 0.881(0.126) & 0.889(0.109) & 2.411(1.108)\\
         & BRHS & 12.060(2.671) & 0.035(0.210) & 0.877(0.132) & 0.886(0.114) & 2.924(1.109)\\
\hline
\end{tabular} }
\end{center}
\centering
\end{table}
\clearpage
\begin{table} [ht!]
\def\arraystretch{1.2}
\begin{center}
\caption{Identification and estimation results of RBHS, RBHS+, RBRHS, BHS, BHS+ and BRHS for the datasets with AR(1) correlation under $non-i.i.d.$ errors, $n = 400$ and $p = 600$. mean(sd) of true positives (TP), false positives (FP), F1 score (F1), MCC and L1 error based on 100 replicates.}\label{id0.7}
\centering
\fontsize{9}{10}\selectfont{
\begin{tabular}{ c c c c c c c c  }
\hline
  &Methods& TP &FP  & F1 & MCC &L1 error                     \\
\hline
Error 1  & RBHS & 15.000(0.000) & 0.010(0.100) & 0.999(0.003) & 0.999(0.003) & 0.833(0.193)\\
   N(0,1)  & RBHS+ & 15.000(0.000) & 0.015(0.122) & 0.999(0.004) & 0.999(0.004) & 0.685(0.175)\\
         & RBRHS & 15.000(0.000) & 0.010(0.100) & 0.999(0.003) & 0.999(0.003) & 0.853(0.196)\\
         & BHS & 14.845(0.390) & 0.045(0.208) & 0.993(0.015) & 0.993(0.015) & 1.583(0.360)\\
         & BHS+ & 14.835(0.398) & 0.055(0.250) & 0.993(0.015) & 0.992(0.015) & 1.217(0.311)\\
         & BRHS & 14.840(0.394) & 0.030(0.171) & 0.993(0.014) & 0.993(0.014) & 1.560(0.342)\\
\hline
Error 2  & RBHS & 14.190(1.114) & 0.000(0.000) & 0.971(0.042) & 0.971(0.040) & 1.114(0.331)\\
   $t$(2)  & RBHS+ & 14.355(0.982) & 0.000(0.000) & 0.977(0.036) & 0.977(0.035) & 0.938(0.287)\\
         & RBRHS & 13.990(1.356) & 0.000(0.000) & 0.963(0.054) & 0.964(0.051) & 1.306(0.414)\\
         & BHS & 4.475(3.514) & 0.055(0.229) & 0.409(0.280) & 0.558(0.195) & 6.648(2.247)\\
         & BHS+ & 4.400(3.364) & 0.055(0.229) & 0.406(0.271) & 0.548(0.192) & 6.184(2.341)\\
         & BRHS & 4.305(3.489) & 0.060(0.238) & 0.395(0.280) & 0.549(0.198) & 6.592(2.209)\\
\hline
Error 3  & RBHS & 14.970(0.171) & 0.005(0.071) & 0.999(0.007) & 0.999(0.007) & 0.843(0.189)\\
   Laplace(0,1)  & RBHS+ & 14.960(0.196) & 0.000(0.000) & 0.999(0.007) & 0.999(0.007) & 0.710(0.167)\\
         & RBRHS & 14.965(0.184) & 0.000(0.000) & 0.999(0.006) & 0.999(0.006) & 0.888(0.185)\\
         & BHS & 12.675(1.785) & 0.050(0.218) & 0.909(0.077) & 0.913(0.069) & 2.660(0.702)\\
         & BHS+ & 12.635(1.799) & 0.040(0.196) & 0.908(0.078) & 0.912(0.070) & 2.122(0.649)\\
         & BRHS & 12.585(1.822) & 0.030(0.171) & 0.906(0.079) & 0.910(0.071) & 2.623(0.692)\\
\hline
Error 4  & RBHS & 14.835(0.423) & 0.000(0.000) & 0.994(0.015) & 0.994(0.015) & 0.979(0.229)\\
   80$\%$N(0,1) +20$\%$N(0,3)  & RBHS+ & 14.850(0.372) & 0.005(0.071) & 0.995(0.013) & 0.995(0.014) & 0.820(0.201)\\
         & RBRHS & 14.820(0.410) & 0.000(0.000) & 0.994(0.014) & 0.994(0.014) & 1.061(0.233)\\
         & BHS & 11.450(2.147) & 0.065(0.247) & 0.856(0.101) & 0.865(0.089) & 3.202(0.870)\\
         & BHS+ & 11.315(2.229) & 0.070(0.292) & 0.849(0.107) & 0.859(0.094) & 2.620(0.867)\\
         & BRHS & 11.240(2.240) & 0.045(0.231) & 0.846(0.107) & 0.857(0.094) & 3.152(0.853)\\
\hline
Error 5  & RBHS & 14.930(0.275) & 1.000(0.000) & 0.965(0.010) & 0.965(0.010) & 1.847(0.251)\\
   Lognormal(0,1)  & RBHS+ & 14.940(0.295) & 1.000(0.000) & 0.966(0.010) & 0.965(0.011) & 1.730(0.228)\\
         & RBRHS & 14.955(0.208) & 1.000(0.000) & 0.966(0.007) & 0.966(0.007) & 1.859(0.249)\\
         & BHS & 7.665(3.918) & 1.080(0.380) & 0.606(0.246) & 0.631(0.226) & 6.746(1.972)\\
         & BHS+ & 7.660(3.909) & 1.090(0.378) & 0.605(0.246) & 0.630(0.225) & 6.128(2.065)\\
         & BRHS & 7.640(3.884) & 1.065(0.302) & 0.605(0.243) & 0.631(0.222) & 6.620(1.955)\\
\hline
\end{tabular} }
\end{center}
\centering
\end{table}
\clearpage
\begin{table} [ht!]
\def\arraystretch{1.2}
\begin{center}
\caption{Identification and estimation results of RBHS, RBHS+, RBRHS, BHS, BHS+ and BRHS for the datasets with banded correlation under $non-i.i.d.$ errors, $n = 400$ and $p = 600$. mean(sd) of true positives (TP), false positives (FP), F1 score (F1), MCC and L1 error based on 100 replicates.}
\label{id0.8}
\centering
\fontsize{9}{10}\selectfont{
\begin{tabular}{ c c c c c c c c  }
\hline
  &Methods& TP &FP  & F1 & MCC &L1 error                     \\
\hline
Error 1 & RBHS   & 15.000(0.000) & 0.000(0.000) & 1.000(0.000) & 1.000(0.000) & 0.853(0.175)\\
N(0,1) & RBHS+  & 15.000(0.000) & 0.000(0.000) & 1.000(0.000) & 1.000(0.000) & 0.696(0.159)\\
        & RBRHS  & 14.995(0.071) & 0.000(0.000) & 0.999(0.002) & 0.999(0.002) & 0.887(0.180)\\
        & BHS    & 14.775(0.588) & 0.060(0.258) & 0.990(0.024) & 0.990(0.024) & 1.615(0.360)\\
        & BHS+   & 14.755(0.563) & 0.055(0.250) & 0.990(0.022) & 0.990(0.022) & 1.240(0.318)\\
        & BRHS   & 14.765(0.567) & 0.040(0.196) & 0.990(0.022) & 0.990(0.022) & 1.598(0.352)\\

\hline
Error 2 & RBHS   & 14.145(1.365) & 0.000(0.000) & 0.968(0.056) & 0.969(0.052) & 1.151(0.409)\\
$t(2)$   & RBHS+  & 14.290(1.214) & 0.000(0.000) & 0.974(0.049) & 0.974(0.045) & 0.981(0.375)\\
        & RBRHS  & 13.875(1.790) & 0.000(0.000) & 0.956(0.087) & 0.958(0.076) & 1.398(0.641)\\
        & BHS    & 4.395(3.455) & 0.070(0.256) & 0.405(0.269) & 0.536(0.199) & 6.542(2.080)\\
        & BHS+   & 4.385(3.474) & 0.070(0.256) & 0.404(0.271) & 0.530(0.206) & 6.088(2.227)\\
        & BRHS   & 4.195(3.475) & 0.050(0.218) & 0.387(0.274) & 0.526(0.204) & 6.494(2.063)\\

\hline
Error 3 & RBHS   & 14.980(0.140) & 0.000(0.000) & 0.999(0.005) & 0.999(0.005) & 0.808(0.181)\\
Laplace(0,1) & RBHS+  & 14.985(0.122) & 0.000(0.000) & 0.999(0.004) & 0.999(0.004) & 0.680(0.159)\\
        & RBRHS  & 14.975(0.157) & 0.000(0.000) & 0.999(0.005) & 0.999(0.005) & 0.893(0.203)\\
        & BHS    & 12.605(1.818) & 0.050(0.218) & 0.907(0.078) & 0.911(0.071) & 2.663(0.761)\\
        & BHS+   & 12.620(1.784) & 0.060(0.238) & 0.907(0.077) & 0.911(0.070) & 2.113(0.724)\\
        & BRHS   & 12.520(1.957) & 0.050(0.218) & 0.902(0.087) & 0.907(0.078) & 2.638(0.760)\\

\hline
Error 4 & RBHS   & 14.820(0.478) & 0.000(0.000) & 0.994(0.017) & 0.994(0.017) & 1.029(0.275)\\
 80$\%$N(0,1) +20$\%$N(0,3)& RBHS+  & 14.835(0.434) & 0.000(0.000) & 0.994(0.015) & 0.994(0.015) & 0.868(0.232)\\
        & RBRHS  & 14.790(0.517) & 0.000(0.000) & 0.993(0.019) & 0.993(0.018) & 1.118(0.296)\\
        & BHS    & 11.150(2.532) & 0.095(0.311) & 0.838(0.131) & 0.849(0.113) & 3.357(1.078)\\
        & BHS+   & 11.160(2.517) & 0.095(0.294) & 0.838(0.128) & 0.849(0.111) & 2.760(1.042)\\
        & BRHS   & 10.930(2.551) & 0.075(0.264) & 0.828(0.134) & 0.841(0.115) & 3.304(1.065)\\

\hline
Error 5 & RBHS   & 14.840(0.515) & 1.000(0.000) & 0.962(0.019) & 0.962(0.019) & 1.910(0.305)\\
Lognormal(0,1) & RBHS+  & 14.885(0.391) & 1.000(0.000) & 0.964(0.014) & 0.963(0.014) & 1.786(0.270)\\
        & RBRHS  & 14.840(0.485) & 1.000(0.000) & 0.962(0.018) & 0.962(0.018) & 1.927(0.282)\\
        & BHS    & 7.400(3.673) & 1.020(0.140) & 0.595(0.239) & 0.617(0.225) & 6.843(2.068)\\
        & BHS+   & 7.480(3.727) & 1.025(0.157) & 0.599(0.244) & 0.620(0.230) & 6.225(2.139)\\
        & BRHS   & 7.365(3.728) & 1.020(0.140) & 0.592(0.244) & 0.614(0.229) & 6.693(2.050)\\
\hline
\end{tabular} }
\end{center}
\centering
\end{table}

\newpage
\subsection{Statistical inference under $(n, p)$= (200, 1000)}
\begin{table}[ht!]
\def\arraystretch{1.2}
\begin{center}
\caption{95\% empirical coverage probabilities assessed over 1,000 replicates for RBHS, RBHS+, RBRHS, BHS, BHS+ and BRHS using the datasets with $i.i.d.$ errors, $n = 200$ and $p = 1000$.}
\label{id3.3}
\centering
\fontsize{9}{11}\selectfont{
\begin{tabular}{ c c c c c c c c }
\hline
&& RBHS & RBHS+ & RBRHS & BHS & BHS+ & BRHS \\
\hline
N(0,1) & Coverage of $\boldsymbol{\beta_1}$ \\
& $\beta_1$ & 0.926 & 0.923 & 0.926 & 0.943 & 0.938 & 0.944 \\
& $\beta_2$ & 0.934 & 0.929 & 0.935 & 0.947 & 0.951 & 0.943 \\
& $\beta_3$ & 0.933 & 0.936 & 0.781 & 0.948 & 0.945 & 0.758 \\
& Average length \\
& $\beta_1$ & 0.334 & 0.332 & 0.318 & 0.331 & 0.330 & 0.315 \\
& $\beta_2$ & 0.370 & 0.369 & 0.349 & 0.368 & 0.367 & 0.346 \\
& $\beta_3$ & 0.335 & 0.334 & 0.323 & 0.332 & 0.331& 0.320 \\
\cline{2-8}
& Coverage of $\boldsymbol{\beta_2}$ & 1.000 & 1.000 & 1.000 & 1.000 & 1.000 & 1.000 \\
& Average length & 0.054& 0.041 & 0.047 & 0.042 & 0.031 & 0.035 \\
\hline
$t$(2) & Coverage of $\boldsymbol{\beta_1}$ \\
& $\beta_1$ & 0.954& 0.957 & 0.941 & 0.830 & 0.828 & 0.846 \\
& $\beta_2$ & 0.967 & 0.964 & 0.949 & 0.891 & 0.888 & 0.917 \\
& $\beta_3$ & 0.971 & 0.968 & 0.649 & 0.936 & 0.929 & 0.905 \\
& Average length \\
& $\beta_1$ & 0.514 & 0.512 & 0.467 & 0.934 & 0.916 & 0.877 \\
& $\beta_2$ & 0.571 & 0.571 & 0.503 & 1.103 & 1.083 & 1.028 \\
& $\beta_3$ & 0.509 & 0.507 & 0.474 & 0.953 & 0.938 & 0.898 \\
\cline{2-8}
& Coverage of $\boldsymbol{\beta_2}$ & 1.000 & 1.000 & 1.000 & 1.000 & 1.000 & 1.000 \\
& Average length & 0.038 & 0.028 & 0.032 & 0.135 & 0.100 & 0.099 \\
\hline
\end{tabular}}
\end{center}
\centering
\end{table}
\begin{figure}[h!]
	\centering
	\includegraphics[angle=0,origin=c,width=0.55\textwidth]{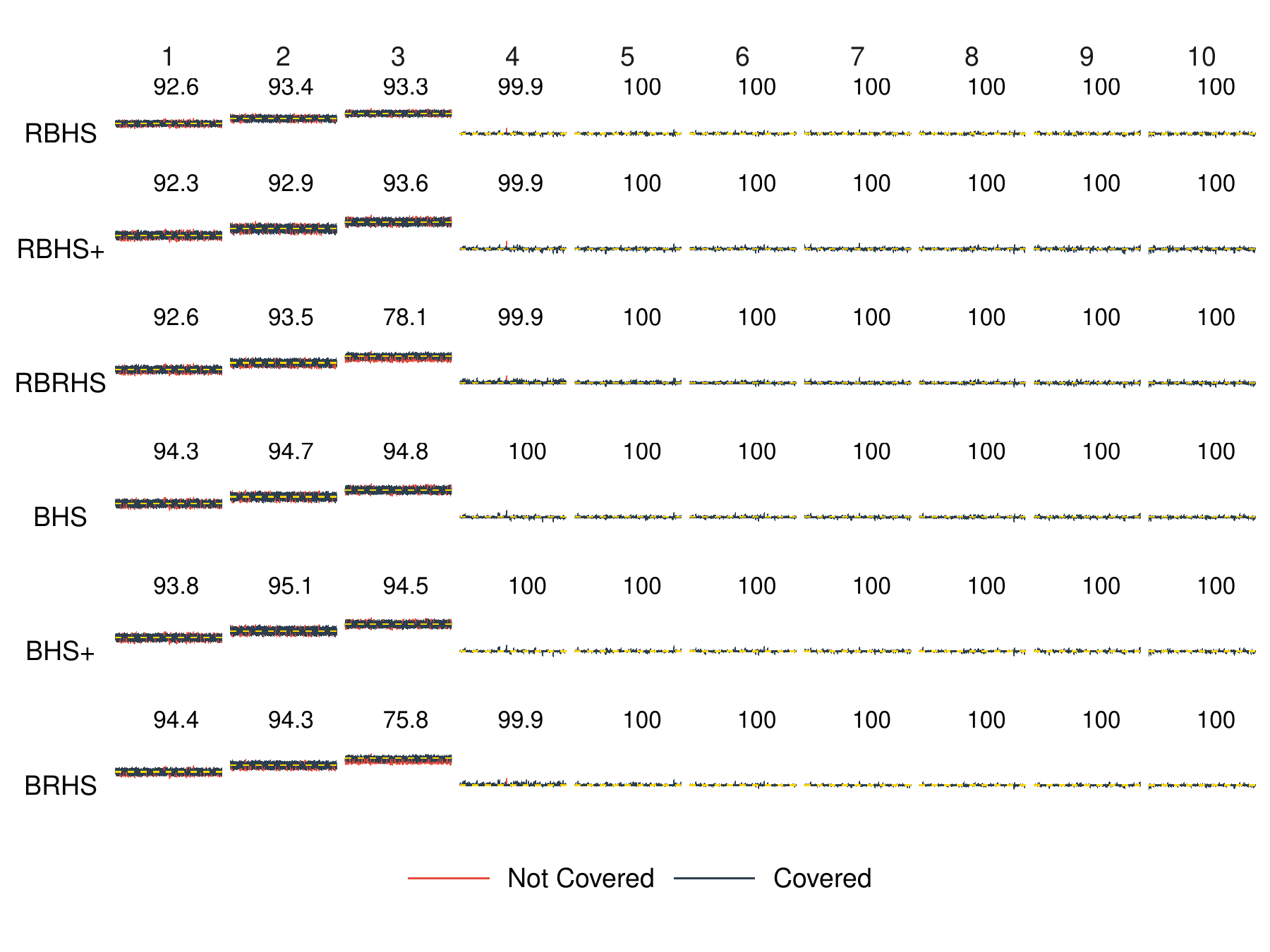}
	\caption{Credible intervals with empirical coverage probabilities under a 95\% nominal level for the first 10 predictors, with the first 3 having true nonzero coefficients. Results are based on 1{,}000 simulated datasets under a homogeneous model with AR(1) covariate correlation, $(n, p) = (200, 1000)$, and N(0,1) error.}
	\label{fig:6}
\end{figure}
\begin{figure}[h!]
	\centering
	\includegraphics[angle=0,origin=c,width=0.55\textwidth]{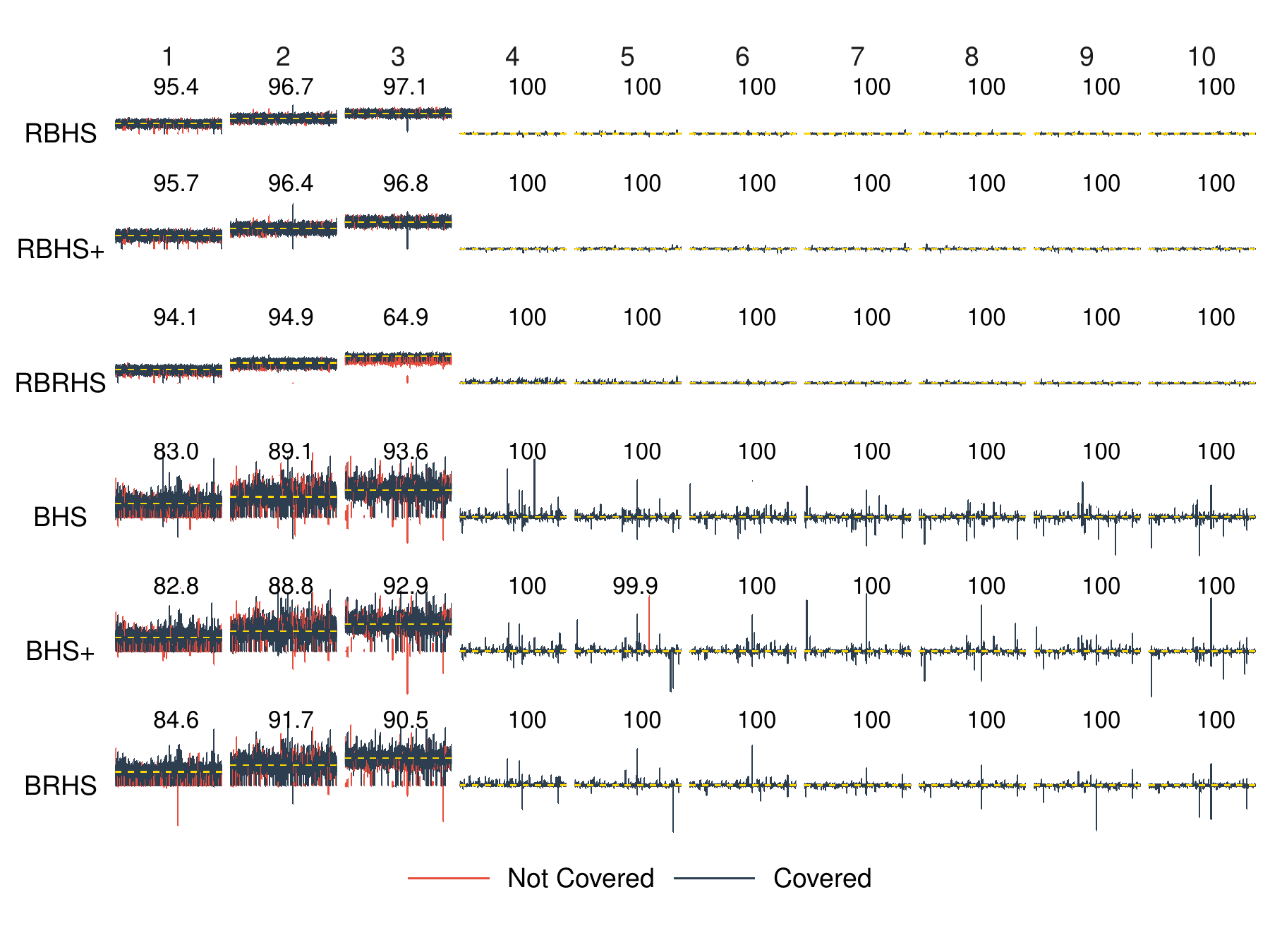}
	\caption{Credible intervals with empirical coverage probabilities under a 95\% nominal level for the first 10 predictors, with the first 3 having true nonzero coefficients. Results are based on 1{,}000 simulated datasets under a homogeneous model with AR(1) covariate correlation, $(n, p) = (200, 1000)$, and $t(2)$ error.}
	\label{fig:7}
\end{figure}
\clearpage
\subsection{Assessment of the convergence of MCMC chains}
\begin{figure}[h!]
	\centering
	\includegraphics[angle=0,origin=c,width=0.8\textwidth]{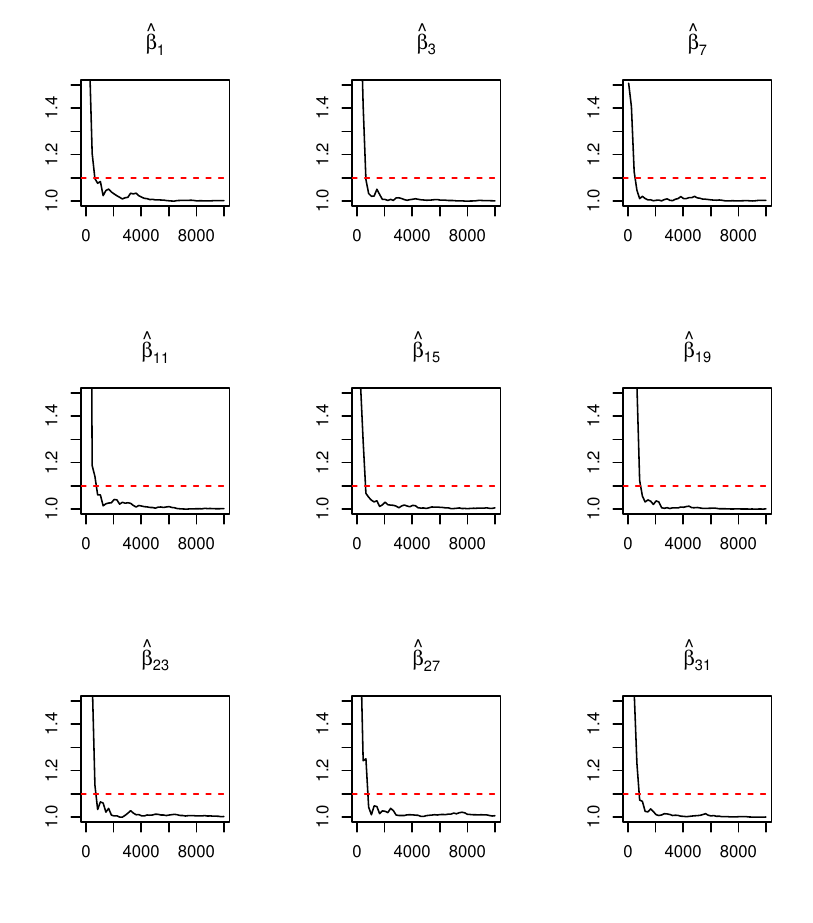}
	\caption{Potential scale reduction factor (PSRF) against iterations for nonzero coefficients in simulation for RBHS with $n = 200$ and $p=600$ under $i.i.d.$ Error 2. Black line: the PSRF. Red line: the threshold of 1.1.}
	\label{fig:curve1}
\end{figure}

\clearpage
\begin{figure}[h!]
	\centering
	\includegraphics[angle=0,origin=c,width=0.8\textwidth]{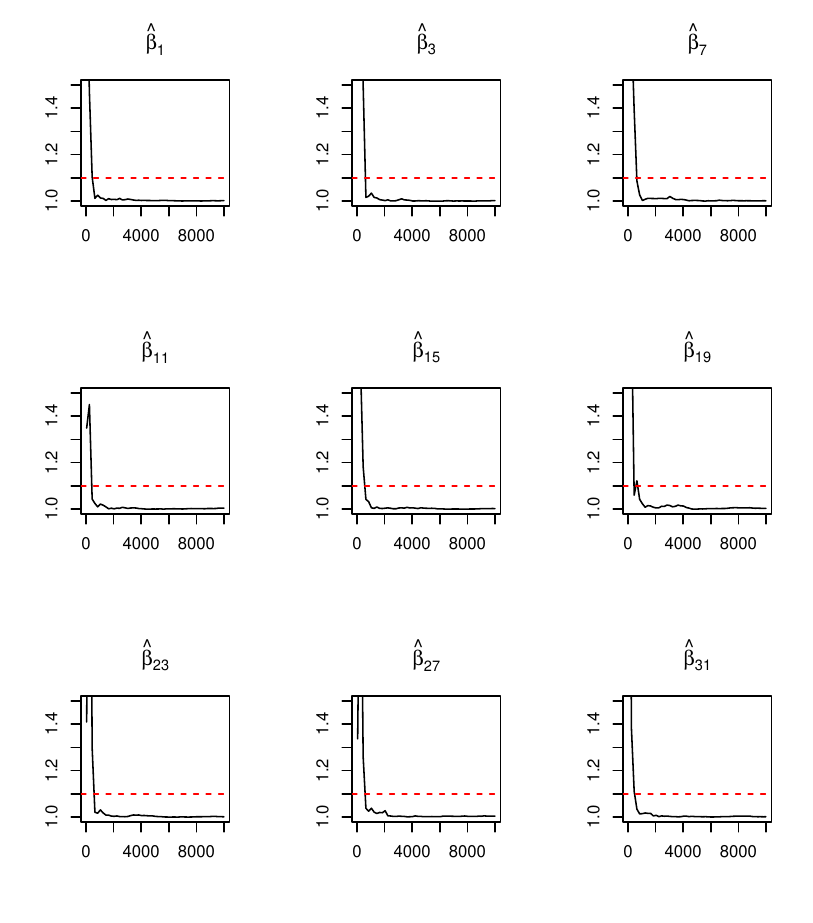}
	\caption{Potential scale reduction factor (PSRF) against iterations for nonzero coefficients in simulation for RBHS+ with $n = 200$ and $p=600$ under $i.i.d.$ Error 2. Black line: the PSRF. Red line: the threshold of 1.1.}
	\label{fig:curve2}
\end{figure}

\clearpage
\begin{figure}[h!]
	\centering
	\includegraphics[angle=0,origin=c,width=0.8\textwidth]{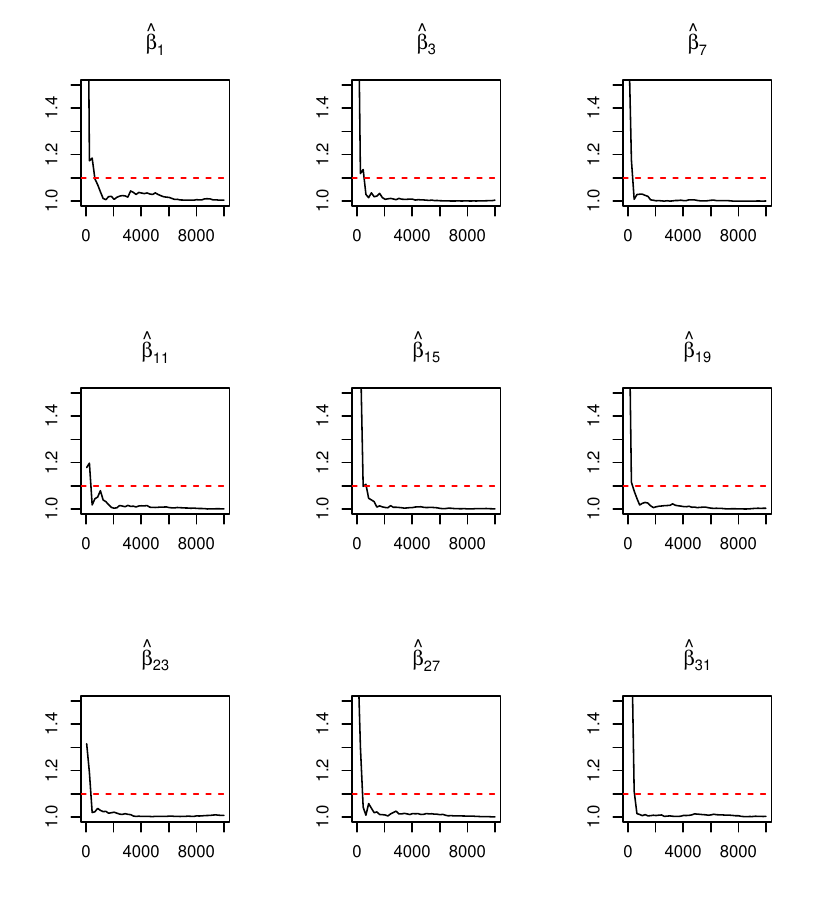}
	\caption{Potential scale reduction factor (PSRF) against iterations for nonzero coefficients in simulation for RBHS+ with $n = 200$ and $p=600$ under $i.i.d.$ Error 2. Black line: the PSRF. Red line: the threshold of 1.1.}
	\label{fig:curve3}
\end{figure}

\clearpage
\section{Comparison to published robust Bayesian variable selection methods}

In literature, robust Bayesian variable selection methods have been developed in several studies, such as the  Horseshoe prior Bayesian quantile regression (termed as HSBQR) \cite{kohns2024horseshoe}, Bayesian quantile adaptive LASSO (termed as BayesQR) \cite{benoit2017bayesqr}, and robust Bayesian variable selection with spike-and-slab priors \cite{ren2023robust,lu2021identifying} among others. Besides, the software paper of \textit{Bayesreg} package mentioned about sampling Laplace distribution using the scale mixture of normals with exponential densities for robust regression with HS and HS+ priors but did not examine their performance \cite{makalic2016high}. 

First, we illustrate the advantage of our proposals over HSBQR in high-dimensional settings from Table \ref{id0.9} which shows that HSBQR consistently yields much lower number of true positives compared to RBHS, RBHS+ and RBRHS. Therefore, its corresponding F1 score, MCC and estimation errors are also inferior to those obtained under the proposed methods.  

Our implementation of Bayesian quantile adaptive LASSO using Package \textit{BayesQR} shows that it constantly report error messages under high-dimensional settings when $p$ is larger than $n$. Therefore, we omit this method for comparison. 

Regarding robust Bayesian regression with the two-group spike-and-slab priors, as the primary goal of our study is to establish the merit of the one-group horseshoe family of priors for high-dimensional robust variable selection and statistical inference, we do not include a comprehensive comparison. Instead, we have assessed the performance of horseshoe family priors using the same setting adopted in Fan et al. (2024) \cite{fan2024seeing} and Song and Liang (2023) \cite{song2023nearly} for spike-and-slab priors, as shown in Table \ref{id1.00}. 

Table \ref{id1.00} shows that the proposed methods are superior over alternatives in statistical inference. For example, HSBQR does not yield nominal coverage probabilities close to 95\% for two out of three nonzero coefficients under $t$(2) errors. Similar observations can be made for approaches implemented using \textit{Bayesreg} package which seems unstable under $t$(2) errors and leads to a few failed MCMC runs or its execution is just stalled as CPU is active, but no output or logs are being generated.

\clearpage
\subsection{Identification and estimation}

\begin{table} [ht!]
\def\arraystretch{1.2}
\begin{center}
\caption{Identification and estimation results of RBHS, RBHS+, RBRHS and HSBQR for the datasets with AR(1) correlation under $i.i.d.$ errors, $n = 200$ and $p = 600$. mean(sd) of true positives (TP), false positives (FP), F1 score (F1), MCC and L1 error based on 100 replicates.}\label{id0.9}
\centering
\fontsize{9}{10}\selectfont{
\begin{tabular}{ c c c c c c c c  }
\hline

  &Methods& TP &FP  & F1 & MCC &L1 error                     \\
\hline
Error 1  & RBHS   &14.600(0.603) &0.050(0.219) &0.984(0.022)&0.984(0.022)&1.902(0.413)\\
   N(0,1)          & RBHS+   &14.600(0.667) & 0.060(0.239) &0.984(0.025)&0.984(0.024)&1.463(0.373)\\ 
     & RBRHS    & 14.560(0.608) &0.070(0.256)&0.982(0.022)&0.982(0.022)&1.876(0.377)\\
         & HSBQR   &10.830(2.025) &0.000(0.000)&0.831(0.094)&0.843(0.082)&2.547(0.808)  \\
         
\hline
Error 2         &     RBHS   &7.340(3.367)&0.030(0.171) &0.624(0.220)&0.682(0.167)&3.909(1.306)\\
   $t$(2)          & RBHS+   &7.980(3.464) & 0.030(0.171) &0.662(0.214)&0.703(0.176)&3.415(1.545)\\ 
     & RBRHS    & 5.700(3.691) &0.000(0.000)&0.505(0.266)&0.601(0.196)&4.821(1.794)\\
         & HSBQR  &7.630(2.529) &0.040(0.243)&0.656(0.158)&0.696(0.131)&4.407(1.380)  \\
         
\hline
Error 3    & RBHS   &12.670(1.798)&0.030(0.171)&0.910(0.074)&0.914(0.068)&2.331(0.727) \\
 Laplace(0,1)   & RBHS+ &12.840(1.739) &0.030(0.171)&0.917(0.071)&0.920(0.066)&1.850(0.618)   \\  
          & RBRHS  &12.450(1.925) &0.010(0.100)  &0.901(0.081)&0.906(0.074)&2.391(0.695)\\
           & HSBQR   &9.290(2.575) &0.000(0.000) &0.751(0.133)&0.776(0.112)&3.299(1.057)\\

\hline
Error 4  &RBHS &11.720(2.184)&0.010(0.100)&0.869(0.102)&0.877(0.090)&2.531(0.724)\\
 80$\%$N(0,1)
 &RBHS+ &12.100(2.018)&0.030(0.171)&0.885(0.091)&0.892(0.081)&2.029(0.662)\\
+20$\%$N(0,3) &RBRHS&11.140(2.234)&0.010(0.100)&0.843(0.108)&0.854(0.093)&2.677(0.785)\\
 &HSBQR&8.430(2.516)&0.070(0.256)&0.702(0.147)&0.733(0.123)&3.777(1.274)\\
 
\hline
Error 5  &RBHS &12.100(2.385)&0.010(0.100)&0.884(0.106)&0.891(0.094)&2.058(0.833)\\
 Lognormal(0,1) &RBHS+&12.320(2.386)&0.010(0.100)&0.892(0.108)&0.899(0.097)&1.766(0.806)\\
 &RBRHS&11.400(2.864)&0.000(0.000)&0.849(0.141)&0.861(0.121)&2.430(1.011)\\
 &HSBQR&1.850(1.572)&0.140(0.403)&0.203(0.159)&0.357(0.134)&8.539(1.565)\\

\hline 
\end{tabular} }
\end{center}
\centering
\end{table}

\clearpage
\subsection{Statistical inference}
\begin{table}[ht!]
\def\arraystretch{1.2}
\begin{center}
\caption{95\% nominal coverage probabilities assessed over 1,000 replicates for RBHS, RBHS+, RBRHS, HSBQR, bayesreg and bayesQR for the datasets with AR(1) covariates, $n = 100$ and $p = 500$.}
\label{id1.00}
\centering
\fontsize{9}{11}\selectfont{
\begin{tabular}{ c c c c c c c c c}
\hline
&&&&& Methods\\
\cline{3-9}
  && RBHS & RBHS+ & RBRHS & HSBQR&bayesreg(HS)&bayesreg(HS+)&bayesQR \\
\hline
N(0,1) & Coverage of $\boldsymbol{\beta_1}$ \\
& $\beta_1$ & 0.936 & 0.940 & 0.935 & 0.890&0.928&0.926&-- \\
& $\beta_2$ & 0.924 & 0.920 & 0.931 & 0.910 &0.916&0.905&-- \\
& $\beta_3$ & 0.943 & 0.936 & 0.856 & 0.880 &0.932&0.927&-- \\
& Average length \\
& $\beta_1$ & 0.499 & 0.498 & 0.473 & 0.496 &0.488&0.485&-- \\
& $\beta_2$ & 0.551 & 0.547 & 0.512 & 0.537 &0.538&0.535&--\\
& $\beta_3$ & 0.494 & 0.491 & 0.471 & 0.484 &0.483&0.480&-- \\
\cline{2-9}
& Coverage of $\boldsymbol{\beta_2}$ & 1.000 & 1.000 & 1.000 & 1.000 &1.000&1.000&-- \\
& Average length & 0.100 & 0.082 & 0.090 & 0.061&0.100&0.084&--  \\
\hline
$t$(2) & Coverage of $\boldsymbol{\beta_1}$ \\
& $\beta_1$ & 0.928 & 0.934 & 0.932 & 0.900 &0.892&0.890&--\\
& $\beta_2$ & 0.952 & 0.956 & 0.966 & 0.930 &0.938&0.930&-- \\
& $\beta_3$ & 0.968 & 0.966 & 0.808 & 0.830 &0.986&0.976&-- \\
& Average length \\
& $\beta_1$ & 0.922 & 0.897 & 0.796 & 0.684 &0.928&0.926&-- \\
& $\beta_2$ & 0.987 & 0.956 & 0.795 & 0.738 &0.988&1.038&--\\
& $\beta_3$ & 0.823 & 0.812 & 0.729 & 0.630 &0.842&0.873&-- \\
\cline{2-9}
& Coverage of $\boldsymbol{\beta_2}$ & 1.000 & 1.000 & 1.000 & 1.000 &1.000&1.000&-- \\
& Average length & 0.101 & 0.079 & 0.082 & 0.101&0.100&0.098&--\\
\hline
\end{tabular}}
\end{center}
\centering
\end{table}

\clearpage
\subsection{Computational cost}
We have evaluated the computational efficiency of multiple robust Bayesian methods under an AR(1) correlation structure with $i.i.d.$ $t(2)$ errors. Table \ref{tab:comp_cost_ar1_t2} reports the total computation time (in seconds) required to complete 10,000 MCMC iterations across various combinations of sample size and dimensionality. As shown in Table \ref{tab:comp_cost_ar1_t2}, the proposed RBHS, RBHS+, and RBRHS demonstrate substantial computational efficiency over HSBQR. For example, under setting $(n, p) = (200, 600)$, all three methods complete 10{,}000 MCMC iterations in less than 4.5 seconds, whereas HSBQR requires over 490 seconds. As the sample size and dimensionality increase, the disparity becomes even more dramatic. For instance, under $(n, p) = (800, 1000)$, RBHS+ takes only 17.6 seconds, while HSBQR requires over 188,000 seconds (approximately 52 hours), making it less competitive for large-scale scenarios. These results further illustrates the merit of our proposals in high-dimensional regression with heavy-tailed model errors.

\begin{table}[htbp]
\def\arraystretch{1.2}
\begin{center}
\caption{Computational time (in seconds) for each method to complete 10{,}000 MCMC iterations under different settings of $(n, p)$ with AR(1) correlation and $i.i.d.$ $t(2)$ error.}
\label{tab:comp_cost_ar1_t2}
\fontsize{9}{11}\selectfont{
\begin{tabular}{cccccccc}
\hline
 $(n, p)$ & RBHS & RBHS+ & RBRHS & HSBQR & bayesreg(HS) & bayesreg(HS+)&bayesQR\\
\hline
(200, 100) &0.749&0.862&0.722&306.651&11.672&13.599&8.857\\
(200, 600)   & 3.483   & 4.386   & 3.486   & 490.624&122.750 &117.803 &-- \\
(400, 800)   & 7.426   & 8.618   & 7.370   & 39319.330 &439.664 &437.382&--\\
(800, 1000)  & 16.118  & 17.628  & 16.312  & 188918.800 &1366.460 &1346.601 &--\\
\hline
\end{tabular}}
\end{center}
\centering
\end{table}
\clearpage



				
				
				

\section{Posterior Inference}
\subsection{RBHS}
\subsubsection{Hierarchical Model Specification}
\begin{equation*}
\begin{aligned}
y_i \mid \beta_0,\boldsymbol{x_i}, \boldsymbol{\beta}, \tilde{v}_i &\sim \text{N}(\beta_0+\boldsymbol{x_i}^\top\boldsymbol{\beta}, \tau^{-1}\xi^2\tilde{v}_i), \\
\beta_0&\sim\text{N}(0,\sigma^2_{\beta_0}),\\
\tilde{v}_{i}|\tau&\sim\text{Exp}(\tau^{-1}), \\
\tau &\sim \text{Gamma}(e, f),\\
\beta_j \mid s_j, \lambda &\sim \text{N}(0, \lambda^2 s_j^2),\\
s_j^2 \mid \nu_j &\sim \text{Inverse-Gamma}(1/2, 1/\nu_j), \\
\lambda^2 \mid \xi_1 &\sim \text{Inverse-Gamma}(1/2, 1/\xi_1), \\
\nu_j &\sim \text{Inverse-Gamma}(1/2, 1),\\
\xi_1 &\sim \text{Inverse-Gamma}(1/2, 1).\\
\end{aligned}
\end{equation*}
\subsubsection{Joint Likelihood}
The joint likelihood of the unknown parameters, given the observed data, is expressed as
\begin{equation*}
\begin{aligned}
\pi(\tilde{v}_i, \beta_0, \boldsymbol{\beta}, \tau, s^2_j, \lambda^2, \nu^2_j, \xi_1 \mid y_i,\boldsymbol{x_i}) \\
&\propto \prod_{i=1}^n \left[\frac{1}{\sqrt{2\pi \tau^{-1}\xi^2 \tilde{v}_i}} 
\exp\left(-\frac{(y_i - \beta_0 - \boldsymbol{x}_i^\top\boldsymbol{\beta})^2}{2 \tau^{-1} \xi^2 \tilde{v}_i}\right)\right] \\
& \times \prod_{i=1}^n \left[\tau \exp(-\tau \tilde{v}_i)\right] \\
& \times \exp\left(-\frac{\beta_0^2}{2\sigma^2_{\beta_0}}\right) \\
& \times \prod_{j=1}^p \left[\frac{1}{\sqrt{2\pi \lambda^2 s_j^2}} \exp\left(-\frac{\beta_j^2}{2\lambda^2 s_j^2}\right)\right] \\
& \times \prod_{j=1}^p \left[ (s_j^2)^{-3/2} \exp\left(-\frac{1}{\nu_j s_j^2}\right) \right] \\
& \times \prod_{j=1}^p \left[ (\nu_j)^{-3/2} \exp\left(-\frac{1}{\nu_j}\right) \right] \\
& \times (\xi_1)^{-3/2} \exp\left(-\frac{1}{\xi_1}\right) \\
& \times (\lambda^2)^{-3/2} \exp\left(-\frac{1}{\xi_1 \lambda^2}\right) \\
& \times \tau^{e - 1} \exp(-f \tau)
\end{aligned}
\end{equation*}
\subsubsection{Gibbs Sampler}
\begin{itemize}
\item The full conditional distribution of $\beta_0$,
\begin{equation*}
\begin{aligned}
\pi(\beta_0 \mid \text{rest}) &\propto \exp\left( - \sum_{i=1}^n \frac{(y_i - \beta_0 - \boldsymbol{x}_i^\top \boldsymbol{\beta})^2}{2 \tau^{-1} \xi^2 \tilde{v}_i} - \frac{\beta_0^2}{2\sigma_{\beta_0}^2} \right) \\
&\propto \exp\left( - \sum_{i=1}^n \frac{-2 y_i \beta_0 + 2 \beta_0 \boldsymbol{x}_i^\top \boldsymbol{\beta} + \beta_0^2}{2 \tau^{-1} \xi^2 \tilde{v}_i} - \frac{\beta_0^2}{2\sigma_{\beta_0}^2} \right) \\
&\propto \exp\left( \beta_0 \sum_{i=1}^n \frac{y_i - \boldsymbol{x}_i^\top \boldsymbol{\beta}}{\tau^{-1} \xi^2 \tilde{v}_i} - \frac{1}{2} \beta_0^2 \left( \sum_{i=1}^n \frac{1}{\tau^{-1} \xi^2 \tilde{v}_i} + \frac{1}{\sigma_{\beta_0}^2} \right) \right) \\
&\propto \exp\left( -\frac{1}{2} \left( \left( \sum_{i=1}^n \frac{1}{\tau^{-1} \xi^2 \tilde{v}_i} + \frac{1}{\sigma_{\beta_0}^2} \right) \beta_0^2 - 2 \beta_0 \sum_{i=1}^n \frac{y_i - \boldsymbol{x}_i^\top \boldsymbol{\beta}}{\tau^{-1} \xi^2 \tilde{v}_i} \right) \right)
\end{aligned}
\end{equation*}
Hence,  \[
    \beta_0 \mid \text{rest} \sim \text{N}(\mu_{\beta_0}, \Sigma_{\beta_0}^2),
    \]
    where
    \[
    \Sigma_{\beta_0}^{-2} = \sum_{i=1}^n \frac{1}{\tau^{-1} \xi^2 \tilde{v}_i} + \frac{1}{\sigma_{\beta_0}^2}, \quad 
    \mu_{\beta_0} = \Sigma_{\beta_0}^2 \sum_{i=1}^n \frac{y_i - \boldsymbol{x}_i^\top \boldsymbol{\beta}}{\tau^{-1} \xi^2 \tilde{v}_i}.
    \]
\item The full conditional distribution of $\beta_j$,
\begin{align*}
\pi(\beta_j \mid \text{rest}) 
&\propto \exp\left( - \sum_{i=1}^n \frac{(y_i - \beta_0 - \sum_{k \neq j} x_{ik} \beta_k - x_{ij} \beta_j)^2}{2 \tau^{-1} \xi^2 \tilde{v}_i} - \frac{\beta_j^2}{2 \lambda^2 s_j^2} \right) \\
&\propto \exp\left( \beta_j \sum_{i=1}^n \frac{x_{ij}(y_i - \beta_0 - \sum_{k \neq j} x_{ik} \beta_k)}{\tau^{-1} \xi^2 \tilde{v}_i} 
- \frac{1}{2} \beta_j^2 \left( \sum_{i=1}^n \frac{x_{ij}^2}{\tau^{-1} \xi^2 \tilde{v}_i} + \frac{1}{\lambda^2 s_j^2} \right) \right)
\end{align*}
Therefore, 
\[
    \beta_j \mid \text{rest} \sim \text{N}(\mu_{\beta_j}, \sigma_{\beta_j}^2),
    \]
    where
    \[
    \sigma_{\beta_j}^{-2} = \sum_{i=1}^n \frac{x_{ij}^2}{\tau^{-1} \xi^2 \tilde{v}_i} + \frac{1}{\lambda^2 s_j^2}, \quad 
    \mu_{\beta_j} = \sigma_{\beta_j}^2 \sum_{i=1}^n \frac{x_{ij} (y_i - \beta_0-\sum_{k \neq j} x_{ik} \beta_k)}{\tau^{-1} \xi^2 \tilde{v}_i}.
    \]
\item The full conditional distribution of $\tilde{v}_i$,
\begin{equation*}
\begin{aligned}
\pi(\tilde{v}_i \mid \text{rest}) &\propto \left( \frac{1}{\sqrt{2\pi \tau^{-1} \xi^2 \tilde{v}_i}} \exp\left( -\frac{(y_i - \beta_0 - \boldsymbol{x}_i^\top \boldsymbol{\beta})^2}{2 \tau^{-1} \xi^2 \tilde{v}_i} \right) \right) \tau \exp(-\tau \tilde{v}_i)\\
&\propto \tilde{v}_i^{-1/2} \exp\left( -\frac{(y_i - \beta_0 - \boldsymbol{x}_i^\top \boldsymbol{\beta})^2}{2 \tau^{-1} \xi^2 \tilde{v}_i} \right) \exp(-\tau \tilde{v}_i) \\
&\propto \tilde{v}_i^{-1/2} \exp\left\{ -\frac{1}{2} \left( \frac{(y_i - \beta_0 - \boldsymbol{x}_i^\top \boldsymbol{\beta})^2}{ \tau^{-1}\xi^2 \tilde{v}_i} + 2 \tau \tilde{v}_i \right) \right\}
\end{aligned}
\end{equation*}
Hence, 
\[
    \tilde{v}_i \mid \text{rest} \sim \text{Inverse-Gaussian}\left(\sqrt{\frac{2\xi^2}{(y_i - \beta_0 - \boldsymbol{x}_i^\top \boldsymbol{\beta})^2}}, 2\tau \right),
    \]
\item The full conditional distribution of $\tau$,
\begin{equation*}
\begin{aligned}
\pi(\tau \mid \text{rest}) 
&\propto \prod_{i=1}^n \left[ \frac{1}{\sqrt{2\pi \tau^{-1} \xi^2 \tilde{v}_i}} 
\exp\left( -\frac{(y_i - \beta_0 - \boldsymbol{x}_i^\top \boldsymbol{\beta})^2}{2 \tau^{-1} \xi^2 \tilde{v}_i} \right) 
\tau \exp(-\tau \tilde{v}_i) \right] 
\tau^{e - 1} \exp(-f \tau) \\
&\propto \prod_{i=1}^n \left[ \tau^{1/2} 
\exp\left( -\frac{\tau (y_i - \beta_0 - \boldsymbol{x}_i^\top \boldsymbol{\beta})^2}{2 \xi^2 \tilde{v}_i} \right) 
\tau \exp(-\tau \tilde{v}_i) \right] 
\tau^{e - 1} \exp(-f \tau) \\
&\propto \tau^{3n/2} 
\exp\left( -\frac{\tau}{2} \sum_{i=1}^n \frac{(y_i - \beta_0 - \boldsymbol{x}_i^\top \boldsymbol{\beta})^2}{\xi^2 \tilde{v}_i} \right) 
\exp\left( -\tau \sum_{i=1}^n \tilde{v}_i \right) 
\tau^{e - 1} 
\exp(-f \tau) \\
&\propto \tau^{e + \frac{3n}{2} - 1} 
\exp\left( -\tau \left( f + \sum_{i=1}^n \tilde{v}_i + \frac{1}{2} \sum_{i=1}^n \frac{(y_i - \beta_0 - \boldsymbol{x}_i^\top \boldsymbol{\beta})^2}{\xi^2 \tilde{v}_i} \right) \right)
\end{aligned}
\end{equation*}
Therefore,
\[
    \tau \mid \text{rest} \sim \text{Gamma}\left(e + \frac{3n}{2}, f + \sum_{i=1}^n \tilde{v}_i + \sum_{i=1}^n \frac{(y_i - \beta_0 - \boldsymbol{x}_i^\top \boldsymbol{\beta})^2}{2 \xi^2 \tilde{v}_i}\right).
    \]
\item The full conditional distribution of $s^2_j$,
\begin{equation*}
\begin{aligned}
\pi(s_j^2 \mid \text{rest}) 
&\propto \frac{1}{\sqrt{2\pi \lambda^2 s_j^2}} \exp\left( -\frac{\beta_j^2}{2 \lambda^2 s_j^2} \right) 
(s_j^2)^{-3/2} \exp\left( -\frac{1}{\nu_j s_j^2} \right) \\
&\propto (s_j^2)^{-1/2} \exp\left( -\frac{\beta_j^2}{2 \lambda^2 s_j^2} \right) 
(s_j^2)^{-3/2} \exp\left( -\frac{1}{\nu_j s_j^2} \right) \\
&\propto (s_j^2)^{-2} 
\exp\left( -\left( \frac{\beta_j^2}{2 \lambda^2} + \frac{1}{\nu_j} \right) \frac{1}{s_j^2} \right)
\end{aligned}
\end{equation*}
Therefore,
\[
    s_j^2 \mid \text{rest} \sim \text{Inverse-Gamma}\left(1, \frac{\beta_j^2}{2\lambda^2} + \frac{1}{\nu_j}\right).
    \]
\item The full conditional distribution of $\nu_j$,
\begin{equation*}
\begin{aligned}
\pi(\nu_j \mid \text{rest}) 
&\propto \nu_j^{-1/2} (s_j^2)^{-3/2} \exp\left( -\frac{1}{\nu_j s_j^2} \right) 
\nu_j^{-3/2} \exp\left( -\frac{1}{\nu_j} \right) \\
&\propto \nu_j^{-2} \exp\left( -\frac{1}{\nu_j} \left( \frac{1}{s_j^2} + 1 \right) \right)
\end{aligned}
\end{equation*}
Hence,
    \[
    \nu_j \mid \text{rest} \sim \text{Inverse-Gamma}\left(1, 1 + \frac{1}{s_j^2}\right).
    \]
\item The full conditional distribution of $\lambda^2$,
\begin{equation*}
\begin{aligned}
\pi(\lambda^2 \mid \text{rest}) 
&\propto \prod_{j=1}^p \frac{1}{\sqrt{2\pi \lambda^2 s_j^2}} \exp\left( -\frac{\beta_j^2}{2 \lambda^2 s_j^2} \right) 
(\lambda^2)^{-3/2} \exp\left( -\frac{1}{\xi_1 \lambda^2} \right) \\
&\propto (\lambda^2)^{-p/2} \exp\left( -\sum_{j=1}^p \frac{\beta_j^2}{2 \lambda^2 s_j^2} \right) 
(\lambda^2)^{-3/2} \exp\left( -\frac{1}{\xi_1 \lambda^2} \right) \\
&\propto (\lambda^2)^{-(p + 3)/2} 
\exp\left( -\frac{1}{\lambda^2} \left( \sum_{j=1}^p \frac{\beta_j^2}{2 s_j^2} + \frac{1}{\xi_1} \right) \right)
\end{aligned}
\end{equation*}
Therefore,
\[
    \lambda^2 \mid \text{rest} \sim \text{Inverse-Gamma}\left(\frac{p + 1}{2}, \frac{1}{\xi_1} + \frac{1}{2} \sum_{j=1}^p \frac{\beta_j^2}{s_j^2}\right).
    \]
\item The full conditional distribution of $\xi_1$,
\begin{align*}
\pi(\xi_1 \mid \text{rest}) 
&\propto \xi_1^{-1/2}(\lambda^2)^{-3/2} \exp\left( -\frac{1}{\xi_1 \lambda^2} \right) 
\xi_1^{-3/2} \exp\left( -\frac{1}{\xi_1} \right) \\
&\propto \xi_1^{-2} \exp\left( -\frac{1}{\xi_1} \left( \frac{1}{\lambda^2} + 1 \right) \right)
\end{align*}
Therefore,
\[
    \xi_1 \mid \text{rest} \sim \text{Inverse-Gamma}\left(1, 1 + \frac{1}{\lambda^2}\right).
    \]
\end{itemize}

\subsection{RBHS+}
\subsubsection{Hierarchical Model Specification}
\begin{equation*}
\begin{aligned}
y_i \mid \beta_0,\boldsymbol{x_i}, \boldsymbol{\beta}, \tilde{v}_i &\sim \text{N}(\beta_0+\boldsymbol{x_i}^\top\boldsymbol{\beta}, \tau^{-1}\xi^2\tilde{v}_i), \\
\beta_0 &\sim \text{N}(0, \sigma^2_{\beta_0}), \\
\tilde{v}_{i} \mid \tau &\sim \text{Exp}(\tau^{-1}), \\
\tau &\sim \text{Gamma}(e, f),\\
\beta_j \mid s_j, \lambda &\sim \text{N}(0, \lambda^2 s_j^2),\\
s_j^2 \mid \nu_j &\sim \text{Inverse-Gamma}(1/2, 1/\nu_j), \\
\nu_j \mid \phi^2_j &\sim \text{Inverse-Gamma}(1/2, 1/\phi^2_j),\\
\phi^2_j \mid \zeta_j &\sim \text{Inverse-Gamma}(1/2, 1/\zeta_j),\\
\zeta_j &\sim \text{Inverse-Gamma}(1/2, 1),\\
\lambda^2 \mid \xi_1 &\sim \text{Inverse-Gamma}(1/2, 1/\xi_1), \\
\xi_1 &\sim \text{Inverse-Gamma}(1/2, 1).\\
\end{aligned}
\end{equation*}
\subsubsection{Joint Likelihood}
The joint likelihood of the unknown parameters, given the observed data, is expressed as
\begin{equation*}
\begin{aligned}
\pi(\tilde{v}_i, \beta_0, \boldsymbol{\beta}, \tau, s^2_j, \nu_j, \phi_j^2, \zeta_j, \lambda^2, \xi_1 \mid y_i, \boldsymbol{x}_i) \\
&\propto \prod_{i=1}^n \left[ \frac{1}{\sqrt{2\pi \tau^{-1} \xi^2 \tilde{v}_i}} \exp\left\{ -\frac{(y_i - \beta_0 - \boldsymbol{x}_i^\top \boldsymbol{\beta})^2}{2 \tau^{-1} \xi^2 \tilde{v}_i} \right\} \right] \\
&\times \prod_{i=1}^n \left[ \tau \exp(-\tau \tilde{v}_i) \right] \\
&\times \exp\left(-\frac{\beta_0^2}{2\sigma^2_{\beta_0}}\right) \\
&\times \prod_{j=1}^p \left[ \frac{1}{\sqrt{2\pi \lambda^2 s_j^2}} \exp\left( -\frac{\beta_j^2}{2 \lambda^2 s_j^2} \right) \right] \\
&\times \prod_{j=1}^p \left[ (s_j^2)^{-3/2} \exp\left(-\frac{1}{\nu_j s_j^2} \right) \right] \\
&\times \prod_{j=1}^p \left[ (\nu_j)^{-3/2} \exp\left(-\frac{1}{\phi_j^2 \nu_j} \right) \right] \\
&\times \prod_{j=1}^p \left[ (\phi_j^2)^{-3/2} \exp\left(-\frac{1}{\zeta_j \phi_j^2} \right) \right] \\
&\times \prod_{j=1}^p \left[ (\zeta_j)^{-3/2} \exp(-\frac{1}{\zeta_j}) \right] \\
&\times (\lambda^2)^{-3/2} \exp\left(-\frac{1}{\xi_1 \lambda^2} \right) \\
&\times (\xi_1)^{-3/2} \exp\left(-\frac{1}{\xi_1} \right) \\
&\times \tau^{e - 1} \exp(-f \tau)
\end{aligned}
\end{equation*}
\subsubsection{Gibbs Sampler}
\begin{itemize}
\item The full conditional distribution of $\beta_0$,
\begin{align*}
\pi(\beta_0 \mid \text{rest}) 
&\propto \exp\left( - \sum_{i=1}^n \frac{(y_i - \beta_0 - \boldsymbol{x}_i^\top \boldsymbol{\beta})^2}{2 \tau^{-1} \xi^2 \tilde{v}_i} - \frac{\beta_0^2}{2\sigma_{\beta_0}^2} \right) \\
&\propto \exp\left( - \sum_{i=1}^n \frac{-2 y_i \beta_0 + 2 \beta_0 \boldsymbol{x}_i^\top \boldsymbol{\beta} + \beta_0^2}{2 \tau^{-1} \xi^2 \tilde{v}_i} - \frac{\beta_0^2}{2\sigma_{\beta_0}^2} \right) \\
&\propto \exp\left( \beta_0 \sum_{i=1}^n \frac{y_i - \boldsymbol{x}_i^\top \boldsymbol{\beta}}{\tau^{-1} \xi^2 \tilde{v}_i}
- \frac{1}{2} \beta_0^2 \left( \sum_{i=1}^n \frac{1}{\tau^{-1} \xi^2 \tilde{v}_i} + \frac{1}{\sigma_{\beta_0}^2} \right) \right)
\end{align*}
Hence,  \[
    \beta_0 \mid \text{rest} \sim \text{N}(\mu_{\beta_0}, \Sigma_{\beta_0}^2),
    \]
    where
    \[
    \Sigma_{\beta_0}^{-2} = \sum_{i=1}^n \frac{1}{\tau^{-1} \xi^2 \tilde{v}_i} + \frac{1}{\sigma_{\beta_0}^2}, \quad 
    \mu_{\beta_0} = \Sigma_{\beta_0}^2 \sum_{i=1}^n \frac{y_i - \boldsymbol{x}_i^\top \boldsymbol{\beta}}{\tau^{-1} \xi^2 \tilde{v}_i}.
    \]
\item The full conditional distribution of $\beta_j$,
\begin{align*}
\pi(\beta_j \mid \text{rest}) 
&\propto \exp\left( - \sum_{i=1}^n \frac{(y_i - \beta_0 - \sum_{k \neq j} x_{ik} \beta_k - x_{ij} \beta_j)^2}{2 \tau^{-1} \xi^2 \tilde{v}_i} - \frac{\beta_j^2}{2 \lambda^2 s_j^2} \right) \\
&\propto \exp\left( \beta_j \sum_{i=1}^n \frac{x_{ij}(y_i - \beta_0 - \sum_{k \neq j} x_{ik} \beta_k)}{\tau^{-1} \xi^2 \tilde{v}_i} 
- \frac{1}{2} \beta_j^2 \left( \sum_{i=1}^n \frac{x_{ij}^2}{\tau^{-1} \xi^2 \tilde{v}_i} + \frac{1}{\lambda^2 s_j^2} \right) \right)
\end{align*}
Therefore, 
\[
    \beta_j \mid \text{rest} \sim \text{N}(\mu_{\beta_j}, \sigma_{\beta_j}^2),
    \]
    where
    \[
    \sigma_{\beta_j}^{-2} = \sum_{i=1}^n \frac{x_{ij}^2}{\tau^{-1} \xi^2 \tilde{v}_i} + \frac{1}{\lambda^2 s_j^2}, \quad 
    \mu_{\beta_j} = \sigma_{\beta_j}^2 \sum_{i=1}^n \frac{x_{ij} (y_i - \beta_0-\sum_{k \neq j} x_{ik} \beta_k)}{\tau^{-1} \xi^2 \tilde{v}_i}.
    \]
\item The full conditional distribution of $\tilde{v}_i$,
\begin{align*}
\pi(\tilde{v}_i \mid \text{rest}) 
&\propto \tilde{v}_i^{-1/2} 
\exp\left( -\frac{(y_i - \beta_0 - \boldsymbol{x}_i^\top \boldsymbol{\beta})^2}{2 \tau^{-1} \xi^2 \tilde{v}_i} \right) 
\exp(-\tau \tilde{v}_i) \\
&\propto \tilde{v}_i^{-1/2} 
\exp\left\{ -\frac{1}{2} \left( \frac{2 \tau (y_i - \beta_0 - \boldsymbol{x}_i^\top \boldsymbol{\beta})^2}{\xi^2 \tilde{v}_i} + 2 \tau \tilde{v}_i \right) \right\}
\end{align*}
Hence, 
\[
    \tilde{v}_i \mid \text{rest} \sim \text{Inverse-Gaussian}\left(\sqrt{\frac{2\xi^2}{(y_i - \beta_0 - \boldsymbol{x}_i^\top \boldsymbol{\beta})^2}}, 2\tau \right),
    \]
\item The full conditional distribution of $\tau$,
\begin{align*}
\pi(\tau \mid \text{rest}) 
&\propto \tau^{3n/2} 
\exp\left( -\frac{\tau}{2} \sum_{i=1}^n \frac{(y_i - \beta_0 - \boldsymbol{x}_i^\top \boldsymbol{\beta})^2}{\xi^2 \tilde{v}_i} \right) 
\exp\left( -\tau \sum_{i=1}^n \tilde{v}_i \right) 
\tau^{e - 1} 
\exp(-f \tau) \\
&\propto \tau^{e + \frac{3n}{2} - 1} 
\exp\left( -\tau \left( f + \sum_{i=1}^n \tilde{v}_i + \frac{1}{2} \sum_{i=1}^n \frac{(y_i - \beta_0 - \boldsymbol{x}_i^\top \boldsymbol{\beta})^2}{\xi^2 \tilde{v}_i} \right) \right)
\end{align*}
Therefore,
\[
    \tau \mid \text{rest} \sim \text{Gamma}\left(e + \frac{3n}{2}, f + \sum_{i=1}^n \tilde{v}_i + \sum_{i=1}^n \frac{(y_i - \beta_0 - \boldsymbol{x}_i^\top \boldsymbol{\beta})^2}{2 \xi^2 \tilde{v}_i}\right).
    \]
\item The full conditional distribution of $s^2_j$,
\begin{align*}
\pi(s_j^2 \mid \text{rest}) 
&\propto (s_j^2)^{-1/2} \exp\left( -\frac{\beta_j^2}{2 \lambda^2 s_j^2} \right) 
(s_j^2)^{-3/2} \exp\left( -\frac{1}{\nu_j s_j^2} \right) \\
&\propto (s_j^2)^{-2} 
\exp\left( -\left( \frac{\beta_j^2}{2 \lambda^2} + \frac{1}{\nu_j} \right) \frac{1}{s_j^2} \right)
\end{align*}
Therefore,
\[
    s_j^2 \mid \text{rest} \sim \text{Inverse-Gamma}\left(1, \frac{\beta_j^2}{2\lambda^2} + \frac{1}{\nu_j}\right).
    \]
\item The full conditional distribution of $\nu_j$,
\begin{align*}
\pi(\nu_j \mid \text{rest}) 
&\propto \nu_j^{-2} \exp\left( -\frac{1}{\nu_j} \left( \frac{1}{s_j^2} + \frac{1}{\phi_j^2} \right) \right)
\end{align*}
Hence,
    \[
    \nu_j \mid \text{rest} \sim \text{Inverse-Gamma}\left(1, \frac{1}{s_j^2}+\frac{1}{\phi_j^2}\right).
    \]
\item The full conditional distribution of $\phi^2_j$,
\begin{align*}
\pi(\phi_j^2 \mid \text{rest}) 
&\propto (\phi_j^2)^{-2} \exp\left( -\frac{1}{\phi_j^2} \left( \frac{1}{\nu_j} + \frac{1}{\zeta_j} \right) \right)
\end{align*}
Hence,
    \[
    \phi^2_j \mid \text{rest} \sim \text{Inverse-Gamma}\left(1, \frac{1}{\nu_j}+\frac{1}{\zeta_j}\right).
    \]
\item The full conditional distribution of $\zeta_j$,
\begin{align*}
\pi(\zeta_j \mid \text{rest}) 
&\propto \zeta_j^{-2} \exp\left( -\frac{1}{\zeta_j} \left( \frac{1}{\phi_j^2} + 1 \right) \right)
\end{align*}
Hence,
    \[
    \zeta_j \mid \text{rest} \sim \text{Inverse-Gamma}\left(1, \frac{1}{\phi^2_j}+1\right).
    \]
\item The full conditional distribution of $\lambda^2$,
\begin{align*}
\pi(\lambda^2 \mid \text{rest}) 
&\propto (\lambda^2)^{-p/2} \exp\left( -\sum_{j=1}^p \frac{\beta_j^2}{2 \lambda^2 s_j^2} \right) 
(\lambda^2)^{-3/2} \exp\left( -\frac{1}{\xi_1 \lambda^2} \right) \\
&\propto (\lambda^2)^{-(p + 3)/2} 
\exp\left( -\frac{1}{\lambda^2} \left( \sum_{j=1}^p \frac{\beta_j^2}{2 s_j^2} + \frac{1}{\xi_1} \right) \right)
\end{align*}
Therefore,
\[
    \lambda^2 \mid \text{rest} \sim \text{Inverse-Gamma}\left(\frac{p + 1}{2}, \frac{1}{\xi_1} + \frac{1}{2} \sum_{j=1}^p \frac{\beta_j^2}{s_j^2}\right).
    \]
\item The full conditional distribution of $\xi_1$,
\begin{align*}
\pi(\xi_1 \mid \text{rest}) 
&\propto \xi_1^{-2} \exp\left( -\frac{1}{\xi_1} \left( \frac{1}{\lambda^2} + 1 \right) \right)
\end{align*}
Hence,
\[
    \xi_1 \mid \text{rest} \sim \text{Inverse-Gamma}\left(1, 1 + \frac{1}{\lambda^2}\right).
    \]
\end{itemize}
\subsection{RBRHS}
\subsubsection{Hierarchical Model Specification}
\begin{equation*}
\begin{aligned}
y_i \mid \beta_0,\boldsymbol{x_i}, \boldsymbol{\beta}, \tilde{v}_i &\sim \text{N}(\beta_0+\boldsymbol{x_i}^\top\boldsymbol{\beta}, \tau^{-1}\xi^2\tilde{v}_i), \\
\beta_0&\sim\text{N}(0,\sigma^2_{\beta_0}),\\
\tilde{v}_{i}|\tau&\sim\text{Exp}(\tau^{-1}), \\
\tau &\sim \text{Gamma}(e, f),\\
\beta_j \mid s_j, \lambda, b &\sim \text{N}(0, \lambda^2 s_j^2) \, \text{N}(0, b^2),\\
s_j^2 \mid \nu_j &\sim \text{Inverse-Gamma}(1/2, 1/\nu_j), \\
\lambda^2 \mid \xi_1 &\sim \text{Inverse-Gamma}(1/2, 1/\xi_1), \\
\nu_j &\sim \text{Inverse-Gamma}(1/2, 1),\\
\xi_1 &\sim \text{Inverse-Gamma}(1/2, 1),\\
b^2 &\sim \text{Inverse-Gamma}(c/2, d/2).\\
\end{aligned}
\end{equation*}
\subsubsection{Joint Likelihood}
The joint likelihood of the unknown parameters, given the observed data, is expressed as
\begin{equation*}
\begin{aligned}
\pi(\tilde{v}_i, \beta_0, \boldsymbol{\beta}, \tau, s^2_j, \lambda^2, b^2, \nu^2_j, \xi_1 \mid y_i,\boldsymbol{x_i}) \\
&\propto \prod_{i=1}^n \left[\frac{1}{\sqrt{2\pi \tau^{-1}\xi^2 \tilde{v}_i}} 
\exp\left(-\frac{(y_i - \beta_0 - \boldsymbol{x}_i^\top\boldsymbol{\beta})^2}{2 \tau^{-1} \xi^2 \tilde{v}_i}\right)\right] \\
& \times \prod_{i=1}^n \left[\tau \exp(-\tau \tilde{v}_i)\right] \\
& \times \exp\left(-\frac{\beta_0^2}{2\sigma^2_{\beta_0}}\right) \\
& \times \prod_{j=1}^p \left[\frac{1}{\sqrt{2\pi \lambda^2 s_j^2}} \exp\left(-\frac{\beta_j^2}{2\lambda^2 s_j^2}\right) \frac{1}{\sqrt{2\pi b^2}} \exp\left(-\frac{\beta_j^2}{2b^2}\right)\right] \\
& \times \prod_{j=1}^p \left[ (s_j^2)^{-3/2} \exp\left(-\frac{1}{\nu_j s_j^2}\right) \right] \\
& \times \prod_{j=1}^p \left[ (\nu_j)^{-3/2} \exp\left(-\frac{1}{\nu_j}\right) \right] \\
& \times (\xi_1)^{-3/2} \exp\left(-\frac{1}{\xi_1}\right) \\
& \times (\lambda^2)^{-3/2} \exp\left(-\frac{1}{\xi_1 \lambda^2}\right) \\
& \times (b^2)^{-(c/2 + 1)} \exp\left(-\frac{d}{2 b^2}\right) \\
& \times \tau^{e - 1} \exp(-f \tau)
\end{aligned}
\end{equation*}
\subsubsection{Gibbs Sampler}
\begin{itemize}
\item The full conditional distribution of $\beta_0$,
\begin{equation*}
\begin{aligned}
\pi(\beta_0 \mid \text{rest}) &\propto \exp\left( - \sum_{i=1}^n \frac{(y_i - \beta_0 - \boldsymbol{x}_i^\top \boldsymbol{\beta})^2}{2 \tau^{-1} \xi^2 \tilde{v}_i} - \frac{\beta_0^2}{2\sigma_{\beta_0}^2} \right) \\
&\propto \exp\left( - \sum_{i=1}^n \frac{y_i^2 - 2 y_i \beta_0 - 2 y_i \boldsymbol{x}_i^\top \boldsymbol{\beta} + \beta_0^2 + 2 \beta_0 \boldsymbol{x}_i^\top \boldsymbol{\beta} + (\boldsymbol{x}_i^\top \boldsymbol{\beta})^2}{2 \tau^{-1} \xi^2 \tilde{v}_i} - \frac{\beta_0^2}{2\sigma_{\beta_0}^2} \right) \\
&\propto \exp\left( - \sum_{i=1}^n \frac{-2 y_i \beta_0 + 2 \beta_0 \boldsymbol{x}_i^\top \boldsymbol{\beta} + \beta_0^2}{2 \tau^{-1} \xi^2 \tilde{v}_i} - \frac{\beta_0^2}{2\sigma_{\beta_0}^2} \right) \\
&\propto \exp\left( \beta_0 \sum_{i=1}^n \frac{y_i - \boldsymbol{x}_i^\top \boldsymbol{\beta}}{\tau^{-1} \xi^2 \tilde{v}_i} - \frac{1}{2} \beta_0^2 \left( \sum_{i=1}^n \frac{1}{\tau^{-1} \xi^2 \tilde{v}_i} + \frac{1}{\sigma_{\beta_0}^2} \right) \right) \\
&\propto \exp\left( -\frac{1}{2} \left( \left( \sum_{i=1}^n \frac{1}{\tau^{-1} \xi^2 \tilde{v}_i} + \frac{1}{\sigma_{\beta_0}^2} \right) \beta_0^2 - 2 \beta_0 \sum_{i=1}^n \frac{y_i - \boldsymbol{x}_i^\top \boldsymbol{\beta}}{\tau^{-1} \xi^2 \tilde{v}_i} \right) \right)
\end{aligned}
\end{equation*}
Hence,  \[
    \beta_0 \mid \text{rest} \sim \text{N}(\mu_{\beta_0}, \Sigma_{\beta_0}^2),
    \]
    where
    \[
    \Sigma_{\beta_0}^{-2} = \sum_{i=1}^n \frac{1}{\tau^{-1} \xi^2 \tilde{v}_i} + \frac{1}{\sigma_{\beta_0}^2}, \quad 
    \mu_{\beta_0} = \Sigma_{\beta_0}^2 \sum_{i=1}^n \frac{y_i - \boldsymbol{x}_i^\top \boldsymbol{\beta}}{\tau^{-1} \xi^2 \tilde{v}_i}.
    \]
\item The full conditional distribution of $\beta_j$,
\begin{equation*}
\begin{aligned}
\pi(\beta_j \mid \text{rest}) &\propto \exp\left( - \sum_{i=1}^n \frac{(y_i - \beta_0 - \boldsymbol{x}_i^\top \boldsymbol{\beta})^2}{2 \tau^{-1} \xi^2 \tilde{v}_i} - \frac{\beta_j^2}{2 \lambda^2 s_j^2} - \frac{\beta_j^2}{2 b^2} \right) \\
&\propto \exp\left( - \sum_{i=1}^n \frac{(y_i - \beta_0 - \sum_{k \neq j} x_{ik} \beta_k - x_{ij} \beta_j)^2}{2 \tau^{-1} \xi^2 \tilde{v}_i} - \left( \frac{1}{2 \lambda^2 s_j^2} + \frac{1}{2 b^2} \right) \beta_j^2 \right) \\
&\propto \exp\left( - \sum_{i=1}^n \frac{(y_i - \beta_0 - \sum_{k \neq j} x_{ik} \beta_k)^2 - 2 x_{ij} (y_i - \beta_0 - \sum_{k \neq j} x_{ik} \beta_k) \beta_j + x_{ij}^2 \beta_j^2}{2 \tau^{-1} \xi^2 \tilde{v}_i} \right. \\
&\qquad\left. - \left( \frac{1}{2 \lambda^2 s_j^2} + \frac{1}{2 b^2} \right) \beta_j^2 \right) \\
&\propto \exp\left( \beta_j \sum_{i=1}^n \frac{x_{ij} (y_i - \beta_0 - \sum_{k \neq j} x_{ik} \beta_k)}{\tau^{-1} \xi^2 \tilde{v}_i}
- \frac{1}{2} \beta_j^2 \left( \sum_{i=1}^n \frac{x_{ij}^2}{\tau^{-1} \xi^2 \tilde{v}_i} + \frac{1}{\lambda^2 s_j^2} + \frac{1}{b^2} \right) \right) \\
&\propto \exp\left( -\frac{1}{2} \left( \left( \sum_{i=1}^n \frac{x_{ij}^2}{\tau^{-1} \xi^2 \tilde{v}_i} + \frac{1}{\lambda^2 s_j^2} + \frac{1}{b^2} \right) \beta_j^2 
- 2 \beta_j \sum_{i=1}^n \frac{x_{ij} (y_i - \beta_0 - \sum_{k \neq j} x_{ik} \beta_k)}{\tau^{-1} \xi^2 \tilde{v}_i} \right) \right)
\end{aligned}
\end{equation*}
Therefore, 
\[
    \beta_j \mid \text{rest} \sim \text{N}(\mu_{\beta_j}, \sigma_{\beta_j}^2),
    \]
    where
    \[
    \sigma_{\beta_j}^{-2} = \sum_{i=1}^n \frac{x_{ij}^2}{\tau^{-1} \xi^2 \tilde{v}_i} + \left(\frac{1}{\lambda^2 s_j^2} + \frac{1}{b^2}\right), \quad 
    \mu_{\beta_j} = \sigma_{\beta_j}^2 \sum_{i=1}^n \frac{x_{ij} (y_i - \beta_0-\sum_{k \neq j} x_{ik} \beta_k)}{\tau^{-1} \xi^2 \tilde{v}_i}.
    \]
\item The full conditional distribution of $\tilde{v}_i$,
\begin{equation*}
\begin{aligned}
\pi(\tilde{v}_i \mid \text{rest}) &\propto \left( \frac{1}{\sqrt{2\pi \tau^{-1} \xi^2 \tilde{v}_i}} \exp\left( -\frac{(y_i - \beta_0 - \boldsymbol{x}_i^\top \boldsymbol{\beta})^2}{2 \tau^{-1} \xi^2 \tilde{v}_i} \right) \right) \tau \exp(-\tau \tilde{v}_i)\\
&\propto \tilde{v}_i^{-1/2} \exp\left( -\frac{(y_i - \beta_0 - \boldsymbol{x}_i^\top \boldsymbol{\beta})^2}{2 \tau^{-1} \xi^2 \tilde{v}_i} \right) \exp(-\tau \tilde{v}_i) \\
&\propto \tilde{v}_i^{-1/2} \exp\left\{ -\frac{1}{2} \left( \frac{(y_i - \beta_0 - \boldsymbol{x}_i^\top \boldsymbol{\beta})^2}{ \tau^{-1}\xi^2 \tilde{v}_i} + 2 \tau \tilde{v}_i \right) \right\}
\end{aligned}
\end{equation*}
Hence, 
\[
    \tilde{v}_i \mid \text{rest} \sim \text{Inverse-Gaussian}\left(\sqrt{\frac{2\xi^2}{(y_i - \beta_0 - \boldsymbol{x}_i^\top \boldsymbol{\beta})^2}}, 2\tau \right),
    \]
\item The full conditional distribution of $\tau$,
\begin{equation*}
\begin{aligned}
\pi(\tau \mid \text{rest}) 
&\propto \prod_{i=1}^n \left[ \frac{1}{\sqrt{2\pi \tau^{-1} \xi^2 \tilde{v}_i}} 
\exp\left( -\frac{(y_i - \beta_0 - \boldsymbol{x}_i^\top \boldsymbol{\beta})^2}{2 \tau^{-1} \xi^2 \tilde{v}_i} \right) 
\tau \exp(-\tau \tilde{v}_i) \right] 
\tau^{e - 1} \exp(-f \tau) \\
&\propto \prod_{i=1}^n \left[ \tau^{1/2} 
\exp\left( -\frac{\tau (y_i - \beta_0 - \boldsymbol{x}_i^\top \boldsymbol{\beta})^2}{2 \xi^2 \tilde{v}_i} \right) 
\tau \exp(-\tau \tilde{v}_i) \right] 
\tau^{e - 1} \exp(-f \tau) \\
&\propto \tau^{3n/2} 
\exp\left( -\frac{\tau}{2} \sum_{i=1}^n \frac{(y_i - \beta_0 - \boldsymbol{x}_i^\top \boldsymbol{\beta})^2}{\xi^2 \tilde{v}_i} \right) 
\exp\left( -\tau \sum_{i=1}^n \tilde{v}_i \right) 
\tau^{e - 1} 
\exp(-f \tau) \\
&\propto \tau^{e + \frac{3n}{2} - 1} 
\exp\left( -\tau \left( f + \sum_{i=1}^n \tilde{v}_i + \frac{1}{2} \sum_{i=1}^n \frac{(y_i - \beta_0 - \boldsymbol{x}_i^\top \boldsymbol{\beta})^2}{\xi^2 \tilde{v}_i} \right) \right)
\end{aligned}
\end{equation*}
Therefore,
\[
    \tau \mid \text{rest} \sim \text{Gamma}\left(e + \frac{3n}{2}, f + \sum_{i=1}^n \tilde{v}_i + \sum_{i=1}^n \frac{(y_i - \beta_0 - \boldsymbol{x}_i^\top \boldsymbol{\beta})^2}{2 \xi^2 \tilde{v}_i}\right).
    \]
\item The full conditional distribution of $s^2_j$,
\begin{equation*}
\begin{aligned}
\pi(s_j^2 \mid \text{rest}) 
&\propto \frac{1}{\sqrt{2\pi \lambda^2 s_j^2}} \exp\left( -\frac{\beta_j^2}{2 \lambda^2 s_j^2} \right) 
(s_j^2)^{-3/2} \exp\left( -\frac{1}{\nu_j s_j^2} \right) \\
&\propto (s_j^2)^{-1/2} \exp\left( -\frac{\beta_j^2}{2 \lambda^2 s_j^2} \right) 
(s_j^2)^{-3/2} \exp\left( -\frac{1}{\nu_j s_j^2} \right) \\
&\propto (s_j^2)^{-2} 
\exp\left( -\left( \frac{\beta_j^2}{2 \lambda^2} + \frac{1}{\nu_j} \right) \frac{1}{s_j^2} \right)
\end{aligned}
\end{equation*}
Therefore,
\[
    s_j^2 \mid \text{rest} \sim \text{Inverse-Gamma}\left(1, \frac{\beta_j^2}{2\lambda^2} + \frac{1}{\nu_j}\right).
    \]
\item The full conditional distribution of $\nu_j$,
\begin{equation*}
\begin{aligned}
\pi(\nu_j \mid \text{rest}) 
&\propto \nu_j^{-1/2} (s_j^2)^{-3/2} \exp\left( -\frac{1}{\nu_j s_j^2} \right) 
\nu_j^{-3/2} \exp\left( -\frac{1}{\nu_j} \right) \\
&\propto \nu_j^{-2} \exp\left( -\frac{1}{\nu_j} \left( \frac{1}{s_j^2} + 1 \right) \right)
\end{aligned}
\end{equation*}
Hence,
    \[
    \nu_j \mid \text{rest} \sim \text{Inverse-Gamma}\left(1, 1 + \frac{1}{s_j^2}\right).
    \]
\item The full conditional distribution of $\lambda^2$,
\begin{equation*}
\begin{aligned}
\pi(\lambda^2 \mid \text{rest}) 
&\propto \prod_{j=1}^p \frac{1}{\sqrt{2\pi \lambda^2 s_j^2}} \exp\left( -\frac{\beta_j^2}{2 \lambda^2 s_j^2} \right) 
(\lambda^2)^{-3/2} \exp\left( -\frac{1}{\xi_1 \lambda^2} \right) \\
&\propto (\lambda^2)^{-p/2} \exp\left( -\sum_{j=1}^p \frac{\beta_j^2}{2 \lambda^2 s_j^2} \right) 
(\lambda^2)^{-3/2} \exp\left( -\frac{1}{\xi_1 \lambda^2} \right) \\
&\propto (\lambda^2)^{-(p + 3)/2} 
\exp\left( -\frac{1}{\lambda^2} \left( \sum_{j=1}^p \frac{\beta_j^2}{2 s_j^2} + \frac{1}{\xi_1} \right) \right)
\end{aligned}
\end{equation*}
Therefore,
\[
    \lambda^2 \mid \text{rest} \sim \text{Inverse-Gamma}\left(\frac{p + 1}{2}, \frac{1}{\xi_1} + \frac{1}{2} \sum_{j=1}^p \frac{\beta_j^2}{s_j^2}\right).
    \]
\item The full conditional distribution of $\xi_1$,
\begin{align*}
\pi(\xi_1 \mid \text{rest}) 
&\propto \xi_1^{-1/2}(\lambda^2)^{-3/2} \exp\left( -\frac{1}{\xi_1 \lambda^2} \right) 
\xi_1^{-3/2} \exp\left( -\frac{1}{\xi_1} \right) \\
&\propto \xi_1^{-2} \exp\left( -\frac{1}{\xi_1} \left( \frac{1}{\lambda^2} + 1 \right) \right)
\end{align*}
Therefore,
\[
    \xi_1 \mid \text{rest} \sim \text{Inverse-Gamma}\left(1, 1 + \frac{1}{\lambda^2}\right).
    \]
\item The full conditional distribution of $b^2$,
\begin{align*}
\pi(b^2 \mid \text{rest}) 
&\propto \prod_{j=1}^p \frac{1}{\sqrt{2\pi b^2}} \exp\left( -\frac{\beta_j^2}{2 b^2} \right) 
(b^2)^{-(c/2 + 1)} \exp\left( -\frac{d}{2 b^2} \right) \\
&\propto (b^2)^{-p/2} \exp\left( -\sum_{j=1}^p \frac{\beta_j^2}{2 b^2} \right) 
(b^2)^{-(c/2 + 1)} \exp\left( -\frac{d}{2 b^2} \right) \\
&\propto (b^2)^{-(p + c)/2 - 1} 
\exp\left( -\frac{1}{2 b^2} \left( \sum_{j=1}^p \beta_j^2 + d \right) \right)
\end{align*}
Hence,
\[
    b^2 \mid \text{rest} \sim \text{Inverse-Gamma}\left(\frac{c + p}{2}, \frac{d + \sum_{j=1}^p \beta_j^2}{2}\right).
    \]
\end{itemize}
\subsection{BHS}
\subsubsection{Hierarchical Model Specification}
\begin{equation*}
\begin{aligned}
y_i \mid \beta_0,\boldsymbol{x_i}, \boldsymbol{\beta} &\sim \text{N}(\beta_0+\boldsymbol{x_i}^\top\boldsymbol{\beta}, \sigma^2), \\
\beta_0 &\sim \text{N}(0,\sigma^2_{\beta_0}),\\
\beta_j \mid s_j, \lambda^2, \sigma^2 &\sim \text{N}(0, \sigma^2\lambda^2 s_j^2),\\
\sigma^2&\sim \text{Inverse-Gamma}(e,f),\\
s_j^2 \mid \nu_j &\sim \text{Inverse-Gamma}(1/2, 1/\nu_j), \\
\lambda^2 \mid \xi_1 &\sim \text{Inverse-Gamma}(1/2, 1/\xi_1), \\
\nu_j &\sim \text{Inverse-Gamma}(1/2, 1),\\
\xi_1 &\sim \text{Inverse-Gamma}(1/2, 1).\\
\end{aligned}
\end{equation*}
\subsubsection{Gibbs Sampler}
\begin{itemize}
\item The full conditional distribution of $\beta_0$,
\begin{align*}
\pi(\beta_0 \mid \text{rest}) 
&\propto \exp\left( -\frac{1}{2\sigma^2} \sum_{i=1}^n (y_i - \beta_0 - \boldsymbol{x}_i^\top \boldsymbol{\beta})^2 - \frac{1}{2\sigma_{\beta_0}^2} \beta_0^2 \right) \\
&\propto \exp\left( -\frac{1}{2\sigma^2} \sum_{i=1}^n \left[ -2 y_i \beta_0 + 2 \beta_0 \boldsymbol{x}_i^\top \boldsymbol{\beta} + \beta_0^2 \right] - \frac{1}{2\sigma_{\beta_0}^2} \beta_0^2 \right) \\
&\propto \exp\left( \beta_0 \sum_{i=1}^n \frac{y_i - \boldsymbol{x}_i^\top \boldsymbol{\beta}}{\sigma^2} - \frac{1}{2} \beta_0^2 \left( \frac{n}{\sigma^2} + \frac{1}{\sigma_{\beta_0}^2} \right) \right) \\
&\propto \exp\left( -\frac{1}{2} \left[ \left( \frac{n}{\sigma^2} + \frac{1}{\sigma_{\beta_0}^2} \right) \beta_0^2 - 2 \beta_0 \sum_{i=1}^n \frac{y_i - \boldsymbol{x}_i^\top \boldsymbol{\beta}}{\sigma^2} \right] \right)
\end{align*}
Hence,
\[
\beta_0 \mid \text{rest} \sim \text{N}(\mu_{\beta_0}, \Sigma_{\beta_0}^2), \quad
\]
where
\[
\Sigma_{\beta_0}^{-2} = \frac{n}{\sigma^2} + \frac{1}{\sigma_{\beta_0}^2}, \quad
\mu_{\beta_0} = \Sigma_{\beta_0}^2 \frac{1}{\sigma^2} \sum_{i=1}^n (y_i - \boldsymbol{x}_i^\top \boldsymbol{\beta})
\]
\item The full conditional distribution of $\beta_j$,
\begin{align*}
\pi(\beta_j \mid \text{rest}) 
&\propto \exp\left( -\frac{1}{2\sigma^2} \sum_{i=1}^n (y_i - \beta_0 - \sum_{k \ne j} x_{ik} \beta_k - x_{ij} \beta_j)^2 
- \frac{\beta_j^2}{2 \sigma^2 \lambda^2 s_j^2}\right) \\
&\propto \exp\left( -\sum_{i=1}^n \frac{(y_i - \beta_0 - \sum_{k \ne j} x_{ik} \beta_k)^2 - 2 x_{ij} (y_i - \beta_0 - \sum_{k \ne j} x_{ik} \beta_k) \beta_j + x_{ij}^2 \beta_j^2}{2 \sigma^2} \right. \\
&\qquad \left. - \frac{\beta_j^2}{2 \sigma^2 \lambda^2 s_j^2} \right) \\
&\propto \exp\left( \beta_j \sum_{i=1}^n \frac{x_{ij}(y_i - \beta_0 - \sum_{k \ne j} x_{ik} \beta_k)}{\sigma^2} 
- \frac{1}{2} \beta_j^2 \left( \sum_{i=1}^n \frac{x_{ij}^2}{\sigma^2} + \frac{1}{\sigma^2 \lambda^2 s_j^2} \right) \right) \\
&\propto \exp\left( -\frac{1}{2} \left( \left( \sum_{i=1}^n \frac{x_{ij}^2}{\sigma^2} + \frac{1}{\sigma^2 \lambda^2 s_j^2} \right) \beta_j^2 
- 2 \beta_j \sum_{i=1}^n \frac{x_{ij}(y_i - \beta_0 - \sum_{k \ne j} x_{ik} \beta_k)}{\sigma^2} \right) \right)
\end{align*}

Therefore,
\[
\beta_j \mid \text{rest} \sim \text{N}(\mu_{\beta_j}, \sigma_{\beta_j}^2), \quad
\]
where
\[
\sigma_{\beta_j}^{-2} = \frac{1}{\sigma^2} \sum_{i=1}^n x_{ij}^2 + \left( \frac{1}{\sigma^2 \lambda^2 s_j^2} \right), \quad
\mu_{\beta_j} = \sigma_{\beta_j}^2 \left( \frac{1}{\sigma^2} \sum_{i=1}^n x_{ij}(y_i - \beta_0 - \sum_{k \ne j} x_{ik} \beta_k) \right)
\]
\item The full conditional distribution of $\sigma^2$,
\begin{align*}
\pi(\sigma^2 \mid \text{rest}) 
&\propto (\sigma^2)^{-n/2} \exp\left( -\frac{1}{2\sigma^2} \sum_{i=1}^n (y_i - \beta_0 - \boldsymbol{x}_i^\top \boldsymbol{\beta})^2 \right)
(\sigma^2)^{-p/2} \exp\left( -\sum_{j=1}^p \frac{\beta_j^2}{2 \sigma^2 \lambda^2 s_j^2} \right)\\
&(\sigma^2)^{-e-1}\exp(-\frac{f}{\sigma^2}) \\
&\propto (\sigma^2)^{-(n + p)/2-e-1} 
\exp\left( -\frac{1}{2\sigma^2} \left[ 2f+\sum_{i=1}^n (y_i - \beta_0 - \boldsymbol{x}_i^\top \boldsymbol{\beta})^2 + \sum_{j=1}^p \frac{\beta_j^2}{\lambda^2 s_j^2} \right] \right)
\end{align*}
Hence,
\[
\sigma^2 \mid \text{rest} \sim \text{Inverse-Gamma}\left(
e+\frac{n + p}{2}, \,
f+\frac{1}{2} \sum_{i=1}^n (y_i - \beta_0 - \boldsymbol{x}_i^\top \boldsymbol{\beta})^2
+ \frac{1}{2} \sum_{j=1}^p \frac{\beta_j^2}{\lambda^2 s_j^2}
\right)
\]
\item The full conditional distribution of $s^2_j$,
\begin{align*}
\pi(s_j^2 \mid \text{rest}) 
&\propto (2\pi \sigma^2 \lambda^2 s_j^2)^{-1/2} \exp\left( -\frac{\beta_j^2}{2 \sigma^2 \lambda^2 s_j^2} \right)
(s_j^2)^{-3/2} \exp\left( -\frac{1}{\nu_j s_j^2} \right) \\
&\propto (s_j^2)^{-1/2} \exp\left( -\frac{\beta_j^2}{2 \sigma^2 \lambda^2 s_j^2} \right)
(s_j^2)^{-3/2} \exp\left( -\frac{1}{\nu_j s_j^2} \right) \\
&\propto (s_j^2)^{-2} \exp\left( -\left( \frac{\beta_j^2}{2 \sigma^2 \lambda^2} + \frac{1}{\nu_j} \right) \frac{1}{s_j^2} \right)
\end{align*}
Therefore,
\[
s_j^2 \mid \text{rest} \sim \text{Inverse-Gamma}\left(1, \frac{\beta_j^2}{2 \sigma^2 \lambda^2} + \frac{1}{\nu_j} \right)
\]
\item The full condtional distribution of $\nu_j$,
\begin{equation*}
\begin{aligned}
\pi(\nu_j \mid \text{rest}) 
&\propto \nu_j^{-1/2}(s_j^2)^{-3/2} \exp\left( -\frac{1}{\nu_j s_j^2} \right) 
\nu_j^{-3/2} \exp\left( -\frac{1}{\nu_j} \right) \\
&\propto \nu_j^{-2} \exp\left( -\frac{1}{\nu_j} \left( \frac{1}{s_j^2} + 1 \right) \right)
\end{aligned}
\end{equation*}
Therefore,
\[
\nu_j \mid \text{rest} \sim \text{Inverse-Gamma}\left(1, 1 + \frac{1}{s_j^2} \right)
\]
\item The full conditional distribution of $\lambda^2$,
\begin{align*}
\pi(\lambda^2 \mid \text{rest}) 
&\propto (\lambda^2)^{-p/2} \exp\left( -\sum_{j=1}^p \frac{\beta_j^2}{2 \sigma^2 \lambda^2 s_j^2} \right)
(\lambda^2)^{-3/2} \exp\left( -\frac{1}{\xi_1 \lambda^2} \right) \\
&\propto (\lambda^2)^{-(p + 3)/2} 
\exp\left( -\frac{1}{\lambda^2} \left( \sum_{j=1}^p \frac{\beta_j^2}{2 \sigma^2 s_j^2} + \frac{1}{\xi_1} \right) \right)
\end{align*}
Hence,
\[
\lambda^2 \mid \text{rest} \sim \text{Inverse-Gamma}\left(
\frac{p + 1}{2}, \frac{1}{\xi_1} + \sum_{j=1}^p \frac{\beta_j^2}{2 \sigma^2 s_j^2}
\right)
\]
\item The full conditional distribution of $\xi_1$,
\begin{align*}
\pi(\xi_1 \mid \text{rest}) 
&\propto \xi_1^{-1/2}(\lambda^2)^{-3/2} \exp\left( -\frac{1}{\xi_1 \lambda^2} \right) 
\xi_1^{-3/2} \exp\left( -\frac{1}{\xi_1} \right) \\
&\propto \xi_1^{-2} \exp\left( -\frac{1}{\xi_1} \left( \frac{1}{\lambda^2} + 1 \right) \right)
\end{align*}
Therefore,
\[
    \xi_1 \mid \text{rest} \sim \text{Inverse-Gamma}\left(1, 1 + \frac{1}{\lambda^2}\right).
    \]
\end{itemize}

\subsection{BHS+}
\subsubsection{Hierarchical Model Specification}
\begin{equation*}
\begin{aligned}
y_i \mid \beta_0,\boldsymbol{x_i}, \boldsymbol{\beta}, \tilde{v}_i &\sim \text{N}(\beta_0+\boldsymbol{x_i}^\top\boldsymbol{\beta}, \sigma^2), \\
\beta_0 &\sim \text{N}(0, \sigma^2_{\beta_0}), \\
\beta_j \mid s_j^2, \lambda^2, \sigma^2&\sim \text{N}(0, \sigma^2\lambda^2 s_j^2),\\
\sigma^2&\sim \text{Inverse-Gamma}(e,f),\\
s_j^2 \mid \nu_j &\sim \text{Inverse-Gamma}(1/2, 1/\nu_j), \\
\nu_j \mid \phi^2_j &\sim \text{Inverse-Gamma}(1/2, 1/\phi^2_j),\\
\phi^2_j \mid \zeta_j &\sim \text{Inverse-Gamma}(1/2, 1/\zeta_j),\\
\zeta_j &\sim \text{Inverse-Gamma}(1/2, 1),\\
\lambda^2 \mid \xi_1 &\sim \text{Inverse-Gamma}(1/2, 1/\xi_1), \\
\xi_1 &\sim \text{Inverse-Gamma}(1/2, 1).\\
\end{aligned}
\end{equation*}
\subsubsection{Gibbs Sampler}
\begin{itemize}
\item The full conditional distribution of $\beta_0$,
\begin{align*}
\pi(\beta_0 \mid \text{rest}) 
&\propto \exp\left( -\frac{1}{2\sigma^2} \sum_{i=1}^n (y_i - \beta_0 - \boldsymbol{x}_i^\top \boldsymbol{\beta})^2 - \frac{1}{2\sigma_{\beta_0}^2} \beta_0^2 \right) \\
&\propto \exp\left( -\frac{1}{2\sigma^2} \sum_{i=1}^n \left[ -2 y_i \beta_0 + 2 \beta_0 \boldsymbol{x}_i^\top \boldsymbol{\beta} + \beta_0^2 \right] - \frac{1}{2\sigma_{\beta_0}^2} \beta_0^2 \right) \\
&\propto \exp\left( \beta_0 \sum_{i=1}^n \frac{y_i - \boldsymbol{x}_i^\top \boldsymbol{\beta}}{\sigma^2} - \frac{1}{2} \beta_0^2 \left( \frac{n}{\sigma^2} + \frac{1}{\sigma_{\beta_0}^2} \right) \right) \\
&\propto \exp\left( -\frac{1}{2} \left[ \left( \frac{n}{\sigma^2} + \frac{1}{\sigma_{\beta_0}^2} \right) \beta_0^2 - 2 \beta_0 \sum_{i=1}^n \frac{y_i - \boldsymbol{x}_i^\top \boldsymbol{\beta}}{\sigma^2} \right] \right)
\end{align*}
Hence,
\[
\beta_0 \mid \text{rest} \sim \text{N}(\mu_{\beta_0}, \Sigma_{\beta_0}^2), \quad
\]
where
\[
\Sigma_{\beta_0}^{-2} = \frac{n}{\sigma^2} + \frac{1}{\sigma_{\beta_0}^2}, \quad
\mu_{\beta_0} = \Sigma_{\beta_0}^2 \frac{1}{\sigma^2} \sum_{i=1}^n (y_i - \boldsymbol{x}_i^\top \boldsymbol{\beta})
\]
\item The full conditional distribution of $\beta_j$,
\begin{align*}
\pi(\beta_j \mid \text{rest}) 
&\propto \exp\left( -\frac{1}{2\sigma^2} \sum_{i=1}^n (y_i - \beta_0 - \sum_{k \ne j} x_{ik} \beta_k - x_{ij} \beta_j)^2 
- \frac{\beta_j^2}{2 \sigma^2 \lambda^2 s_j^2}\right) \\
&\propto \exp\left( -\sum_{i=1}^n \frac{(y_i - \beta_0 - \sum_{k \ne j} x_{ik} \beta_k)^2 - 2 x_{ij} (y_i - \beta_0 - \sum_{k \ne j} x_{ik} \beta_k) \beta_j + x_{ij}^2 \beta_j^2}{2 \sigma^2} \right. \\
&\qquad \left. - \frac{\beta_j^2}{2 \sigma^2 \lambda^2 s_j^2} \right) \\
&\propto \exp\left( \beta_j \sum_{i=1}^n \frac{x_{ij}(y_i - \beta_0 - \sum_{k \ne j} x_{ik} \beta_k)}{\sigma^2} 
- \frac{1}{2} \beta_j^2 \left( \sum_{i=1}^n \frac{x_{ij}^2}{\sigma^2} + \frac{1}{\sigma^2 \lambda^2 s_j^2} \right) \right) \\
&\propto \exp\left( -\frac{1}{2} \left( \left( \sum_{i=1}^n \frac{x_{ij}^2}{\sigma^2} + \frac{1}{\sigma^2 \lambda^2 s_j^2} \right) \beta_j^2 
- 2 \beta_j \sum_{i=1}^n \frac{x_{ij}(y_i - \beta_0 - \sum_{k \ne j} x_{ik} \beta_k)}{\sigma^2} \right) \right)
\end{align*}

Therefore,
\[
\beta_j \mid \text{rest} \sim \text{N}(\mu_{\beta_j}, \sigma_{\beta_j}^2), \quad
\]
where
\[
\sigma_{\beta_j}^{-2} = \frac{1}{\sigma^2} \sum_{i=1}^n x_{ij}^2 + \left( \frac{1}{\sigma^2 \lambda^2 s_j^2} \right), \quad
\mu_{\beta_j} = \sigma_{\beta_j}^2 \left( \frac{1}{\sigma^2} \sum_{i=1}^n x_{ij}(y_i - \beta_0 - \sum_{k \ne j} x_{ik} \beta_k) \right)
\]
\item The full conditional distribution of $\sigma^2$,
\begin{align*}
\pi(\sigma^2 \mid \text{rest}) 
&\propto (\sigma^2)^{-n/2} \exp\left( -\frac{1}{2\sigma^2} \sum_{i=1}^n (y_i - \beta_0 - \boldsymbol{x}_i^\top \boldsymbol{\beta})^2 \right)
(\sigma^2)^{-p/2} \exp\left( -\sum_{j=1}^p \frac{\beta_j^2}{2 \sigma^2 \lambda^2 s_j^2} \right)\\
&(\sigma^2)^{-e}\exp(-\frac{f}{\sigma^2}) \\
&\propto (\sigma^2)^{-(n + p)/2-e-1} 
\exp\left( -\frac{1}{2\sigma^2} \left[ 2f+\sum_{i=1}^n (y_i - \beta_0 - \boldsymbol{x}_i^\top \boldsymbol{\beta})^2 + \sum_{j=1}^p \frac{\beta_j^2}{\lambda^2 s_j^2} \right] \right)
\end{align*}
Hence,
\[
\sigma^2 \mid \text{rest} \sim \text{Inverse-Gamma}\left(
e+\frac{n + p}{2}, \,
f+\frac{1}{2} \sum_{i=1}^n (y_i - \beta_0 - \boldsymbol{x}_i^\top \boldsymbol{\beta})^2 + \frac{1}{2} \sum_{j=1}^p \frac{\beta_j^2}{\lambda^2 s_j^2}
\right)
\]
\item The full conditional distribution of $s^2_j$,
\begin{align*}
\pi(s_j^2 \mid \text{rest}) 
&\propto (2\pi \sigma^2 \lambda^2 s_j^2)^{-1/2} \exp\left( -\frac{\beta_j^2}{2 \sigma^2 \lambda^2 s_j^2} \right)
(s_j^2)^{-3/2} \exp\left( -\frac{1}{\nu_j s_j^2} \right) \\
&\propto (s_j^2)^{-1/2} \exp\left( -\frac{\beta_j^2}{2 \sigma^2 \lambda^2 s_j^2} \right)
(s_j^2)^{-3/2} \exp\left( -\frac{1}{\nu_j s_j^2} \right) \\
&\propto (s_j^2)^{-2} \exp\left( -\left( \frac{\beta_j^2}{2 \sigma^2 \lambda^2} + \frac{1}{\nu_j} \right) \frac{1}{s_j^2} \right)
\end{align*}
Therefore,
\[
s_j^2 \mid \text{rest} \sim \text{Inverse-Gamma}\left(1, \frac{\beta_j^2}{2 \sigma^2 \lambda^2} + \frac{1}{\nu_j} \right)
\]
\item The full conditional distribution of $\nu_j$,
\begin{align*}
\pi(\nu_j \mid \text{rest}) 
&\propto \nu_j^{-2} \exp\left( -\frac{1}{\nu_j} \left( \frac{1}{s_j^2} + \frac{1}{\phi_j^2} \right) \right)
\end{align*}
Hence,
    \[
    \nu_j \mid \text{rest} \sim \text{Inverse-Gamma}\left(1, \frac{1}{s_j^2}+\frac{1}{\phi_j^2}\right).
    \]
\item The full conditional distribution of $\phi^2_j$,
\begin{align*}
\pi(\phi_j^2 \mid \text{rest}) 
&\propto (\phi_j^2)^{-2} \exp\left( -\frac{1}{\phi_j^2} \left( \frac{1}{\nu_j} + \frac{1}{\zeta_j} \right) \right)
\end{align*}
Hence,
    \[
    \phi^2_j \mid \text{rest} \sim \text{Inverse-Gamma}\left(1, \frac{1}{\nu_j}+\frac{1}{\zeta_j}\right).
    \]
\item The full conditional distribution of $\zeta_j$,
\begin{align*}
\pi(\zeta_j \mid \text{rest}) 
&\propto \zeta_j^{-2} \exp\left( -\frac{1}{\zeta_j} \left( \frac{1}{\phi_j^2} + 1 \right) \right)
\end{align*}
Hence,
    \[
    \zeta_j \mid \text{rest} \sim \text{Inverse-Gamma}\left(1, \frac{1}{\phi^2_j}+1\right).
    \]
\item The full conditional distribution of $\lambda^2$,
\begin{align*}
\pi(\lambda^2 \mid \text{rest}) 
&\propto (\lambda^2)^{-p/2} \exp\left( -\sum_{j=1}^p \frac{\beta_j^2}{2 \sigma^2 \lambda^2 s_j^2} \right)
(\lambda^2)^{-3/2} \exp\left( -\frac{1}{\xi_1 \lambda^2} \right) \\
&\propto (\lambda^2)^{-(p + 3)/2} 
\exp\left( -\frac{1}{\lambda^2} \left( \sum_{j=1}^p \frac{\beta_j^2}{2 \sigma^2 s_j^2} + \frac{1}{\xi_1} \right) \right)
\end{align*}
Hence,
\[
\lambda^2 \mid \text{rest} \sim \text{Inverse-Gamma}\left(
\frac{p + 1}{2}, \frac{1}{\xi_1} + \sum_{j=1}^p \frac{\beta_j^2}{2 \sigma^2 s_j^2}
\right)
\]
\item The full conditional distribution of $\xi_1$,
\begin{align*}
\pi(\xi_1 \mid \text{rest}) 
&\propto \xi_1^{-2} \exp\left( -\frac{1}{\xi_1} \left( \frac{1}{\lambda^2} + 1 \right) \right)
\end{align*}
Hence,
\[
    \xi_1 \mid \text{rest} \sim \text{Inverse-Gamma}\left(1, 1 + \frac{1}{\lambda^2}\right).
    \]
\end{itemize}
\subsection{BRHS}
\subsubsection{Hierarchical Model Specification}
\begin{equation*}
\begin{aligned}
y_i \mid \beta_0,\boldsymbol{x_i}, \boldsymbol{\beta} &\sim \text{N}(\beta_0+\boldsymbol{x_i}^\top\boldsymbol{\beta}, \sigma^2), \\
\beta_0 &\sim \text{N}(0,\sigma^2_{\beta_0}),\\
\beta_j \mid s_j, \lambda^2, \sigma^2, b^2 &\sim \text{N}(0, \sigma^2\lambda^2 s_j^2)\, \text{N}(0, b^2),\\
\sigma^2&\sim\text{Inverse-Gamma}(e,f),\\
s_j^2 \mid \nu_j &\sim \text{Inverse-Gamma}(1/2, 1/\nu_j), \\
\lambda^2 \mid \xi_1 &\sim \text{Inverse-Gamma}(1/2, 1/\xi_1), \\
\nu_j &\sim \text{Inverse-Gamma}(1/2, 1),\\
\xi_1 &\sim \text{Inverse-Gamma}(1/2, 1),\\
b^2 &\sim \text{Inverse-Gamma}(c/2, d/2).
\end{aligned}
\end{equation*}
\subsubsection{Gibbs Sampler}
\begin{itemize}
\item The full conditional distribution of $\beta_0$,
\begin{align*}
\pi(\beta_0 \mid \text{rest}) 
&\propto \exp\left( -\frac{1}{2\sigma^2} \sum_{i=1}^n (y_i - \beta_0 - \boldsymbol{x}_i^\top \boldsymbol{\beta})^2 - \frac{1}{2\sigma_{\beta_0}^2} \beta_0^2 \right) \\
&\propto \exp\left( -\frac{1}{2\sigma^2} \sum_{i=1}^n \left[ -2 y_i \beta_0 + 2 \beta_0 \boldsymbol{x}_i^\top \boldsymbol{\beta} + \beta_0^2 \right] - \frac{1}{2\sigma_{\beta_0}^2} \beta_0^2 \right) \\
&\propto \exp\left( \beta_0 \sum_{i=1}^n \frac{y_i - \boldsymbol{x}_i^\top \boldsymbol{\beta}}{\sigma^2} - \frac{1}{2} \beta_0^2 \left( \frac{n}{\sigma^2} + \frac{1}{\sigma_{\beta_0}^2} \right) \right) \\
&\propto \exp\left( -\frac{1}{2} \left[ \left( \frac{n}{\sigma^2} + \frac{1}{\sigma_{\beta_0}^2} \right) \beta_0^2 - 2 \beta_0 \sum_{i=1}^n \frac{y_i - \boldsymbol{x}_i^\top \boldsymbol{\beta}}{\sigma^2} \right] \right)
\end{align*}
Hence,
\[
\beta_0 \mid \text{rest} \sim \text{N}(\mu_{\beta_0}, \Sigma_{\beta_0}^2), \quad
\]
where
\[
\Sigma_{\beta_0}^{-2} = \frac{n}{\sigma^2} + \frac{1}{\sigma_{\beta_0}^2}, \quad
\mu_{\beta_0} = \Sigma_{\beta_0}^2 \frac{1}{\sigma^2} \sum_{i=1}^n (y_i - \boldsymbol{x}_i^\top \boldsymbol{\beta})
\]
\item The full conditional distribution of $\beta_j$,
\begin{align*}
\pi(\beta_j \mid \text{rest}) 
&\propto \exp\left( -\frac{1}{2\sigma^2} \sum_{i=1}^n (y_i - \beta_0 - \sum_{k \ne j} x_{ik} \beta_k - x_{ij} \beta_j)^2 
- \frac{\beta_j^2}{2 \sigma^2 \lambda^2 s_j^2} - \frac{\beta_j^2}{2 b^2} \right) \\
&\propto \exp\left( -\sum_{i=1}^n \frac{(y_i - \beta_0 - \sum_{k \ne j} x_{ik} \beta_k)^2 - 2 x_{ij} (y_i - \beta_0 - \sum_{k \ne j} x_{ik} \beta_k) \beta_j + x_{ij}^2 \beta_j^2}{2 \sigma^2} \right. \\
&\qquad \left. - \frac{\beta_j^2}{2 \sigma^2 \lambda^2 s_j^2} - \frac{\beta_j^2}{2 b^2} \right) \\
&\propto \exp\left( \beta_j \sum_{i=1}^n \frac{x_{ij}(y_i - \beta_0 - \sum_{k \ne j} x_{ik} \beta_k)}{\sigma^2} 
- \frac{1}{2} \beta_j^2 \left( \sum_{i=1}^n \frac{x_{ij}^2}{\sigma^2} + \frac{1}{\sigma^2 \lambda^2 s_j^2} + \frac{1}{b^2} \right) \right) \\
&\propto \exp\left( -\frac{1}{2} \left( \left( \sum_{i=1}^n \frac{x_{ij}^2}{\sigma^2} + \frac{1}{\sigma^2 \lambda^2 s_j^2} + \frac{1}{b^2} \right) \beta_j^2 
- 2 \beta_j \sum_{i=1}^n \frac{x_{ij}(y_i - \beta_0 - \sum_{k \ne j} x_{ik} \beta_k)}{\sigma^2} \right) \right)
\end{align*}

Therefore,
\[
\beta_j \mid \text{rest} \sim \text{N}(\mu_{\beta_j}, \sigma_{\beta_j}^2), \quad
\]
where
\[
\sigma_{\beta_j}^{-2} = \frac{1}{\sigma^2} \sum_{i=1}^n x_{ij}^2 + \left( \frac{1}{\sigma^2 \lambda^2 s_j^2} + \frac{1}{b^2} \right), \quad
\mu_{\beta_j} = \sigma_{\beta_j}^2 \left( \frac{1}{\sigma^2} \sum_{i=1}^n x_{ij}(y_i - \beta_0 - \sum_{k \ne j} x_{ik} \beta_k) \right)
\]
\item The full conditional distribution of $\sigma^2$,
\begin{align*}
\pi(\sigma^2 \mid \text{rest}) 
&\propto (\sigma^2)^{-n/2} \exp\left( -\frac{1}{2\sigma^2} \sum_{i=1}^n (y_i - \beta_0 - \boldsymbol{x}_i^\top \boldsymbol{\beta})^2 \right)
(\sigma^2)^{-p/2} \exp\left( -\sum_{j=1}^p \frac{\beta_j^2}{2 \sigma^2 \lambda^2 s_j^2} \right)\\
&(\sigma^2)^{-e-1}\text{exp}(-\frac{f}{\sigma^2}) \\
&\propto (\sigma^2)^{-(n + p)/2-e-1} 
\exp\left( -\frac{1}{2\sigma^2} \left[ 2f+\sum_{i=1}^n (y_i - \beta_0 - \boldsymbol{x}_i^\top \boldsymbol{\beta})^2 + \sum_{j=1}^p \frac{\beta_j^2}{\lambda^2 s_j^2} \right] \right)
\end{align*}
Hence,
\[
\sigma^2 \mid \text{rest} \sim \text{Inverse-Gamma}\left(
e+\frac{n + p}{2}, \,
f+\frac{1}{2} \sum_{i=1}^n (y_i - \beta_0 - \boldsymbol{x}_i^\top \boldsymbol{\beta})^2 + \frac{1}{2} \sum_{j=1}^p \frac{\beta_j^2}{\lambda^2 s_j^2}
\right)
\]
\item The full conditional distribution of $s^2_j$,
\begin{align*}
\pi(s_j^2 \mid \text{rest}) 
&\propto (2\pi \sigma^2 \lambda^2 s_j^2)^{-1/2} \exp\left( -\frac{\beta_j^2}{2 \sigma^2 \lambda^2 s_j^2} \right)
(s_j^2)^{-3/2} \exp\left( -\frac{1}{\nu_j s_j^2} \right) \\
&\propto (s_j^2)^{-1/2} \exp\left( -\frac{\beta_j^2}{2 \sigma^2 \lambda^2 s_j^2} \right)
(s_j^2)^{-3/2} \exp\left( -\frac{1}{\nu_j s_j^2} \right) \\
&\propto (s_j^2)^{-2} \exp\left( -\left( \frac{\beta_j^2}{2 \sigma^2 \lambda^2} + \frac{1}{\nu_j} \right) \frac{1}{s_j^2} \right)
\end{align*}
Therefore,
\[
s_j^2 \mid \text{rest} \sim \text{Inverse-Gamma}\left(1, \frac{\beta_j^2}{2 \sigma^2 \lambda^2} + \frac{1}{\nu_j} \right)
\]
\item The full condtional distribution of $\nu_j$,
\begin{equation*}
\begin{aligned}
\pi(\nu_j \mid \text{rest}) 
&\propto  \nu_j^{-1/2}(s_j^2)^{-3/2} \exp\left( -\frac{1}{\nu_j s_j^2} \right) 
\nu_j^{-3/2} \exp\left( -\frac{1}{\nu_j} \right) \\
&\propto \nu_j^{-2} \exp\left( -\frac{1}{\nu_j} \left( \frac{1}{s_j^2} + 1 \right) \right)
\end{aligned}
\end{equation*}
Therefore,
\[
\nu_j \mid \text{rest} \sim \text{Inverse-Gamma}\left(1, 1 + \frac{1}{s_j^2} \right)
\]
\item The full conditional distribution of $\lambda^2$,
\begin{align*}
\pi(\lambda^2 \mid \text{rest}) 
&\propto (\lambda^2)^{-p/2} \exp\left( -\sum_{j=1}^p \frac{\beta_j^2}{2 \sigma^2 \lambda^2 s_j^2} \right)
(\lambda^2)^{-3/2} \exp\left( -\frac{1}{\xi_1 \lambda^2} \right) \\
&\propto (\lambda^2)^{-(p + 3)/2} 
\exp\left( -\frac{1}{\lambda^2} \left( \sum_{j=1}^p \frac{\beta_j^2}{2 \sigma^2 s_j^2} + \frac{1}{\xi_1} \right) \right)
\end{align*}
Hence,
\[
\lambda^2 \mid \text{rest} \sim \text{Inverse-Gamma}\left(
\frac{p + 1}{2}, \frac{1}{\xi_1} + \sum_{j=1}^p \frac{\beta_j^2}{2 \sigma^2 s_j^2}
\right)
\]
\item The full conditional distribution of $\xi_1$,
\begin{align*}
\pi(\xi_1 \mid \text{rest}) 
&\propto \xi_1^{-1/2}(\lambda^2)^{-3/2} \exp\left( -\frac{1}{\xi_1 \lambda^2} \right) 
\xi_1^{-3/2} \exp\left( -\frac{1}{\xi_1} \right) \\
&\propto \xi_1^{-2} \exp\left( -\frac{1}{\xi_1} \left( \frac{1}{\lambda^2} + 1 \right) \right)
\end{align*}
Therefore,
\[
    \xi_1 \mid \text{rest} \sim \text{Inverse-Gamma}\left(1, 1 + \frac{1}{\lambda^2}\right).
    \]
\item The full conditional distribution of $b^2$,
\begin{align*}
\pi(b^2 \mid \text{rest}) 
&\propto \prod_{j=1}^p \frac{1}{\sqrt{2\pi b^2}} \exp\left( -\frac{\beta_j^2}{2 b^2} \right) 
(b^2)^{-(c/2 + 1)} \exp\left( -\frac{d}{2 b^2} \right) \\
&\propto (b^2)^{-p/2} \exp\left( -\sum_{j=1}^p \frac{\beta_j^2}{2 b^2} \right) 
(b^2)^{-(c/2 + 1)} \exp\left( -\frac{d}{2 b^2} \right) \\
&\propto (b^2)^{-(p + c)/2 - 1} 
\exp\left( -\frac{1}{2 b^2} \left( \sum_{j=1}^p \beta_j^2 + d \right) \right)
\end{align*}
Hence,
\[
    b^2 \mid \text{rest} \sim \text{Inverse-Gamma}\left(\frac{c + p}{2}, \frac{d + \sum_{j=1}^p \beta_j^2}{2}\right).
    \]
\end{itemize}
\section{Proofs for propositions}
\subsection{Proposition 1.1}
We obtain a closed-form expression for the posterior mean:
\[
\mu_{\beta_j}
= \left( \frac{ \sum_{i=1}^n \frac{x_{ij}^2}{\tau^{-1} \xi^2 \tilde{v}_i} }
{ \sum_{i=1}^n \frac{x_{ij}^2}{\tau^{-1} \xi^2 \tilde{v}_i} + \frac{1}{\lambda^2 s_j^2} } \right)
\hat{\beta}_j
= (1 - \kappa_j) \hat{\beta}_j
\]
where
\begin{equation*}\label{eq}
\kappa_j = \left( 1 + \lambda^2 s_j^2 \sum_{i=1}^n \frac{x_{ij}^2}{\tau^{-1} \xi^2 \tilde{v}_i} \right)^{-1}.
\end{equation*}

\subsection{Proposition 1.2}
Solving for $s_j$ in terms of $\kappa_j$, we get
\[
s_j^2 = \frac{1 - \kappa_j}{\lambda^2 \kappa_j \sum_{i=1}^n \frac{x_{ij}^2}{\tau^{-1} \xi^2 \tilde{v}_i}},
\quad
s_j = \sqrt{ \frac{1 - \kappa_j}{\lambda^2 \kappa_j \sum_{i=1}^n \frac{x_{ij}^2}{\tau^{-1} \xi^2 \tilde{v}_i}} }.
\]
Using the change-of-variable formula,
\[
p(\kappa_j \mid \lambda) = p(s_j) \left| \frac{ds_j}{d\kappa_j} \right|.
\]
We compute the derivative:
\[
\left| \frac{ds_j}{d\kappa_j} \right|
= \frac{1}{2} \left( \frac{1 - \kappa_j}{\lambda^2 \kappa_j \sum_{i=1}^n \frac{x_{ij}^2}{\tau^{-1} \xi^2 \tilde{v}_i}} \right)^{-1/2} \left| \frac{d}{d\kappa_j} \left( \frac{1 - \kappa_j}{\lambda^2 \kappa_j \sum_{i=1}^n \frac{x_{ij}^2}{\tau^{-1} \xi^2 \tilde{v}_i}} \right) \right|.
\]
The inner derivative is
\[
\frac{d}{d\kappa_j} \left( \frac{1 - \kappa_j}{\kappa_j} \right)
= \frac{-\kappa_j - (1 - \kappa_j)}{\kappa_j^2}
= -\frac{1}{\kappa_j^2}.
\]
So,
\[
\left| \frac{ds_j}{d\kappa_j} \right|
=
\frac{1}{2}\sqrt{\frac{ \lambda^2 \kappa_j \sum_{i=1}^n\frac{x_{ij}^2}{\tau^{-1} \xi^2 \tilde{v}_i}}{1 - \kappa_j}} \frac{1}{\kappa_j^2 \lambda^2 \sum_{i=1}^n\frac{x_{ij}^2}{\tau^{-1} \xi^2 \tilde{v}_i}}.
\]
Now substitute into the density:
\[
p(\kappa_j \mid \lambda,\tau^{-1},\tilde{v}_i)
= \frac{2}{\pi \left(1 + \frac{1 - \kappa_j}{\lambda^2 \kappa_j \sum_{i=1}^n \frac{x_{ij}^2}{\tau^{-1} \xi^2 \tilde{v}_i}} \right)}\frac{1}{2}\sqrt{\frac{ \lambda^2 \kappa_j \sum_{i=1}^n\frac{x_{ij}^2}{\tau^{-1} \xi^2 \tilde{v}_i}}{1 - \kappa_j}} \frac{1}{\kappa_j^2 \lambda^2 \sum_{i=1}^n \frac{x_{ij}^2}{\tau^{-1} \xi^2 \tilde{v}_i}}.
\]
Simplify:
\[
p(\kappa_j \mid \lambda,\tau^{-1},\tilde{v}_i)
=
\frac{1}{\pi} \frac{ \lambda \sqrt{ \sum_{i=1}^n \dfrac{x_{ij}^2}{\tau^{-1} \xi^2 \tilde{v}_i} } }{ \left( \lambda^2 \sum_{i=1}^n \dfrac{x_{ij}^2}{\tau^{-1} \xi^2 \tilde{v}_i} - 1 \right) \kappa_j + 1 }\frac{1}{ \sqrt{ \kappa_j (1 - \kappa_j) } },
\quad \kappa_j \in (0, 1).
\]
\subsection{Proposition 1.3}
Solving for $s_j$ in terms of $\kappa_j$, we get
\[
s_j^2 = \frac{1 - \kappa_j}{\lambda^2 \kappa_j \sum_{i=1}^n \frac{x_{ij}^2}{\tau^{-1} \xi^2 \tilde{v}_i}}, \quad s_j = \sqrt{ \frac{1 - \kappa_j}{\lambda^2 \kappa_j \sum_{i=1}^n \frac{x_{ij}^2}{\tau^{-1} \xi^2 \tilde{v}_i}} }.
\]
We now use the change-of-variable formula to derive the marginal density of $\kappa_j$:
\[
p(\kappa_j \mid \lambda) = p(s_j) \left| \frac{ds_j}{d\kappa_j} \right|.
\]
Substitute \(s_j\) into the density:
\[
p(s_j) =
\frac{4}{\pi^2} 
\frac{ \log \left( \sqrt{ \frac{1 - \kappa_j}{\lambda^2 \kappa_j \sum_{i=1}^n \frac{x_{ij}^2}{\tau^{-1} \xi^2 \tilde{v}_i}} } \right) }
{ \frac{1 - \kappa_j}{\lambda^2 \kappa_j \sum_{i=1}^n\frac{x_{ij}^2}{\tau^{-1} \xi^2 \tilde{v}_i}} - 1 }.
\]
Next, compute the Jacobian:
\[
\left| \frac{ds_j}{d\kappa_j} \right|
=
\frac{1}{2}
\left( \frac{1 - \kappa_j}{\lambda^2 \kappa_j \sum_{i=1}^n \frac{x_{ij}^2}{\tau^{-1} \xi^2 \tilde{v}_i}} \right)^{-1/2}
\left| \frac{d}{d\kappa_j} \left( \frac{1 - \kappa_j}{\lambda^2 \kappa_j \sum_{i=1}^n \frac{x_{ij}^2}{\tau^{-1} \xi^2 \tilde{v}_i}} \right) \right|.
\]
The inner derivative is:
\[
\frac{d}{d\kappa_j} \left( \frac{1 - \kappa_j}{\kappa_j} \right)
= - \frac{1}{\kappa_j^2}.
\]
Putting it all together:
\[
\left| \frac{ds_j}{d\kappa_j} \right|
=
\frac{1}{2}\sqrt{\frac{ \lambda^2 \kappa_j \sum_{i=1}^n\frac{x_{ij}^2}{\tau^{-1} \xi^2 \tilde{v}_i}}{1 - \kappa_j}} \frac{1}{\kappa_j^2 \lambda^2 \sum_{i=1}^n\frac{x_{ij}^2}{\tau^{-1} \xi^2 \tilde{v}_i}}.
\]
Now plug back into the change-of-variable formula:
\begin{equation*}
p(\kappa_j \mid \lambda,\tau^{-1},\tilde{v}_i)
=
\frac{2}{\pi^2}
\frac{
\log \left( \sqrt{\frac{1 - \kappa_j}{\kappa_j \lambda^2 \sum_{i=1}^n \frac{x_{ij}^2}{\tau^{-1} \xi^2 \tilde{v}_i}} }\right)
}{
1 - \kappa_j \left(1 + \lambda^2 \sum_{i=1}^n \frac{x_{ij}^2}{\tau^{-1} \xi^2 \tilde{v}_i} \right)
}
\frac{\lambda \sqrt{\sum_{i=1}^n \frac{x_{ij}^2}{\tau^{-1} \xi^2 \tilde{v}_i}}}{
\sqrt{\kappa_j(1 - \kappa_j) }
}.
\end{equation*}

\end{document}